\documentclass{article}
\PassOptionsToPackage{numbers, compress}{natbib}
\usepackage[preprint]{neurips_2026}

\usepackage[utf8]{inputenc}
\usepackage[T1]{fontenc}
\usepackage[colorlinks=true,linkcolor=blue!70!black,citecolor=blue!70!black,urlcolor=blue!70!black]{hyperref}
\hypersetup{
  pdftitle={Variance-Corrected Multi-Asset Equity Simulation with Hybrid Hidden Markov Marginals},
  pdfauthor={Abdulrahman Alswaidan and Jeffrey D. Varner}
}
\usepackage{url}
\usepackage{booktabs}
\usepackage{amsfonts}
\usepackage{nicefrac}
\usepackage{microtype}
\usepackage{xcolor}
\usepackage{graphicx}
\usepackage{subcaption}
\usepackage{amsmath}
\usepackage{amssymb}
\usepackage{amsthm}
\usepackage{algorithm}
\usepackage{algorithmic}
\usepackage{bm}
\usepackage{multirow}
\usepackage{lineno}
\usepackage{needspace}
\usepackage{tikz}
\usetikzlibrary{arrows.meta, decorations.pathreplacing, positioning}

\usepackage[section]{placeins} % keep arXiv floats within their source section
\usepackage{flafter} % never place a float before its source location

\theoremstyle{definition}

\theoremstyle{remark}

\newcommand{\R}{\mathbb{R}}
\newcommand{\E}{\mathbb{E}}
\newcommand{\Var}{\operatorname{Var}}
\newcommand{\Cov}{\operatorname{Cov}}
\newcommand{\sign}{\operatorname{sign}}
\newcommand{\eps}{\varepsilon}
\newcommand{\tg}{\tilde{g}}
\newcommand{\gen}{\mathrm{gen}}
\newcommand{\real}{\mathrm{real}}
\newcommand{\eff}{\mathrm{eff}}

\title{Variance-Corrected Multi-Asset Equity Simulation with Hybrid Hidden Markov Marginals}

\author{%
  Abdulrahman Alswaidan \\
  Robert Frederick Smith School of Chemical and Biomolecular Engineering \\
  Cornell University, Ithaca, NY 14850 \\
  \texttt{aa2725@cornell.edu} \\
  \And
  Jeffrey D.~Varner\thanks{Corresponding author.} \\
  Robert Frederick Smith School of Chemical and Biomolecular Engineering \\
  Cornell University, Ithaca, NY 14850 \\
  \texttt{jdv27@cornell.edu} \\
}

\begin{document}
\raggedbottom
\clubpenalty=10000
\widowpenalty=10000
\displaywidowpenalty=10000

\maketitle

\begin{abstract}
Synthetic multi-asset equity data must reproduce each asset's return
distribution and its relationship with the market. Reusing a generator
fitted to full asset returns creates a problem: adding its draws to a
market factor counts market variance twice. We derived a correction
that centers and rescales each draw over a fixed horizon before adding
the market factor,
allowing reuse without fitting a second generator to regression
residuals. We tested the correction on $423$ non-market assets in a
$424$-asset United States equity and exchange-traded-fund universe,
using hidden Markov generators with heavy-tailed emissions. The
corrected paths retained heavy tails and recovered the calibrated
market loadings with low error. On $416$ complete asset histories held
out from $2025$, the correction improved the mean Kolmogorov--Smirnov
pass rate over naive composition and brought the median ratio of
synthetic to observed variance close to one. Its one-day left-tail
$99\%$ Value-at-Risk exceedance rate, using thresholds pooled across
separate simulated paths, was $1.12\%$, compared with $1.16\%$ for
the residual-fit hidden Markov comparator and a nominal rate of
$1\%$. We also
tested a jump-duration mechanism that extended visits to extreme-return
states and improved volatility clustering for the broad-market
exchange-traded fund with ticker SPY. The multi-asset correction
remained effective with jumps enabled, but transferring the SPY jump
settings improved temporal fit in training and worsened it in the
$2025$ holdout. The method provides a way to reuse fitted asset
generators for daily-fit comparisons. Its zero-sum residual constraint
removes terminal residual uncertainty, and unmodeled dependence among
asset-specific residuals can lead to overstated rebalancing returns. Code,
cached inputs, result summaries, and instructions for fitting the
per-asset models accompany the paper.

\end{abstract}

\vspace{0.5em}

\section{Introduction}\label{sec:introduction}
Researchers and practitioners use synthetic equity paths to test risk
models, evaluate portfolio strategies, and examine market scenarios
beyond those observed in a historical sample~\cite{assefaEtAl2020,
jordonEtAl2022}. A multi-asset simulator must represent both the
behavior of each asset and the relationships among assets. Factor
models link many assets through a shared market path, a time series of
market returns~\cite{sharpe1963, rossAPT1976, famaFrench1993}. The
simplest of these, the single-index model, separates each asset's
return into an intercept, a market component scaled by the asset's
fitted loading, and a zero-mean asset-specific residual. A synthetic
asset path can then be composed by adding a simulated residual to the
intercept and market component. If single-asset generators have
already been fitted to each asset's full returns, their paths might
be reused as residuals. Each generator, however, already represents
its asset's total variance, including the part driven by the market.
Adding its paths to the market component therefore counts market
variance a second time. Subtracting each path's mean removes its mean
shift but leaves this excess variance. Fitting a second generator to
each asset's regression residuals avoids the problem but repeats the
fitting for every asset. Reusing the existing generators would avoid
that second fit, provided their paths could be transformed to supply
only the residual contribution. The question is whether such a
transformation can control the mean and variance of the composed paths
while retaining the generator's heavy tails and volatility clustering.

The transformation is useful only if each generator reproduces the
behavior of its own asset. We illustrate the features a generator must
capture with the broad-market exchange-traded fund with ticker SPY,
which also supplied the market path in our multi-asset experiments.
Its $2014$--$2024$ returns show heavy tails and volatility clustering
(Fig.~\ref{fig:empirical_motivation}). Most returns lie close to zero,
but extreme values occur much more often than a Gaussian model
predicts (Fig.~\ref{fig:empirical_motivation}a,b), a widely documented
feature of financial returns~\cite{mandelbrot1963, cont2001}. Raw
returns show little correlation across days, while absolute returns
remain positively correlated across many lags
(Fig.~\ref{fig:empirical_motivation}c,d). Large and small absolute
returns therefore cluster in time, a property that synthetic
financial data often miss even when their return histogram appears
realistic~\cite{stengerEtAl2024}. To represent heavy tails and
volatility clustering, a hidden Markov model (HMM) draws returns from
state-dependent distributions
and uses transitions between states to represent temporal
behavior~\cite{rabinerJuang1986, rydenTerasvirtaAsbrink1998}. Each state
in the model used here has a heavy-tailed Student-$t$ return
distribution, and the next state depends only on the current state,
not on earlier returns. Fitted to SPY, the model leaves states with
the most negative or positive mean returns quickly, and the
autocorrelation of its absolute returns decays faster than in the
data. A duration
mechanism that occasionally keeps drawing from extreme-return states
for consecutive days can slow that decay~\cite{bullaBulla2006}.
Adding the market component, however, changes the distribution and
temporal behavior of the composed returns. An improvement in a
single-asset generator must therefore be checked again after
composition.

% Main figure fig:empirical_motivation is collected after the references.

In this study, we developed a variance correction to reuse full-return
generators in a single-index model without counting market variance
twice. The correction centered and rescaled each simulated path to
remove the mean shift and account for the variance supplied by the
market component
(Fig.~\ref{fig:preservation}). We applied it to heavy-tailed hidden
Markov generators for the $423$ non-market assets in a $424$-asset
universe, using $2014$--$2024$ for fitting and $2025$ as a holdout.
We compared the correction with naive composition, which centered the
paths without rescaling, with Gaussian residuals, and with generators
fitted directly to regression residuals. The correction improved
agreement with observed return distributions over naive composition
while retaining heavy tails in training (Table~\ref{tab:aggregate}).
With all fits held fixed, the correction also improved distributional
fit over naive composition and
brought the median ratio of synthetic to observed variance close to
one on the $416$ assets with complete holdout histories
(Table~\ref{tab:oos_composition}). The residual-fit hidden Markov
generator and the correction gave similar agreement with observed
return distributions in both periods. Without the duration mechanism, the correction left absolute-return
autocorrelation error essentially unchanged from naive composition (Table~\ref{tab:jump_ablation}).
Transferring SPY's duration settings to the composed paths improved
their autocorrelation agreement in training but worsened it in the
holdout (Table~\ref{tab:jump_ablation}).
The correction imposed a zero residual sum over each simulated path,
removing cumulative residual uncertainty at the path endpoint. Neither reuse nor residual
fitting modeled dependence among different assets' residuals, which
portfolio evaluation also requires~\cite{embrechtsMcNeilStraumann2002,
boucheyNemtchinovWong2015}. These results support reuse as a practical
option for daily-return comparisons when full-return generators
already exist.

\section{Related Work}\label{sec:related-work}
Single-asset return models treat changing volatility, market regimes, and
extreme moves in different ways. Autoregressive conditional heteroskedasticity
(ARCH) and generalized autoregressive conditional heteroskedasticity (GARCH)
models allow conditional
variance to change over time~\cite{engle1982,bollerslev1986}. Even with
Gaussian innovations, the changing variance can produce an unconditional
return distribution with excess kurtosis. Heavy-tailed innovations can improve
the tail fit when Gaussian GARCH understates the frequency of extreme returns.
Stochastic-volatility models make volatility a latent
process~\cite{steinStein1991,heston1993,
kimShephardChib1998}. Jump-diffusion models add discontinuous price
moves~\cite{merton1976,kou2002}, but a jump component alone does not explain
persistent volatility. Markov-switching ARCH and GARCH models combine discrete
regimes with changing conditional variance~\cite{hamiltonSusmel1994,
klaassen2002}.
Hidden-state models represent returns with a small number of market regimes.
In a hidden Markov model (HMM), each latent state has a separate return distribution.
Gaussian-emission HMMs can reproduce heavy tails
and weak raw-return autocorrelation, but may still miss volatility
clustering~\cite{hamilton1989,rydenTerasvirtaAsbrink1998}. Hidden semi-Markov
models allow more flexible state durations and can improve volatility
persistence~\cite{bullaBulla2006}. An HMM does not create a jump process by
itself. Rare extreme moves must come from the state-dependent return
distributions or from a transition rule designed for such events.
Deep generative models learn return distributions and temporal features
without specifying a small set of market regimes or a parametric return
process. Generative adversarial networks (GANs) are one class of deep
generative model. Quant GANs have
reproduced several financial time-series features, including volatility
clustering and leverage effects~\cite{wieseEtAl2020}.
Sig-Wasserstein GANs use path signatures to learn features that depend on the
order of observations~\cite{niEtAl2021}. In contrast, evaluations of GAN
paths, including paths from the time-series generative adversarial network
(TimeGAN), have found too little autocorrelation in absolute
returns~\cite{takahashiEtAl2019,yoonEtAl2019timegan,kwonLee2024}. The
conflicting results show that flexible architectures do not guarantee a
match to financial stylized facts. Model design, training, and evaluation all
matter~\cite{assefaEtAl2020,stengerEtAl2024}.

Multi-asset generation also requires a model of cross-asset dependence.
Multivariate GARCH models estimate time-varying covariance matrices. The
dynamic conditional correlation model reduces the number of parameters by
separating individual volatility models from correlation
dynamics~\cite{engle2002dcc,bauwensLaurentRombouts2006}. Copulas separate each
asset's return distribution from the dependence model. Dynamic copulas allow
cross-asset dependence to change over time, while vine copulas provide flexible
high-dimensional constructions~\cite{patton2006,
aasCzadoFrigessiBakken2009}. Factor models reduce dimension by linking each
asset to a small number of shared
factors~\cite{sharpe1963,rossAPT1976,famaFrench1993}. Shared-state HMMs and
multivariate GANs can also generate joint market
moves~\cite{diasVermuntRamos2010,rizzatoEtAl2023}. A standard factor model can
resample historical residuals or fit a separate residual generator for every
asset. This paper addresses a narrower problem. We start with single-asset
generators already fitted to full returns and ask whether they can be reused in
a factor model.
The relevant comparison is with methods fitted directly to regression
residuals: those methods remove the market component before fitting,
whereas reuse adjusts an existing full-return draw during composition.
Neither approach specifies cross-asset residual dependence unless that
dependence is modeled separately.

\section{Methods}\label{sec:method}

We fitted a single-asset generator for each asset and combined the generated paths with a variance-corrected single-index model. The training data contained $424$ United States-listed equities and
exchange-traded funds observed from $2014$ through $2024$, and data
from $2025$ served as a holdout. The
single-asset analysis focused on the broad-market exchange-traded fund
with ticker SPY. It used
$2{,}767$ daily volume-weighted average prices for fitting, which
yielded $2{,}766$ growth rates, and the next $250$ prices from $2025$,
which yielded $249$ holdout growth rates. The multi-asset analysis used daily closing prices for all assets,
with SPY as the market, so SPY had two fitted generators: one on
volume-weighted prices for the single-asset analysis and one on
closing prices for the market path. Of
the other $423$ assets in the training universe, $416$ had complete
$2025$ histories and were included in the multi-asset holdout
evaluation. We defined the excess growth rate
$G_{i,j}$ for ticker $i$ between trading days $j-1$ and $j$ from prices
$P_{i,j-1}$ and $P_{i,j}$.
\begin{equation}
    G_{i,j} \;\equiv\; \left(\frac{1}{\Delta t}\right)\,
    \ln\!\left( \frac{P_{i,j}}{P_{i,j-1}} \right) \;-\; r_f.
    \label{eq:growth_rate}
\end{equation}
Here, $\Delta t = 1/252$ is the fraction of a trading year spanned by
one trading day, and $r_f$ is a constant continuously compounded
risk-free rate. The excess growth rate has units of
$\mathrm{year}^{-1}$; multiplied by $\Delta t$, it gives a daily log
excess return, which is additive across days under continuous
compounding. We used $r_f=0.043$ for the SPY training period and
$r_f=0.0421$ for its $2025$ holdout, based on Treasury Separate Trading
of Registered Interest and Principal of Securities (STRIPS) bond yields.
For the multi-asset closing-price experiments, we set $r_f=0$ in both
windows, so their growth rates were annualized log returns rather than excess
growth rates. Because
$r_f$ is constant, this choice shifts each series by a constant and leaves its variance
and the regression slopes unchanged.

\subsection{Per-asset hidden Markov marginal generator}\label{sec:method-hmm}
% method_hmm.tex
% Subsection 3.1 body. Four narrative paragraphs: model and state
% partition; emissions and counted transitions; jump mechanism and
% initialization; and hyperparameter calibration. The bridge to the
% multi-asset composition lives at the top of method_sim.tex.

For each asset, the hidden Markov generator divided the range of growth rates, which we also call returns, into $N=100$ ordered intervals and treated the interval that contained a day's
return as that day's state (Table~\ref{tab:hyperparams}). We write
$t$ for the day index $j$ and omit the ticker index $i$ when
discussing one asset. At each simulated time step, the model
first selected a state and then drew a return from that state's
distribution (Fig.~\ref{fig:model_architecture}). We represented the
model as
$\mathcal{M} = (\mathcal{S}, \mathcal{O}, \mathbf{T}, \mathbf{E}, \bar{\pi})$
~\cite{rabinerJuang1986}.
The hidden state $S_t \in \mathcal{S} = \{1,\dots,N\}$ identifies the
interval on trading day $t$, and the observation space
$\mathcal{O}\subset\mathbb{R}$ contains the excess growth rates $G_t$
of Eq.~\eqref{eq:growth_rate}. The transition
matrix $\mathbf{T}$ sets the probabilities of moving between states,
$\mathbf{E}$ gives the return distribution within each state, and
$\bar{\pi}$ gives the initial state probabilities. We ordered the
states from the most negative to the most positive returns and placed their boundaries at the quantiles of a Laplace distribution,
fitted to the asset's in-sample returns, at the equally spaced
probability levels $k/N$. We used the Laplace
distribution because its sharp peak matches the concentration of
small price movements in Fig.~\ref{fig:empirical_motivation}a
~\cite{kotzKozubowskiPodgorski2001, tothJones2019}. Its quantile
function also has a closed form, so the boundaries for all $424$
tickers required no iterative fitting~\cite{bilmes1998}. Specifically,
$Q_k = F_L^{-1}(k/N;\, \mu_L, b_L)$, where $F_L^{-1}$ is the Laplace
quantile function and $\mu_L$ and $b_L$ are the fitted location and
scale. We set $Q_0=-\infty$ and $Q_N=+\infty$ and assigned return
$G_t$ to state $k$ when $Q_{k-1}\le G_t<Q_k$. Because each training
return determines its state, the fitted state sequence is known; the
states are hidden only in simulated paths, where the model generates
the state sequence and reports only the returns.

The within-state emission followed a location-scale Student-$t$
distribution:
\begin{equation}
    G_t \mid S_t = k \;\sim\; \mu_k + \sigma_k \cdot t_{\nu},
    \qquad \nu = 5,
    \label{eq:emission}
\end{equation}
where $t_\nu$ denotes a standard Student-$t$ random variable with
$\nu$ degrees of freedom. For observations assigned to state $k$,
we used the sample mean for $\mu_k$ and the sample standard
deviation for $\sigma_k$, subject to two fallback rules. A state
with fewer than two observations used the mean and standard
deviation of that asset's full training series. A state with at
least two observations and sample standard deviation below $10^{-12}\,\mathrm{yr}^{-1}$
retained its state mean but used the full-series standard deviation
as its scale. These exceptions kept emission parameters defined but could broaden
states containing repeated returns (Online Appendix~\ref{sec:supp-fallback});
neither applied to SPY, whose states all held many observations. The same rules applied to the
generators that the multi-asset comparison fitted to regression
residuals. A standard $t_\nu$ has variance $\nu/(\nu-2)$, so
$\sigma_k$ is a scale parameter rather than the emission standard
deviation. At $\nu=5$, each state's emission standard deviation is
$\sigma_k\sqrt{5/3}$, larger than the sample standard deviation of
the returns assigned to that state. The emission is not truncated to
the state's interval, so a simulated return can fall outside it. We
selected $\nu = 5$ from a
sensitivity sweep and used $\nu = \infty$, which gives Normal
emissions, as a baseline; Table~\ref{tab:hyperparams} gives the
explored range. For SPY, holding the rest of the model fixed, the
Student-$t$ emission raised simulated kurtosis toward the observed value and
raised the Kolmogorov--Smirnov (KS) and Anderson--Darling (AD) pass
rates relative to the Normal emission (Table~\ref{tab:emission_hsmm}).
A pass rate is the percentage of simulated paths whose two-sample
test against the observed returns gave $p>0.05$.

We estimated the transition matrix $\mathbf{T}$ by counting observed
transitions between consecutive states and dividing each row by its
total count. A state with no observed outgoing transition used a
uniform row, $\mathbf{T}_{kj}=1/N$ for every destination state $j$.
Because the training state sequence is known, these row frequencies
are the maximum-likelihood estimates, and counting avoided the
initialization sensitivity of iterative
expectation-maximization~\cite{bilmes1998}. We obtained the
initial-state distribution $\bar{\pi}$ by solving
$\bar{\pi}\mathbf{T}=\bar{\pi}$ with
$\sum_k\bar{\pi}_k=1$~\cite{hamilton2018regime, grinsteadSnell1997}.
Because $\mathbf{T}$ was estimated by counting, $\bar{\pi}$ equals
the vector of in-sample state frequencies up to end effects, so the
model's stationary growth-rate distribution is the
frequency-weighted mixture of the state emission distributions. The counted probability of remaining in the same state was low
(Online Appendix~\ref{sec:supp-gen-nojump}), but volatility clustering
requires the chain to stay in the tail region, not in one state. Counting reproduces the observed
frequencies of consecutive state pairs. At longer lags, however, the
chain's absolute-return autocorrelation decays geometrically at a rate
set by the counted matrix, much faster than the slow decay in
Fig.~\ref{fig:empirical_motivation}d. The counted chain therefore
forgets within a few steps whether it was in the tails and cannot
reproduce the observed volatility clustering~\cite{bullaBulla2006}. Online
Appendix~\ref{sec:supp-generator} derives the fitted generator's
stationary law and autocorrelation structure and quantifies both this
rapid mixing and the autocorrelation added by the jump mechanism.

We added jump episodes to extend visits to extreme-return states and
call the resulting generator the hidden Markov model with jumps
(HMM-WJ).
When no jump episode was active, a jump was triggered with
probability $\epsilon$ at each step, and its duration was drawn as
$K\sim\mathrm{Poisson}(\lambda)$, where $\lambda$ is the mean episode
length~\cite{glasserman2003}. A draw of $K=0$ produced no episode,
and the chain made an ordinary transition. For $K>0$, the episode
consisted of $K$ \emph{forced tail steps}, beginning with the step
that triggered it. At each forced step, the model replaced the
ordinary transition by first choosing one of two tail sets and then
drawing a state uniformly from that set. The set
$\mathcal{S}_{-}=\{1,\dots,N_{\rm tail}\}$ contained the most negative
states, and $\mathcal{S}_{+}=\{N-N_{\rm tail}+1,\dots,N\}$ contained the
most positive states. The model chose the negative set with
probability $p_{\rm neg}$ and the positive set otherwise. Because
$p_{\rm neg}$ is slightly above one half, negative extreme returns
occurred slightly more often than positive ones. The choice was
repeated at each forced step, so one episode can contain
returns of both signs. Table~\ref{tab:hyperparams} lists $\epsilon$, $\lambda$,
$p_{\rm neg}$, and $N_{\rm tail}$; we tuned $\epsilon$ and $\lambda$
and fixed $p_{\rm neg}$ and $N_{\rm tail}$. Jump-diffusion models
use Poisson arrivals to add instantaneous price
jumps~\cite{merton1976, kou2002}. Our model instead uses the Poisson distribution for the episode
length, and an episode is a
sustained visit to the tail states rather than a single price move.
Setting $\epsilon=0$ removes the mechanism; we call this variant the
hidden Markov model with no jumps (HMM-NJ). Two closed-form approximations describe how much of a path, and how
many paths, the mechanism affects (Online
Appendix~\ref{sec:supp-generator}). Forced steps occupy a long-run
fraction of approximately $\epsilon\lambda/(1+\epsilon\lambda)$ of all
steps. A path of $T$ steps contains at least one episode with
probability approximately $1 - e^{-\epsilon T}$, so $\epsilon$ sets
how many paths contain an episode, while $\lambda$ sets how long an
episode lasts. The jump-augmented process is Markov only in the pair
formed by the state and the number of remaining forced steps. The solved $\bar{\pi}$ is therefore the exact stationary distribution
of the ordinary chain but only an approximate stationary start for
the jump-augmented process. The full simulation
procedure appears in Algorithm~\ref{alg:hmmwj}. Each step followed
$\mathbf{T}$ or continued a jump episode and then drew a return from
the selected state's emission distribution.

% --- Architecture figure ----------------------------------------------------
% Main figure fig:model_architecture is collected after the references.

\begin{algorithm}[t]
\caption{Hidden Markov model with jumps (HMM-WJ) single-asset marginal generator.}
\label{alg:hmmwj}
\begin{algorithmic}[1]
  \REQUIRE Fitted parameters
    $(\mathbf{T}, \mathbf{E}, \bar{\pi}, \epsilon, \lambda,
    p_{\rm neg}, N_{\rm tail})$; horizon $T$;
    tail sets $\mathcal{S}_{-} = \{1, \dots, N_{\rm tail}\}$ and
    $\mathcal{S}_{+} = \{N - N_{\rm tail} + 1, \dots, N\}$.
  \ENSURE State sequence $\{S_t\}_{t=1}^{T}$ and growth-rate path
    $\{G_t\}_{t=1}^{T}$.
  \STATE Draw $S_1 \sim \bar{\pi}$;\quad
    draw $G_1 \sim \mu_{S_1} + \sigma_{S_1}\, t_{5}$;\quad
    $k \gets 0$
    \hfill \COMMENT{stationary start; $k$: remaining forced-step counter}
  \FOR{$t = 2, \dots, T$}
    \IF{$k > 0$}
      \STATE $\mathcal{J} \gets \mathcal{S}_{-}$ w.p.\ $p_{\rm neg}$, else $\mathcal{S}_{+}$;\quad
        $S_t \sim \mathrm{Uniform}(\mathcal{J})$;\quad $k \gets k - 1$
        \hfill \COMMENT{forced tail step; sign drawn per step}
    \ELSIF{$\mathrm{Uniform}(0,1) > \epsilon$}
      \STATE $S_t \sim \mathbf{T}_{S_{t-1}, \cdot}$
        \hfill \COMMENT{standard Markov transition}
    \ELSE
      \STATE $K \sim \mathrm{Poisson}(\lambda)$
        \hfill \COMMENT{trigger jump; $K$ is the episode length}
      \IF{$K = 0$}
        \STATE $S_t \sim \mathbf{T}_{S_{t-1}, \cdot}$
          \hfill \COMMENT{$K = 0$: no forced steps, standard transition}
      \ELSE
        \STATE $\mathcal{J} \gets \mathcal{S}_{-}$ w.p.\ $p_{\rm neg}$, else $\mathcal{S}_{+}$;\quad
          $S_t \sim \mathrm{Uniform}(\mathcal{J})$;\quad
          $k \gets \min(K - 1, T - t)$
          \hfill \COMMENT{first forced step; $K$ steps total, sign per step}
      \ENDIF
    \ENDIF
    \STATE Draw $G_t \sim \mu_{S_t} + \sigma_{S_t}\, t_{5}$
      \hfill \COMMENT{Eq.~\eqref{eq:emission}}
  \ENDFOR
\end{algorithmic}
\end{algorithm}

We tuned the jump probability $\epsilon$ and mean duration $\lambda$
on the volume-weighted SPY training series to match its absolute-return autocorrelation and excess
kurtosis~\cite{mandelbrot1963, cont2001, rydenTerasvirtaAsbrink1998};
the multi-asset comparisons either disabled jumps or transferred the
SPY values to the closing-price generators
(Table~\ref{tab:hyperparams}). A path with no episode follows HMM-NJ, so its autocorrelation and
kurtosis errors do not depend on $(\epsilon,\lambda)$. We scored only \emph{jump-active} paths, those containing at least one
episode~\cite{glasserman2003}, so the objective measures how well an
episode reproduces the observed statistics, not how often episodes
occur. At each point of a grid of
$(\epsilon,\lambda)$ values, we simulated $200$ independent paths of
$2{,}766$ trading days. We minimized an objective that averaged, over
the jump-active paths, the squared errors in the autocorrelation
function (ACF) of absolute returns through lag $L$ plus a weighted
squared error in excess kurtosis:
\begin{equation}
    J(\epsilon, \lambda) =
    \frac{1}{|\mathcal{R}_J|}\sum_{r\in\mathcal{R}_J}
    \left[
    \sum_{\tau=1}^{L} \left( \mathrm{ACF}_{\rm obs}(\tau)
    - \mathrm{ACF}_{r}(\tau) \right)^2
    + w_\kappa \left( \kappa_{\rm obs} - \kappa_{r} \right)^2
    \right],
    \label{eq:grid_search}
\end{equation}
where $\mathcal{R}_J$ is the set of jump-active simulated paths,
$\mathrm{ACF}_{\rm obs}(\tau)$ and $\kappa_{\rm obs}$ are the empirical
absolute-growth autocorrelation at lag $\tau$ and excess kurtosis,
$\mathrm{ACF}_{r}(\tau)$ and $\kappa_r$ are the corresponding values
for path $r$, and $w_\kappa$ sets the relative weight of the kurtosis
error. When $\epsilon$ is small, as at the selected value, most
jump-active paths contain a single episode, whose forced
steps make up a fraction of about $\lambda/T$ of the path. The episode
adds a slowly decaying component to the absolute-return
autocorrelation, and $\lambda$ sets both its height and the number of
lags over which it persists. The kurtosis penalty limits $\lambda$, because forced steps make tail
returns common within a path rather than rare; they raise the variance
proportionally more than the fourth moment and so lower the path
kurtosis. The parameter $\epsilon$ acts mainly through the share
of paths that contain an episode, which the restriction to jump-active paths removes. The objective
therefore constrains $\epsilon$ only weakly, and the selected
$\epsilon$ lay at the lower edge of its search range
(Table~\ref{tab:hyperparams}; Online
Appendix~\ref{sec:supp-gen-tuning}).
The search ranges, $w_\kappa$, $L$, and the remaining hyperparameters
are reported in Table~\ref{tab:hyperparams}. We fixed the selected
hyperparameters before running the generator comparisons and the
holdout evaluation.

Beyond the Normal-emission variant, we compared HMM-WJ with two
benchmarks that model persistence differently. A hidden semi-Markov
model (HSMM) baseline adapted from Bulla and
Bulla~\cite{bullaBulla2006} lengthens state visits through explicit
dwell-time distributions instead of jump episodes
(Table~\ref{tab:emission_hsmm}; Online
Appendix~\ref{sec:supp-emission-hsmm}). The Normal-emission comparison
isolated the effect of heavy-tailed emissions, and the HSMM
comparison contrasted explicit dwell times with the jump-duration
mechanism, although the HSMM also used $8$ states rather than $100$ and so
differed in more than its duration model. To compare the hidden Markov
generator with a neural autoregressive model, we
included a gated recurrent unit (GRU) benchmark
(Table~\ref{tab:model_comparison}). The network took a $50$-return input window and had two recurrent
layers of $64$ hidden units each. We standardized the training series with its mean and standard
deviation, and separate output heads predicted the mean and log
variance of the next standardized return. We fitted the network on training SPY only,
selected weights by training loss, and sampled recursively from
the predicted Gaussian distributions. Online
Appendix~\ref{sec:supp-gru} specifies the optimizer, stopping rule,
initialization, and recorded computational environment. We scored all single-asset generators,
including these benchmarks, by their KS and AD pass rates and by the mean absolute error between the observed and the
ensemble-average simulated absolute-return autocorrelation functions
(ACF-MAE), over lags $1$--$252$ in training and $1$--$248$ in the
$249$-day holdout (Eq.~\eqref{eq:acf-ensemble-error} in Online
Appendix~\ref{sec:supp-validation}).

\subsection{Variance-corrected multi-asset composition}\label{sec:method-sim}
% method_sim.tex
% Method: variance-corrected SIM composition.
% Simplified 2026-07-14 around a direct variance explanation. The
% section now proceeds from the composition goal, to naive double
% counting, to the variance correction, the high-R^2 exception, and
% finally the evaluation protocol. Detailed verification remains in
% the Online Appendix.

The single-asset model generated each ticker separately and did not
specify cross-asset dependence. For the multi-asset model, we added a
shared market path as a common factor, using the single-index model
(SIM) of Sharpe~\cite{sharpe1963}. We write $g_i(t)$ for the growth rate $G_{i,t}$ of asset $i$ on days
$t=1,\dots,T$. The horizon $T$ is $2{,}766$ days in the training
comparisons and $249$ days in the holdout. The same symbols denote
observed paths when we estimate parameters and synthetic paths when
we compose them:
\begin{equation}
  g_i(t) \;=\; \alpha_i \;+\; \beta_i\, g_m(t) \;+\; \eps_i(t),
  \qquad t = 1,\dots,T.
  \label{eq:sim-composition}
\end{equation}
Here $\alpha_i$ is the intercept for asset $i$, $g_m$ is the
growth-rate path of the market, SPY, $\beta_i$ is the asset's market
loading, and $\eps_i$ is its asset-specific residual path. We use
$\eps_i$ for the residual path and $\epsilon$ for the jump probability
in the single-asset model. We estimated the intercept and market
loading $(\alpha_i,\beta_i)$ from observed data by ordinary least
squares (OLS). The choice of $\eps_i$ determined the composition
method. A standard SIM draws independent Gaussian residuals with the
variance estimated by OLS. These Gaussian residuals preserve the fitted market relationship but
lack the heavy tails, and with jumps enabled the volatility
clustering, that the single-asset generator produces. We called the Gaussian-residual method
\emph{Gaussian SIM} and used it as one of the compared methods.

\paragraph{Naive full-path composition.}

Using a full synthetic growth-rate path $\tg_i$ drawn from the asset's
own single-asset generator, the \emph{generator draw}, directly as the
residual would give:
\begin{equation}\label{eq:naive-composition}
g_i = \alpha_i + \beta_i\, g_m + \tg_i.
\end{equation}
The fitted intercept and market term already set the conditional mean
of asset $i$, while the generator draw generally has
$\E[\tg_i]\ne 0$. Equation~\eqref{eq:naive-composition} therefore has
conditional mean $\alpha_i+\E[\tg_i]+\beta_i g_m$, not
$\alpha_i+\beta_i g_m$.
We therefore centered every generator draw before composition,
subtracting its realized path mean:
\begin{equation}
  \overline{\tg}_i \equiv \frac{1}{T}\sum_{t=1}^{T}\tg_i(t),
  \qquad
  \tg_i^{\mathrm c}(t) \equiv \tg_i(t)-\overline{\tg}_i.
  \label{eq:center-generator}
\end{equation}
Therefore, $T^{-1}\sum_t\tg_i^{\mathrm c}(t)=0$ exactly. Subtracting
the realized rather than the population mean makes the centering exact on every path. Subtracting a
constant leaves both the sample variance and the sample covariance with
$g_m$ unchanged. We used the centered construction
$g_i=\alpha_i+\beta_i g_m+\tg_i^{\mathrm c}$, which corrects the mean
but not the variance, as the \emph{naive} benchmark in all
comparisons. The sample variance of each draw,
$\sigma_{\gen,i}^2=\Var(\tg_i)=\Var(\tg_i^{\mathrm c})$, is the
\emph{generator variance}; it is recomputed for every draw. It already represents the asset's total
variance, so adding the market term adds market variance a second
time. During composition, $g_m$ was the observed SPY path in the training
comparison and a path simulated from SPY's closing-price generator in the holdout
and jump comparisons, with one market path per replication shared by
all assets, and
in both cases it was independent of the asset draws. The covariance
of a centered draw with the market path is zero in population and close to zero on a long realized
path. We approximated the naive composed variance accordingly:
\begin{equation}
  \Var(g_i)
  \;\approx\; \beta_i^2\sigma_m^2 + \sigma_{\gen,i}^2,
  \qquad \Cov(\tg_i^{\mathrm c},g_m)\approx 0,
  \label{eq:naive-inflation}
\end{equation}
where $\sigma_m^2=\Var(g_m)$ is the sample variance of the market path
used in composition. The naive composed variance therefore exceeds the generator variance
by $\beta_i^2\sigma_m^2$, an excess that grows with the square of the
market loading.

\paragraph{Variance correction.}

We set $\sigma_{\gen,i}^2$ as the target total variance for asset $i$.
The market term contributed $\beta_i^2\sigma_m^2$, so the residual term
could contribute only the remaining variance. We defined $s_i$ as the
nonnegative asset-specific residual scale, set
$\eps_i=s_i\tg_i^{\mathrm c}$, and required
$\Var(\beta_i g_m+s_i\tg_i^{\mathrm c})=\sigma_{\gen,i}^2$. Under the
zero-covariance assumption, solving for $s_i$ gives the correction:
\begin{equation}
  \boxed{\;s_i^2 \;=\; 1 \;-\; \frac{\beta_i^2\, \sigma_m^2}{\sigma_{\gen,i}^2}\;}
  \label{eq:scale}
\end{equation}
We call the market term's share of the target variance the
\emph{market loading ratio}:
\begin{equation}
  \rho_i \;\equiv\; \frac{\beta_i^2\, \sigma_m^2}{\sigma_{\gen,i}^2},
  \label{eq:rho}
\end{equation}
so that $s_i^2=1-\rho_i$ under the zero-covariance assumption. At
$\rho_i=1$, the rule of Eq.~\eqref{eq:scale} leaves no asset-specific
variance; for $\rho_i>1$, it has no real-valued solution.

With the observed variances, $\rho_i$ would equal the coefficient of
determination of the regression, which is below one. Because $\rho_i$
instead uses the generator and market-path variances, it can reach or exceed one when the generator
variance is low or the market path is more volatile than the observed
SPY training series, particularly for assets with large $\beta_i$. We therefore required the asset-specific draw to
contribute at least a fraction $f$ of the target variance, the
\emph{floor}, and set
$\rho_i^{\max}=1-f$. We applied the correction through an
\emph{effective loading} $\beta_i^{\eff}$, the loading used in
composition, which equals $\beta_i$ unless $\rho_i>\rho_i^{\max}$:
\begin{equation}
  \beta_i^{\eff} \;=\;
  \begin{cases}
    \beta_i & \text{if } \rho_i \le \rho_i^{\max} \\[4pt]
    \sign(\beta_i)\,\sqrt{(1 - f)\,\dfrac{\sigma_{\gen,i}^2}{\sigma_m^2}} & \text{if } \rho_i > \rho_i^{\max}
  \end{cases}
  \qquad
  s_i^2 \;=\;
  \begin{cases}
    1 - \rho_i & \text{if } \rho_i \le \rho_i^{\max} \\[4pt]
    f          & \text{if } \rho_i > \rho_i^{\max}
  \end{cases}
  \label{eq:clip}
\end{equation}
We call the second case of Eq.~\eqref{eq:clip} the \emph{clipped}
case. In it, the correction reduced the magnitude of the loading but
kept the sign of the calibrated $\beta_i$. We used $f=0.10$, so the asset-specific draw contributed
at least $10\%$ of the target variance for every asset under this rule
(Table~\ref{tab:hyperparams}). We retained the fitted intercept
$\alpha_i$ when clipping changed the loading. Because the scaled
residual had zero sample mean, the composed mean was
$\overline{g}_i=\alpha_i+\beta_i^{\eff}\overline{g}_m$, where
$\overline{g}_m$ is the mean of the market path. Clipping therefore
changed the mean by $(\beta_i^{\eff}-\beta_i)\overline{g}_m$ relative
to the calibrated linear mean on that same market path. The clipped case thus kept the intercept and the target variance and
set the residual share to exactly $f$, but changed the loading and,
unless $\overline{g}_m=0$, the mean. Online
Appendix~\ref{sec:supp-residual-choice} examines the substitution of
a full-return draw for the residual: it derives what the construction
preserves exactly and how much of the generator's excess kurtosis
survives composition as $\rho_i$ grows.

\paragraph{High-$R^2$ index trackers.}

The coefficient of determination $R^2_{i,\real}$, from the OLS
regression of the observed asset on observed SPY, measures the
fraction of the asset's variance explained by the market. The rule of
Eq.~\eqref{eq:clip} targets total generator variance, not
$R^2_{i,\real}$. Under the zero-covariance
assumption, the composed $R^2$ under that rule equals $\rho_i$, capped
at $1-f$ by the floor. It therefore cannot exceed $1-f$, and it matches $R^2_{i,\real}$ only
when $\sigma_m^2/\sigma_{\gen,i}^2$ equals the ratio of the observed SPY
variance to the observed asset variance. Index-tracking funds, such as the Nasdaq-100 fund QQQ and the
S\&P~500 growth fund SPYG, have $R^2$ against SPY close to or above
the cap $1-f$, so that rule could not reliably reproduce their co-movement with the
market. We therefore used two branches: the rule of
Eq.~\eqref{eq:clip}, the \emph{variance-preserving branch}, for most
assets, and a separate \emph{$R^2$-preserving branch} when
$R^2_{i,\real}$ was at least $R^2_{\mathrm{preserve}}$. We set $R^2_{\mathrm{preserve}}=0.80$,
which lies in a wide gap of the observed $R^2$ distribution between
the index trackers and all other assets in our universe
(Online Appendix~\ref{sec:supp-sim-derivations}). This branch keeps $\beta_i^{\eff}=\beta_i$ and uses the same
market-path variance $\sigma_m^2$ as the variance-preserving branch.
We computed the residual variance required to recover $R^2_{i,\real}$
under the zero-covariance
assumption:
\begin{equation}
  \boxed{\;\sigma_{\eps,\mathrm{target}}^2 \;=\; \beta_i^2\, \sigma_m^2 \cdot
    \frac{1 - R^2_{i,\real}}{R^2_{i,\real}}\;}
  \label{eq:r2-target}
\end{equation}
We rescaled the generator draw to the target residual variance:
\begin{equation}
  \eps_i \;=\; \left(\sqrt{\frac{\sigma_{\eps,\mathrm{target}}^2}{\sigma_{\gen,i}^2}}\,\right)\tg_i^{\mathrm c}.
  \label{eq:r2-rescale}
\end{equation}
This branch preserves $R^2_{i,\real}$ rather than the generator
variance: the composed variance becomes
$\beta_i^2\sigma_m^2/R^2_{i,\real}$. The scale factor in Eq.~\eqref{eq:r2-rescale} exceeds one when the generator variance is below the target residual
variance; the composed variance then exceeds the generator variance.

\paragraph{Complete composition rule.}

The complete rule has three outcomes, recorded by the branch flag
$b_i$ in Algorithm~\ref{alg:hybrid}. Index trackers used the
$R^2$-preserving branch; all other assets used the
variance-preserving branch, with the calibrated loading or, in the
clipped case, a reduced loading. Each composed path is then
$g_i=\alpha_i+\beta_i^{\eff}g_m+\eps_i$. We call this complete rule
the \emph{hybrid} composition, the label used in the tables and
figures, and all results reported for the variance correction refer
to it. Membership in the $R^2$-preserving branch depends only on
$R^2_{i,\real}$ and is fixed for each asset. Everything else is
computed per draw: because $\sigma_{\gen,i}^2$ is recomputed for each
draw, so are $s_i$, $\rho_i$, and $\beta_i^{\eff}$, and one asset can be
clipped in some draws and not in others. Because the draws are independent of the market path, OLS estimates
from composed paths converge, as the horizon grows, to the intercept
$\alpha_i$ and the effective loading $\beta_i^{\eff}$, which equals
$\beta_i$ unless the draw is clipped. The variance target and the loading recovery are exact only when the
sample covariance between the draw and the market path is zero; on a
finite path, they deviate by that covariance term (Online
Appendix~\ref{sec:supp-sim-derivations}). Because centering and
scaling fix the residual sum, every branch imposes a zero residual sum over the horizon $T$, which is stronger than a zero population
mean. For a fixed market path and effective loading, the terminal
cumulative log return therefore has no asset-specific residual
uncertainty, although daily and intermediate returns still vary. In
the clipped case, the effective loading depends on the variances of
each draw, so it can vary between draws and produce terminal
variation through the market term. Because centering and scaling use
statistics of the entire draw, the construction also depends on the
chosen horizon. We therefore evaluated it as a constrained
finite-horizon construction for daily distributional fit and recovery
of the market loading; it does not preserve cumulative residual uncertainty.

The construction links assets only through same-day market returns.
A large market move can therefore affect several assets together, but
the rule does not separately fit lead--lag effects or a common
volatility factor in the residuals. Algorithm~\ref{alg:hybrid} also
allows an optional rank reorder that adds residual dependence across
assets; the six-method comparison did not use it. The reorder draws
one joint vector per day from a copula fitted to the observed OLS
residuals and permutes each asset's residual values over days so that
their ranks over time match the ranks of that asset's copula
coordinate. The reorder imposes cross-asset residual
dependence and keeps each path's values and variance, but it replaces the generator's day-to-day ordering and with it any
volatility clustering in the draw.

\begin{algorithm}[t]
\caption{Hybrid single-index model (SIM) composition with variance correction.}
\label{alg:hybrid}
\small
\begin{algorithmic}[1]
  \REQUIRE Nonconstant $g_m,\tg_i\in\R^T$, $i=1,\dots,N$; each
    $\tg_i$ is independent of $g_m$.
  \REQUIRE Parameters $(\alpha_i,\beta_i,R^2_{i,\real})$; floor
    $f\in(0,1)$; threshold $R^2_{\mathrm{preserve}}\in(0,1]$.
  \ENSURE Composed paths $g_i\in\R^T$ and branch flags $b_i$.
  \STATE $\sigma_m^2 \gets \Var(g_m)$.
  \FOR{each asset $i = 1,\dots,N$}
    \STATE $\sigma_{\gen,i}^2 \gets \Var(\tg_i)$;
      $\tg_i^{\mathrm c} \gets \tg_i-\overline{\tg}_i$.
    \IF{$R^2_{i,\real} \ge R^2_{\mathrm{preserve}}$}
      \STATE $v_i \gets \beta_i^2\sigma_m^2(1-R^2_{i,\real})/R^2_{i,\real}$.
      \STATE $s_i \gets \sqrt{v_i/\sigma_{\gen,i}^2}$;
        $\beta_i^{\eff} \gets \beta_i$; $b_i\gets\textsc{r2-preserve}$.
    \ELSE
      \STATE $\rho_i \gets \beta_i^2 \sigma_m^2 / \sigma_{\gen,i}^2$
      \IF{$\rho_i \le 1 - f$}
        \STATE $s_i \gets \sqrt{1-\rho_i}$;
          $\beta_i^{\eff}\gets\beta_i$; $b_i\gets\textsc{variance}$.
      \ELSE
        \STATE $s_i\gets\sqrt{f}$;
          $\beta_i^{\eff}\gets\sign(\beta_i)
          \sqrt{(1-f)\sigma_{\gen,i}^2/\sigma_m^2}$.
        \STATE $b_i\gets\textsc{clipped}$.
      \ENDIF
    \ENDIF
    \STATE $\eps_i \gets s_i\tg_i^{\mathrm c}$.
  \ENDFOR
  \STATE \textit{Optional:} jointly rank-reorder $\{\eps_i\}_{i=1}^{N}$.
  \FOR{each asset $i = 1,\dots,N$}
    \STATE $g_i \gets \alpha_i + \beta_i^{\eff}\, g_m + \eps_i$
      \hfill \COMMENT{Eq.~\eqref{eq:sim-composition}}
  \ENDFOR
\end{algorithmic}
\end{algorithm}

% --- Table: Hyperparameters ---
% Main table tab:hyperparams is collected after the references.
\FloatBarrier

\paragraph{Evaluation protocol.}

To test variance correction against alternative residual models, we
compared six composition methods using a dataset of $424$ United
States equities and exchange-traded funds from Polygon.io (Table~\ref{tab:aggregate}).
Each ticker had $2{,}767$ daily closing prices from January $2014$
through December $2024$, yielding $2{,}766$ growth rates; this
complete-case selection conditioned the analysis on survival and data
coverage. We fitted the same hidden Markov generator to each asset,
including SPY (Table~\ref{tab:hyperparams}; $r_f=0$), and regressed
the other $423$ assets on observed SPY to estimate $\alpha_i$,
$\beta_i$, $R^2_{i,\real}$, and residual variance. All generators
had jumps disabled to isolate composition from jump calibration. The
naive and hybrid methods shared a centered full-return draw, with
variance correction applied only by the hybrid method
(Algorithm~\ref{alg:hybrid}). Four alternatives generated residuals
from a Gaussian distribution, a residual-fit hidden Markov model
(labeled JumpHMM-on-residuals), a stationary block bootstrap, or a
GARCH(1,1)-$t$ model, and added them to $\alpha_i+\beta_i g_m$
without centering or rescaling. Comparator fitting and eligibility
rules are given in Online Appendix~\ref{sec:supp-composition-evaluation};
$30$ non-stationary GARCH fits left $393$ eligible assets for that
method. Using observed SPY as the common market path, we ran $100$
replications per asset with seed $1234$ and pooled training scores
across tickers. We measured factor recovery by the absolute OLS
errors $|\hat\alpha-\alpha_i|$, $|\hat\beta-\beta_i|$, and
$|\hat R^2-R^2_{i,\real}|$, and naive and hybrid variance preservation
by the composed-to-generator variance ratio. Distributional scores
were KS and AD pass rates, Wasserstein-1 distance, excess kurtosis,
and the Hill index of absolute returns (Online
Appendix~\ref{sec:supp-validation}). Because only naive and hybrid
forced the residual mean to zero, a paired centering control on the
common $393$ assets separated this constraint from model fitting
(Table~\ref{tab:centering_control}; Online
Appendix~\ref{sec:supp-centering-control}).

To test performance without refitting, we froze all $2014$--$2024$
fits and evaluated $100$ simulated paths per asset against the $249$
growth rates observed in $2025$ (Table~\ref{tab:oos_composition}).
Complete holdout histories were available for $416$ assets, including
$386$ with accepted GARCH fits. Each replication used one simulated
$249$-day no-jump SPY path shared across asset compositions. The
correction used the observed training market variance in
Eqs.~\eqref{eq:scale}--\eqref{eq:r2-rescale}, while each asset's
generator variance remained draw-specific. Consequently, variation
in simulated market variance could move composed variance away from
the generator target even at zero sample cross-covariance. Observed
$2025$ returns were used only for scoring. We repeated the KS, AD,
Wasserstein-1, and kurtosis evaluations and measured variance for all
six methods relative to observed $2025$ variance. Because goodness-of-fit tests have less power on short samples, we
also compared the same simulated paths with randomly selected
contiguous $249$-day blocks of observed training returns as a
length-matched training reference. We averaged
replications within ticker before summarizing across tickers, giving
each asset equal weight.

To estimate one-day Value-at-Risk (VaR) thresholds without relying on
a single short path, we pooled daily log returns, $\Delta t$ times
the growth rates, from $5{,}000$ separately composed $249$-day paths
per asset and method (Table~\ref{tab:var}). Each replication shared
one simulated SPY path across assets and methods and followed the
holdout centering and scaling conventions. We fixed the $95\%$ and
$99\%$ coverage thresholds before scoring observed $2025$ returns;
no holdout data entered fitting or threshold calibration. These
unconditional thresholds describe a uniformly selected day within the
$249$-day construction, retain its horizon dependence, and do not
remove the zero-sum residual constraint of naive and hybrid paths.
We reported mean exceedance rates and cross-ticker standard deviations
descriptively, without treating dependent days or assets sharing a
market as independent replications. Pool-size convergence and an
independent simulation check are reported in Online
Appendix~\ref{sec:supp-var-ensemble}.

\paragraph{Jump-enabled multi-asset comparison.}

To test whether the temporal benefit of the single-asset jump
mechanism survived composition, we compared naive and hybrid
composition under three jump settings (Table~\ref{tab:jump_ablation}).
Unlike the six-method comparison, which used observed SPY in
training, this comparison simulated the SPY market path in both the
$2{,}766$-day training window and the $249$-day holdout, so a jump
setting could apply to the market generator. Jumps were disabled,
enabled only in the SPY market generator, or enabled in both SPY and
every asset generator. We froze the closing-price hidden Markov generators and SIM
calibrations of the six-method comparison. Enabled models
received the SPY settings of Table~\ref{tab:hyperparams},
$(\epsilon,\lambda)=(10^{-4},90)$, $p_{\rm neg}=0.52$, and
$N_{\rm tail}=5$, without per-asset tuning. These settings were tuned on volume-weighted SPY prices and applied
here to generators fitted to closing prices. At these settings,
most paths contain no episode, especially over the $249$-day holdout,
so the jump effects are carried by a minority of paths. In both windows and for all three jump settings, the correction used
each simulated market path's variance and each asset draw's variance.
This differs from the six-method holdout, which fixed the market
variance at its training value.

For each jump setting, naive and hybrid composition received the same
asset and market draws, and one market path was
shared across all assets in each replication. We generated $250$
paths for each of four seeds ($1234$, $2345$, $3456$, and $4567$),
giving $1{,}000$ replications per asset, setting, and method. We used a path-level version of ACF-MAE, computed on each path and
then averaged across paths (Eq.~\eqref{eq:acf-path-error} in Online
Appendix~\ref{sec:supp-validation}); the single-asset ACF-MAE instead
averages the path ACFs before measuring the error. We used
lags $1$--$25$, the lag window of the tuning objective, with lags
$1$--$60$ as a secondary check, and repeated the marginal and variance
diagnostics. The temporal scores of the single-asset and jump comparisons
therefore have different estimands and cannot be compared directly. We included all paths, not only jump-active ones, in the summaries;
jump-active subsets only described the mechanism
(Online Appendix~\ref{sec:supp-jump-composition}). For Monte Carlo uncertainty, we averaged each metric across tickers
within each replication, which kept assets together under their
shared market path. Jump settings were paired by replication, sharing
the draws of any component whose jump setting did not change, and the
approximate $95\%$ confidence intervals were the mean of the $1{,}000$
replication-level differences plus or minus $1.96$ standard errors
(Table~\ref{tab:jump_ablation_uncertainty}). These intervals describe simulation
uncertainty conditional on the fitted models and observed histories,
not uncertainty across market regimes.

\section{Results}\label{sec:results}
\subsection{Single-asset generator validation}

\begin{figure}[H]
  \centering
  \includegraphics[width=\textwidth]{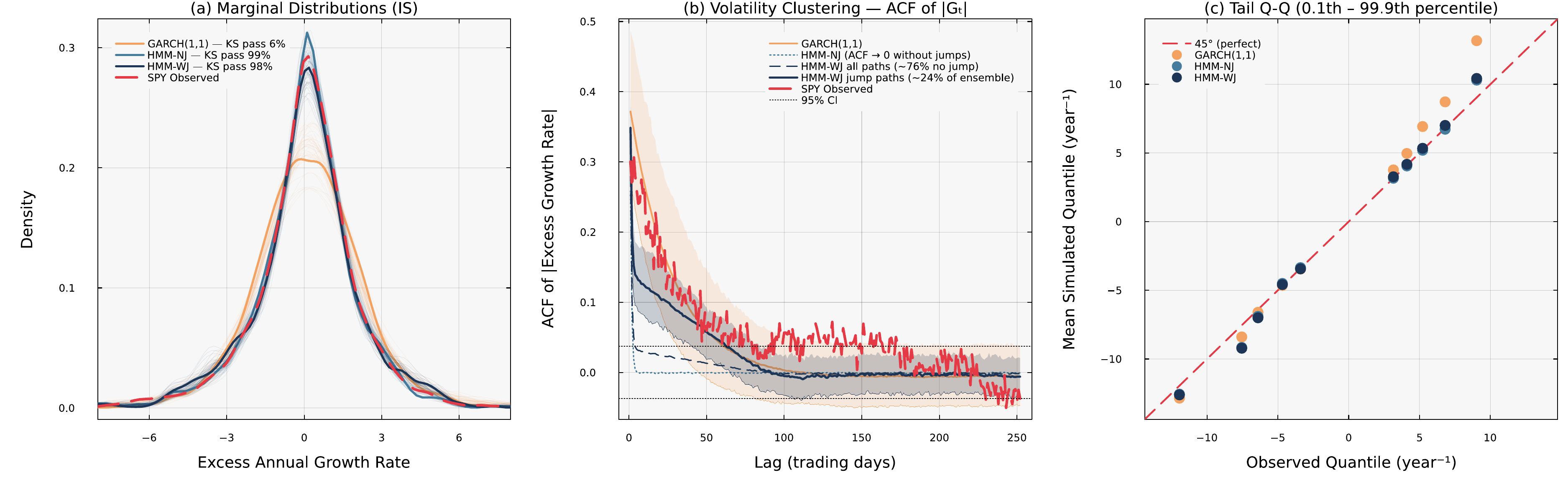}
  \caption{\textbf{Head-to-head in-sample model comparison for SPY}
    ($N = 100$, $1{,}000$ simulated paths). \emph{Panel (a):} marginal
    density of excess growth rates with in-sample Kolmogorov--Smirnov
    (KS) pass rates
    annotated per generator. \emph{Panel
    (b):} mean autocorrelation function (ACF) of $|G_t|$ at lags
    $1$--$252$. For the hidden Markov model with jumps (HMM-WJ),
    the solid curve and its $10$th--$90$th percentile band use
    jump-active paths; the dashed curve averages all paths. The
    generalized autoregressive conditional heteroskedasticity (GARCH)
    band also gives path percentiles. Horizontal dotted lines are the white-noise
    references $\pm1.96/\sqrt{2{,}766}$. \emph{Panel (c):}
    tail quantile-quantile plot at the $0.1$st--$99.9$th percentile region per
    generator.}
  \label{fig:model_comparison}
\end{figure}

The SPY returns had heavy tails and persistent absolute-return
autocorrelation in both the training and holdout windows, providing
two targets for generator validation
(Fig.~\ref{fig:empirical_motivation};
Online Appendix~\ref{sec:supp-generator}). Individual raw-return correlations
were small, although the Ljung--Box test detected joint serial
dependence. We compared HMM-WJ with seven alternatives on the training
window, using $1{,}000$ paths of $2{,}766$ trading days
(Table~\ref{tab:model_comparison}; Fig.~\ref{fig:model_comparison}).
The alternatives were independent bootstrap, Gaussian, and Laplace
draws; HMM-NJ; GARCH(1,1) with Gaussian innovations; the GRU
benchmark~\cite{choEtAl2014gru}; and the HSMM baseline adapted from
Bulla and Bulla~\cite{bullaBulla2006}. Both HMM variants had KS pass
rates within about one percentage point of the fitted Laplace
baseline but came much closer to the observed excess kurtosis, which
Laplace underestimated at $3.0$ against $7.7$. Both also had lower
Wasserstein-1 distances than Laplace, whereas Laplace had the lowest
Hellinger distance among the fitted generators. Bootstrap resampling
passed every KS test while reusing the observed returns, and Gaussian
sampling passed neither distribution test. The GARCH model had the
lowest mean absolute error between the observed and ensemble-average
absolute-return autocorrelation functions (ACF-MAE) but rarely passed
the distribution tests. The GRU also had lower ACF-MAE than the
HMM variants but rarely passed the distribution tests. The HSMM
passed $82\%$ of KS tests, underestimated excess kurtosis by more than
one third, and had the same ACF-MAE as HMM-NJ despite its flexible
state durations. A controlled emission ablation confirmed that
Student-$t$ emissions brought HMM-WJ excess kurtosis closer to the
observed value than Normal emissions
(Table~\ref{tab:emission_hsmm}). The HMM variants therefore captured
heavy tails well, but no fitted generator led on every metric.

Jump episodes improved temporal fit at a small cost in marginal fit
in training (Table~\ref{tab:model_comparison};
Fig.~\ref{fig:model_comparison}b). Under the counted transition
matrix, every state had an expected residence time of $1$--$2$ steps,
and absolute-return autocorrelation decayed to near zero by lag $5$
(Online Appendix~\ref{sec:supp-gen-nojump}). Calibration selected mean episode
duration $\lambda=90$ and jump probability $\epsilon=10^{-4}$
(Table~\ref{tab:hyperparams}). About $24\%$ of training-length HMM-WJ
paths contained an episode; paths without one were equivalent to
HMM-NJ (Online Appendix~\ref{sec:supp-gen-renewal}). Across all paths,
ACF-MAE fell from $0.059$ for HMM-NJ to $0.053$ for HMM-WJ, with
small reductions in KS and AD pass rates. The episode frequency and
jump-active autocorrelation agreed with the generator calculations
(Table~\ref{tab:supp-generator-quantities}; Online Appendix~\ref{sec:supp-gen-tuning}). The calibration objective scored
only jump-active paths, so it weakly identified $\epsilon$ and
selected the smallest grid value; the roughly one-quarter episode
share followed from that value and the path length, rather than being
a calibration target. Likewise, $\lambda$ was a fitted duration,
not an estimate of observed high-volatility spell lengths. Supporting sweeps showed that this tradeoff depended on both the
jump-active share and the state count: increasing either improved
autocorrelation fit but reduced distribution-test pass rates (Online
Appendix~\ref{sec:supp-gen-tuning}; Online Appendix~\ref{sec:supp-sensitivity}).
The primary choice $N=100$ balanced the training metrics, but the
state-count sweep used SPY alone and did not establish a suitable
resolution for other assets or shorter histories (Online
Appendix~\ref{sec:supp-sensitivity}).

The frozen HMM variants retained high distribution-test pass rates on
the $249$ held-out SPY returns from $2025$, although the short sample
limited discrimination among generators
(Table~\ref{tab:model_comparison};
Online Appendix~\ref{sec:supp-gen-renewal}). We compared those returns
with $249$-day simulated paths. The KS pass rates were $96.6\%$ for
HMM-NJ and $95.4\%$ for HMM-WJ, but weaker generators also passed
more often than in the longer training comparison. Few paths
contained an episode at $\epsilon=10^{-4}$, so the holdout mainly
tested the no-jump dynamics shared by the two variants (Online
Appendix~\ref{sec:supp-gen-renewal}). The fitted Laplace baseline
matched or exceeded both on KS and AD pass rates and had equal or
smaller Wasserstein-1 and Hellinger distances. Its excess kurtosis of
$2.8$, however, remained well below the observed $6.9$, whereas both
HMM variants reached $6.2$. Repeating HMM-WJ fitting and calibration
for NVDA, JNJ, and JPM gave training KS pass rates near $99\%$ for
all three, but holdout rates ranged from $74.0\%$ for JNJ to
$97.7\%$ for NVDA (Online Appendix~\ref{sec:supp-cross-asset}). The single-asset evidence supported heavy-tail representation, but
did not establish uniform holdout fit across assets or metrics. The six-method composition comparison used
the no-jump variant to avoid per-ticker jump calibration, and a
separate multi-asset comparison tested the transfer of SPY's jump
settings.

% Main table tab:model_comparison and figure fig:model_comparison are collected after the references.

\subsection{Variance-corrected multi-asset composition}

We tested whether hybrid composition preserved generator variance
while recovering the calibrated market loading across the $423$
non-market assets of the training universe
(Table~\ref{tab:aggregate}; Fig.~\ref{fig:preservation}). The six-method
comparison included GARCH(1,1)-$t$ on the $393$ assets with eligible
fits. With observed training SPY as the market path, the index trackers QQQ
and SPYG used the $R^2$-preserving branch, and the other $421$ assets,
which we call non-tracker assets, used the variance-preserving branch
without clipping (Online Appendix~\ref{sec:supp-sim-derivations}). The two trackers targeted
the calibrated $R^2$ rather than the generator variance, so we marked
them separately (Fig.~\ref{fig:preservation}). We paired each composed
path with its uncomposed generator draw to measure the ratio of
composed to generator variance directly. Across the non-tracker
assets, the median of the per-ticker median ratios was $1.3171$ for
naive composition and $1.0001$ for hybrid composition, and the hybrid
ticker medians ranged from $0.9933$ to $1.0072$
(Fig.~\ref{fig:preservation}b). Individual hybrid paths still varied
around the target, because the correction assumes zero covariance
between the scaled asset draw and the market path, whereas a finite
path has a small nonzero sample covariance. Across non-tracker hybrid
paths, the median absolute departure of the variance ratio from one
was $1.12\%$, and its $95$th percentile was $3.41\%$
(Fig.~\ref{fig:preservation}b). The correlation between generator
draws and the observed market path averaged near zero for every asset
(Online Appendix~\ref{sec:supp-cov}), which supported the zero-covariance
assumption on average but not on each path.

% Main figure fig:preservation is collected after the references.

% Main table tab:aggregate is collected after the references.

The methods differed more in distributional fit than in factor
recovery (Table~\ref{tab:aggregate}; Fig.~\ref{fig:tails}). All six
methods had median absolute $\beta$ errors between $0.017$ and
$0.022$. Naive and hybrid composition had a median absolute intercept
error of $0.002\,\mathrm{yr}^{-1}$, an order of magnitude below the
other methods, because their centered residual paths reproduced the
observed training asset means exactly when combined with observed SPY
without clipping. Refitting OLS left an intercept error equal to the
loading error times the market mean. The small intercept errors thus
followed from centering, not from predicting asset means. Gaussian SIM had the
lowest median $R^2$ error but almost never passed the distribution
tests, and its median excess kurtosis was an order of magnitude below
the observed median of about $13$. Hybrid composition passed
$73.6\%$ of KS comparisons, against $7.1\%$ for naive composition,
and its advantage held across the calibrated $\beta$ range. Its own
pass rate nonetheless fell from $87.6\%$ in the lowest $\beta$
quartile to $65.9\%$ in the highest
(Fig.~\ref{fig:preservation}a; Table~\ref{tab:beta_bucket}). A larger market variance share may help explain this trend by
diluting the asset generator's shape (Online
Appendix~\ref{sec:supp-residual-choice}). Hybrid composition
retained heavy tails, with median excess kurtosis of $7.2$ and a
median Hill estimate of $0.37$, reported as the reciprocal tail index
so that larger values indicate heavier tails (Fig.~\ref{fig:tails}).
The excess-kurtosis shortfall was also present under naive
composition, whose median was similar. Hybrid composition's median
$R^2$ error fell between those of the naive and Gaussian methods,
because its variance-preserving branch targeted generator variance
rather than observed $R^2$.

Direct residual fitting was competitive with reuse, and controlling
for residual centering favored the residual-fit HMM and block
bootstrap (Tables~\ref{tab:aggregate}
and~\ref{tab:centering_control}). Under each method's native mean
treatment, JumpHMM-on-residuals passed $75.8\%$ of KS comparisons,
slightly above hybrid composition; block bootstrap passed $73.3\%$
and GARCH(1,1)-$t$ passed $67.4\%$. The residual-fit HMM and block
bootstrap had slightly lower Wasserstein distances than hybrid
composition. The residual-fit HMM also closely matched hybrid composition
in excess kurtosis and Hill estimate, whereas block bootstrap and GARCH
produced lighter tails. Only naive and hybrid composition centered
their residual draws, so we repeated the comparison after centering
all six methods on the common $393$-asset set
(Table~\ref{tab:centering_control}). The hybrid KS pass rate stayed
at $74.4\%$ because its paths were already centered; the
JumpHMM-on-residuals rate rose from $75.8\%$ to $78.4\%$, and block
bootstrap rose from $73.5\%$ to $76.7\%$. Both residual methods
also had higher AD pass rates and lower Wasserstein distances than
hybrid composition after centering. The control brought every
method's median intercept error to $0.002\,\mathrm{yr}^{-1}$ while
leaving variance, recovered loading, $R^2$, and excess kurtosis
unchanged. A separate training diagnostic found that emission fallback rules
also disadvantaged the full-return methods on $39$ assets but affected
none of the residual-fit models (Table~\ref{tab:fallback_sensitivity}; Online Appendix~\ref{sec:supp-fallback}). Using empirical scales in their
populated zero-return states implied an aggregate hybrid KS gain of
about one percentage point, less than the gap to JumpHMM-on-residuals.
The main comparisons retained the original rule because this
alternative had no holdout validation (Online
Appendix~\ref{sec:supp-fallback}). Together, these
checks supported variance correction over naive reuse without a
second fit, but did not establish an advantage over direct residual
fitting.

\subsection{Holdout performance and VaR coverage}

\begin{figure}[H]
  \centering
  \includegraphics[width=\textwidth]{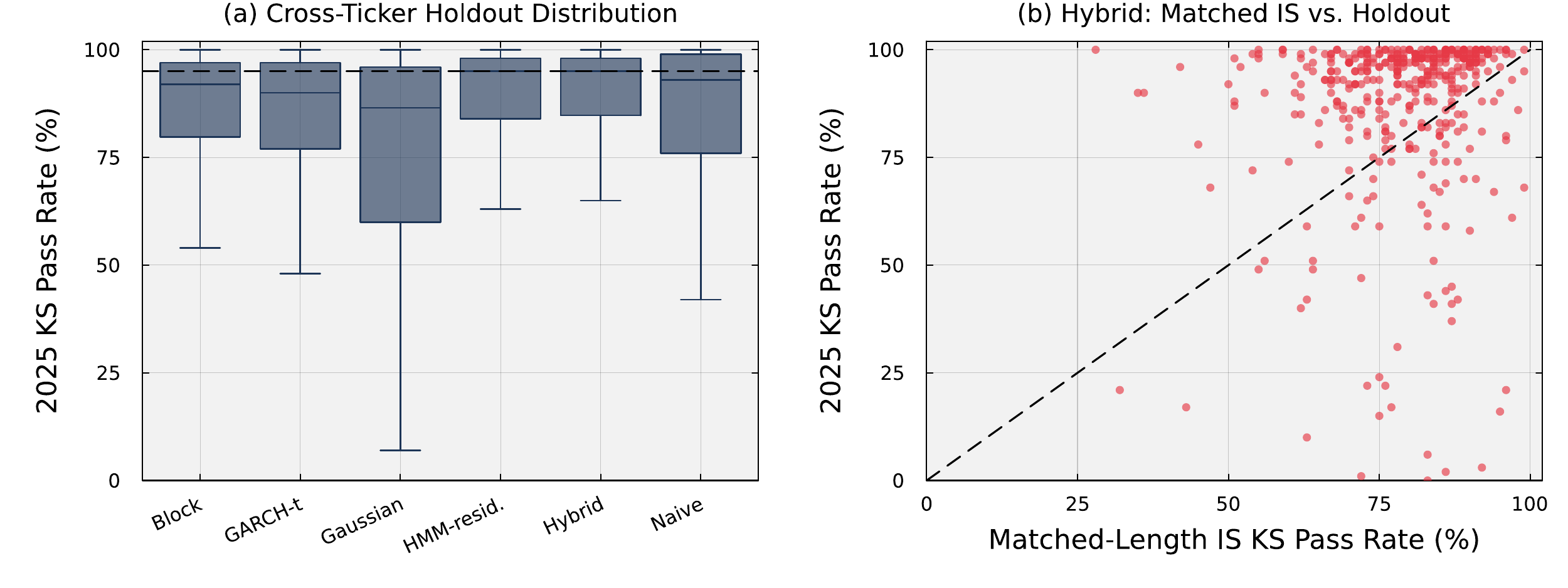}
  \caption{\textbf{Cross-ticker distribution of frozen-parameter
    out-of-sample marginal fit.} Panel~(a) shows the distribution of
    per-ticker $2025$ Kolmogorov--Smirnov (KS) pass rates across $100$ replications; the
    dashed line marks $95\%$. Panel~(b) compares each ticker's hybrid
    KS pass rate against a matched-length training block with its
    pass rate against the $2025$ holdout. The diagonal denotes equal
    performance.}
  \label{fig:oos_composition}
\end{figure}

To test out-of-sample generalization without refitting or look-ahead
bias, we froze the fitted models and SIM parameters before evaluating
them against $416$ complete asset histories from $2025$
(Table~\ref{tab:oos_composition}). Hybrid composition reached an
$87.0\%$ mean cross-ticker KS pass rate, against $83.5\%$ for naive
composition. Its median ratio of synthetic to observed $2025$ variance
was $0.97$, whereas naive composition inflated that ratio to $1.29$.
The JumpHMM-on-residuals method reached an $86.2\%$ KS pass rate and a slightly lower median $W_1$ than hybrid composition. As in
training, these holdout comparisons used each method's native mean
treatment, and no common-centering control was run on the holdout. Higher holdout pass rates did not by themselves establish better
generation: all methods passed more often than in training, and the
gap between hybrid and naive composition narrowed (Tables~\ref{tab:aggregate}
and~\ref{tab:oos_composition}). For every method, pass rates against
matched-length $249$-day training blocks lay between the full-length
training and $2025$ rates, consistent with lower test power in shorter
samples contributing to the rise. Those blocks also used simulated
rather than observed SPY, so the comparison did not isolate sample
length or fully explain the $2025$ increase. Nor did improvement hold
for every asset: about three quarters of hybrid tickers passed more
often in $2025$ than against their matched training blocks and one
quarter passed less often, and every method retained a lower tail of
weakly fitted tickers (Fig.~\ref{fig:oos_composition}).

To test whether the holdout improvement over naive composition
extended to reliable tail thresholds, we estimated per-ticker one-day
left-tail Value-at-Risk (VaR) thresholds from $5{,}000$ separate
$249$-day paths and counted exceedances on the observed $2025$
histories (Table~\ref{tab:var}). The coverage level
$\alpha=0.95$ has a nominal exceedance rate of $1-\alpha=5\%$ and gave
each asset about $12$ expected exceedances. At that level, hybrid
composition had a mean
exceedance rate of $5.48\%$, closer to nominal than the $4.04\%$ of
naive composition. At $\alpha=0.99$, the hybrid rate of $1.12\%$ and
the naive rate of $0.88\%$ lay equally close to the $1\%$ nominal rate
on opposite sides. The JumpHMM-on-residuals method had rates of
$5.62\%$ and $1.16\%$, close to hybrid composition, and Gaussian SIM
had the largest $99\%$ rate at $1.72\%$. Doubling the threshold pool from
$2{,}500$ to $5{,}000$ paths changed every method's mean exceedance
rate by less than $0.006$ percentage points at either coverage level
(Online Appendix~\ref{sec:supp-var-ensemble}). The advantage of hybrid over
naive composition thus appeared at the $95\%$ level but not at the
$99\%$ level. Neither comparison established a ranking, because each
asset contributed few exceedances, only about $2$--$3$ expected at the
$99\%$ level, and all assets shared one market period, so their
exceedances were not independent. The hybrid mean exceedance rates were above nominal at both
coverage levels, and this descriptive check could not confirm
threshold calibration. The
cross-asset standard deviations of the exceedance rates described how
realized rates varied among assets; they did not test miscalibration
or validate coverage in other periods.

% Main table tab:oos_composition is collected after the references.

% Main figure fig:oos_composition is collected after the references.

% Main table tab:var is collected after the references.

\subsection{Jump-enabled multi-asset composition}
\label{sec:results-jump-composition}

Transferring SPY's jump settings improved temporal fit in training
but worsened it in the $2025$ holdout
(Table~\ref{tab:jump_ablation}, where hybrid composition is labeled
Corrected). We compared jumps disabled, market-only jumps, and
market-plus-asset jumps under naive and hybrid composition, without
per-asset tuning. This experiment simulated the market in both
windows and used each market path's variance in the correction, so
its baseline differed from the six-method comparisons. Its ACF-MAE
was calculated for each path and then averaged, unlike the
single-asset error calculated from the ensemble-average
autocorrelation. Under hybrid composition, market-only jumps reduced
the training $25$-lag ACF-MAE by $5.3\%$; the $95\%$ Monte Carlo
interval excluded zero. The KS pass rate fell by $0.3$ percentage
points, with an interval that included zero
(Table~\ref{tab:jump_ablation_uncertainty}). Adding asset jumps
reduced ACF-MAE by $14.6\%$ relative to no jumps but lowered the KS
pass rate from $73.9\%$ to $61.7\%$, a much larger marginal-fit
cost than for SPY, whose settings were tuned. In the holdout,
market-only and market-plus-asset jumps instead raised the $25$-lag
ACF-MAE by $2.7\%$ and $6.9\%$. Both $95\%$ Monte Carlo
intervals excluded zero, and the $60$-lag errors also rose
(Tables~\ref{tab:jump_ablation}
and~\ref{tab:jump_ablation_uncertainty}). Holdout KS pass rates
changed little with market-only jumps; with both market and asset
jumps, they fell by about $1.5$ percentage points relative to no jumps. Across all settings
and both windows, the mean ratio of composed to generator variance
under hybrid composition remained within $0.05\%$ of one. This
aggregate included the trackers, whose branch targeted $R^2$ rather
than generator variance, and did not establish preservation for every
asset. Only $2.2\%$ of simulated one-year market paths contained an
episode (Online Appendix~\ref{sec:supp-jump-composition}), limiting evidence
about market jumps; the single observed holdout year also could not
establish performance in other market periods.

% Main table tab:jump_ablation is collected after the references.

\subsection{Sensitivity analyses}

We examined how the branch-selection threshold
$R^2_{\mathrm{preserve}}$ assigned assets to branches in the observed
universe and whether any value of the variance floor $f$ triggered
clipping (Online Appendix~\ref{sec:supp-sim-derivations}). The
$R^2$-preserving branch applied only to QQQ and SPYG, whose hybrid
compositions passed every KS test in training. The highest-$R^2$ asset
outside these trackers, the individual stock BLK, fell about $0.16$
below the $0.80$ threshold, so the assignment did not depend on a
narrow margin. Across the explored thresholds from $0.70$ to $0.90$ at
the default floor, only the QQQ assignment changed, moving to the
variance-preserving branch at $0.90$. At the default threshold, no
value of $f$ from $0.05$ to $0.30$ caused any asset to be clipped. The
empirical evaluation of the $R^2$-preserving branch was therefore
limited to two assets, and the observed market variance never
activated clipping.

A synthetic Gaussian check supplemented the two real assets using
the $R^2$-preserving branch: across the tested grid, median recovered
loadings and $R^2$ differed from their supplied targets by at most
$0.0011$ and $0.0017$, respectively (Online Appendix~\ref{sec:supp-sim-derivations}).
Each reference series combined the observed training SPY path with an
independent Gaussian residual whose variance gave the target $R^2$
under Eq.~\eqref{eq:r2-target}. The composer received the true
loading and $R^2$, a second independent Gaussian draw, and the target
residual variance, which made the residual scale in
Eq.~\eqref{eq:r2-rescale} exactly one. Every composed series passed
the KS comparison with its reference, but their shared SPY path made
the samples dependent, so the pass rates were descriptive rather than
calibrated tests. The grid checked implementation and finite-sample
accuracy under this restricted construction; it did not exercise
full-return HMM fitting, non-unit residual scaling, or simulated
market paths.

Increasing market volatility activated clipping for most assets,
although no path clipped under the observed market variance
(Online Appendix~\ref{sec:supp-sim-derivations}). A larger market variance raises the market
loading ratio $\rho_i$, the market's share of the target variance;
on the variance-preserving branch, clipping begins when
$\rho_i>1-f$. To measure when clipping activates, we
multiplied deviations of the observed training SPY path from its mean
by $\gamma\in\{1,2,3\}$, which preserved the market mean and
multiplied its standard deviation by $\gamma$. We then recomposed all
$423$ assets with the scaled path's variance in the correction,
holding the fitted marginals, intercepts, and calibrated loadings
fixed (Online Appendix~\ref{sec:supp-sim-derivations}). Clipping applied to no composed path
at $\gamma=1$ and to $76.4\%$ and $95.2\%$ of composed paths at
$\gamma=2$ and $3$. Because clipped paths retained their fitted
intercepts while the market mean stayed unchanged and nonzero, their
means shifted by $(\beta_i^{\eff}-\beta_i)\overline{g}_m$ by
construction. The stress test therefore characterized when clipping
activated and how it moved asset means; it did not evaluate fit under
the scaled markets.

\section{Discussion}\label{sec:discussion}
Synthetic generators of equity returns should reproduce three
stylized facts: heavy tails, weak autocorrelation in raw returns, and
volatility clustering (Fig.~\ref{fig:empirical_motivation}). We used
heavy-tailed hidden Markov generators with optional jump episodes to
represent these features, then examined their reuse in a multi-asset
model with a shared market path. We developed a correction that
centers and rescales each generator's draws to avoid counting market
variance twice, without refitting to regression residuals
(Fig.~\ref{fig:preservation}; Table~\ref{tab:aggregate}). Training and
$2025$ holdout comparisons tested distributional fit and whether
single-asset improvements in volatility clustering survived
composition.

The correction's practical value was avoiding a separate residual
fit while improving variance and distributional agreement over naive
composition (Fig.~\ref{fig:preservation};
Table~\ref{tab:aggregate}). The corrected method passed $73.6\%$ of
full-length training KS comparisons and retained heavy tails. On the
$2025$ holdout, it raised the mean cross-ticker KS pass rate from
$83.5\%$ under centered naive composition to $87.0\%$ and brought
the median ratio of synthetic to observed variance to $0.97$
(Table~\ref{tab:oos_composition}). The improvement therefore
persisted beyond the fitting period. Direct residual fitting remained
competitive: JumpHMM-on-residuals had a slightly higher pass rate in
training and a slightly lower rate in the holdout. When all methods'
residual draws were centered over the same training horizon on the
common asset set, JumpHMM-on-residuals and block bootstrap both
outperformed the corrected method on distribution-test pass rates and
Wasserstein distance (Table~\ref{tab:centering_control}). Before centering was matched, the comparison with residual-fit
methods therefore partly reflected different constraints on simulated
residual means. For corrected composition, some assets had lower pass rates in the
holdout than against matched-length training blocks. Reuse thus offered a way to improve
on naive composition without refitting, but did not establish a
distributional advantage over fitting the residuals themselves.

The jump comparison showed that improving the single-asset generator
did not automatically improve the composed paths
(Tables~\ref{tab:model_comparison} and~\ref{tab:jump_ablation}). On
SPY, the generator with jumps reduced absolute-return autocorrelation
error relative to its no-jump counterpart while retaining high
distributional pass rates. Other generators achieved lower
autocorrelation error or higher pass rates, so temporal and marginal
fit still required separate assessment. With multi-asset composition,
market-only jumps improved training temporal fit with little change
in the KS pass rate, whereas adding asset-level episodes produced a
larger temporal improvement at a substantial marginal-fit cost. Both
settings increased autocorrelation error in $2025$, despite the mean
composed-to-generator variance ratio remaining near one
(Table~\ref{tab:jump_ablation}). Variance control therefore did not
ensure that the transferred temporal behavior matched the data. These
tests transferred one SPY calibration from volume-weighted average
prices to closing-price models. The asset episodes also came from
full-return models and were not identified separately as idiosyncratic
events. Separate calibrations for NVDA, JNJ, and JPM had strong
in-sample fit, but three assets did not establish a general calibration
rule (Online Appendix~\ref{sec:supp-cross-asset}). Per-asset tuning or distinct shared
and asset-specific episode models could change the tradeoff, but
neither was tested in the composition comparison.

Preserving total variance did not preserve the complete return
distribution, and corrected fit varied with market exposure
(Fig.~\ref{fig:preservation}; Table~\ref{tab:beta_bucket}). Corrected
KS pass rates fell from $87.6\%$ in the lowest $\beta$ quartile to
$65.9\%$ in the highest, consistent with a larger market contribution
diluting the asset generator's shape. The correction accounted for variance
supplied by the market, but combining two independent
return distributions still changed the shape of the composed returns.
Clipping, the reduction of the market loading to retain some residual
variance, was unnecessary at observed market variance but activated
for most assets when market volatility was doubled or tripled (Online Appendix~\ref{sec:supp-sim-derivations}). That stress transformation
preserved the market mean. Reducing the loading while retaining the fitted intercept changed
the asset mean by
$(\beta_i^{\eff}-\beta_i)\overline{g}_m$, where $\overline{g}_m$
is the market mean. The clipped branch therefore targeted total variance while retaining
a minimum residual contribution, at the cost of changing the
calibrated loading and mean. The stress test did not establish realistic return distributions
under the scaled markets. Variance correction therefore required separate checks of
distributional shape and calibrated market exposure.

The correction supported daily-return comparisons, but broader
simulation uses require additional modeling and validation. The alternative
correction for market trackers targeted their calibrated $R^2$, the
fraction of return variance explained by the market. Only two
observed assets tested it; a Gaussian simulation checked its formula
without fitting full-return generators (Online Appendix~\ref{sec:supp-sim-derivations}). Allowing states whose training observations were identical to
generate nearly constant returns improved training fit (Table~\ref{tab:fallback_sensitivity}). For those states, the main comparisons retained emissions scaled
by the asset's full-series spread, and the alternative still needs holdout validation. Improved
marginal fit did not establish reliable tail coverage: the corrected
method and JumpHMM-on-residuals had pooled-threshold one-day left-tail
$99\%$ VaR exceedance rates of $1.12\%$ and $1.16\%$, against a
nominal $1\%$ (Table~\ref{tab:var}). More simulated paths improved
threshold precision without supplying more observed extreme events;
the $249$-day holdout was too short for a precise ranking and covered
only one shared market period. Serial dependence also limited the
interpretation of KS and AD pass rates, whose tests assume independent
observations. Variance accounting also assumed independent asset and market
draws; nonzero covariance would require an additional cross-term.
For portfolio applications,
a shared market path did not supply the sector and style dependence
omitted by the evaluated independent residual draws
(Algorithm~\ref{alg:hybrid}; Online Appendix~\ref{sec:supp-bias}).
Omitting positive residual covariance can overstate the
diversification return from periodic rebalancing~\cite{boothFama1992,
boucheyNemtchinovWong2015, markowitz1976longRun}, including for
residual-fit baselines that draw residuals independently. A
synchronized multivariate block bootstrap, residual factor model, or
copula could address that omission, but the resulting dependence
needs direct evaluation. Per-path centering imposed a separate
limitation: each residual sum was zero at the chosen horizon, leaving
no residual uncertainty in terminal cumulative log returns conditional on
the market path and effective loading (Online
Appendix~\ref{sec:supp-sim-derivations}). Portfolio and cumulative-risk
applications therefore require both a joint residual model and a
construction that retains cumulative residual variation.

\section{Conclusion}\label{sec:conclusion}
We developed a variance-corrected single-index method that centers and
rescales full-return draws for reuse without counting market variance
twice or fitting separate residual generators
(Fig.~\ref{fig:preservation}; Table~\ref{tab:aggregate}). The method
retained heavy tails and improved distributional fit over centered
naive composition in training and the $2025$ holdout
(Table~\ref{tab:oos_composition}), but did not establish an advantage
over direct residual fitting. Transferring SPY's jump settings
improved temporal fit in training but worsened it in the holdout
(Table~\ref{tab:jump_ablation}), showing that single-asset temporal
gains did not transfer reliably. These findings support reuse for
daily-return comparisons, without establishing calibrated tail
coverage (Table~\ref{tab:var}) or validating broader risk applications.
Independent residual draws omitted cross-asset dependence, and
per-path centering removed residual uncertainty from terminal
cumulative log returns conditional on the market path and the loading
used in composition. Portfolio and cumulative-risk applications therefore require
a joint residual model and a construction that retains cumulative
residual variation.

\section*{Acknowledgments}
The authors thank the eCornell artificial intelligence (AI) in Finance program participants for
discussions that motivated this work,
and Polygon.io for access to the 2014 to 2024 United States equity and
exchange-traded-fund daily
price histories used in the experiments. Computations were run on a
single laptop; no shared-resource allocation was required.

\paragraph{Author contributions.}
A.A.\ designed the hybrid hidden Markov marginal generator, performed the
single-asset experiments and validation analyses, and contributed to the
manuscript. J.D.V.\ designed the variance-corrected single-index
composition, performed the multi-asset experiments and downstream
Value-at-Risk validation, released the code, and contributed to the
manuscript.

\paragraph{Competing interests.}
The authors declare no competing interests.

\needspace{12\baselineskip}
\paragraph{Code and data availability.}
The experiment scripts, evaluation code (six-method comparison and
Value-at-Risk coverage check), cached inputs and
result summaries, and instructions for regenerating the fitted
per-asset models are provided in the paper repository
(\url{https://github.com/varnerlab/HMM-w-jumps-paper}); the
underlying model-fitting library is released separately as
\texttt{JumpHMM.jl}
(\url{https://github.com/varnerlab/JumpHMM.jl}), pinned as a
dependency of the experiment code.

\paragraph{Generative artificial intelligence use.}
During the preparation of this work, the authors used OpenAI Codex to
assist with language editing, consistency checks, LaTeX formatting,
and the implementation and checking of experiment code. The authors
reviewed and edited the assisted work and take full responsibility for
the content of the publication.

\clearpage
\bibliographystyle{unsrtnat}
\bibliography{References_v1}

\clearpage
\section*{Tables}
% Main-manuscript tables, in their original numerical order.
% Kept separate from SI tables and from the journal layout.

\begin{table}[H]
\centering
\caption{\textbf{Hidden Markov model with jumps (HMM-WJ) and
    single-index model (SIM) composition hyperparameters.} Values
    used for the main SPY analysis, explored ranges, and locations of
    sensitivity checks. For the primary six-method SIM comparison, we set
    $\epsilon = 0$ instead of the SPY value. The separate jump-enabled
    comparison transferred the listed SPY settings to enabled models. The cross-asset
    validation used per-ticker $(\epsilon^{*}, \lambda^{*})$ values
    for NVDA, JNJ, and JPM (Online Appendix~\ref{sec:supp-cross-asset}).}
\label{tab:hyperparams}
\small
\begin{tabular}{@{}llll@{}}
\toprule
\textbf{Symbol} & \textbf{Description} & \textbf{SPY value} & \textbf{Explored range} \\ \midrule
\multicolumn{4}{l}{\textit{State partition and emission}} \\[2pt]
$N$            & Laplace-quantile states                      & $100$     & $\{30,\dots,200\}$, Online Appendix~\ref{sec:supp-sensitivity} \\
$\nu$          & Student-$t$ emission degrees of freedom      & $5$       & $\{3,\dots,30,\infty\}$ \\[4pt]
\multicolumn{4}{l}{\textit{Jump-duration override}} \\[2pt]
$\epsilon$     & jump probability per step                    & $10^{-4}$ & $[10^{-4}, 2.5\times10^{-2}]$, Eq.~\eqref{eq:grid_search} \\
$\lambda$      & mean jump-episode duration (steps)           & $90$      & $[10, 160]$, Eq.~\eqref{eq:grid_search} \\
$p_{\rm neg}$  & probability a jump targets $\mathcal{S}_{-}$ & $0.52$    & fixed \\
$N_{\rm tail}$ & quantile states per tail set                 & $5$       & fixed (at $N = 100$) \\
$w_\kappa$     & kurtosis-penalty weight, Eq.~\eqref{eq:grid_search} & $0.20$ & fixed \\
$L$            & autocorrelation lag window, Eq.~\eqref{eq:grid_search} & $25$ days & fixed \\[4pt]
\multicolumn{4}{l}{\textit{SIM composition}} \\[2pt]
$f$                        & idiosyncratic variance floor & $0.10$ & $\{0.05,\dots,0.30\}$ \\
$R^2_{\mathrm{preserve}}$  & branch-selection threshold   & $0.80$ & $\{0.70,\dots,0.90\}$ \\
\bottomrule
\end{tabular}
\end{table}

% Laplace-only correction: 1,000 paths per window; paired seeds 20260914/20260915.
% Provenance: audits/2026-09-14-review/laplace-correction/metadata.toml.
% Other model values are retained from the preceding Table 2.
\begin{table}[H]
\centering
\caption{\textbf{In-sample and out-of-sample model comparison for SPY.}
Results summarize $1{,}000$ simulated paths. Values in parentheses are
Monte Carlo standard errors conditional on the fitted model and observed
history, and bold entries are best among the fitted generative
models within each panel; the empirical bootstrap is excluded from
this highlighting. Metric
definitions and standard-error calculations are given in Online
Appendix~\ref{sec:supp-validation}.}
\label{tab:model_comparison}
\footnotesize
\setlength{\tabcolsep}{1.8pt}
\renewcommand{\arraystretch}{1.08}
\resizebox{\textwidth}{!}{%
\begin{tabular}{@{}lccccccc@{}}
\toprule
\textbf{Model} & \textbf{KS (\%)} & \textbf{AD (\%)} & $\boldsymbol{\kappa}$ &
\textbf{ACF-MAE} & \textbf{Coverage (\%)} & $\boldsymbol{W_1}$ & \textbf{Hellinger} \\
\midrule
\multicolumn{8}{@{}l}{\textit{In sample: $2014$--$2024$ ($2{,}766$ days; observed $\kappa=7.715$)}} \\
Bootstrap & 100.0 ($<$0.1) & 99.7 (0.2) & 7.6 (0.08) & 0.060 ($<$0.001) & 100.0 ($<$0.1) & 0.062 (0.001) & 0.042 ($<$0.001) \\
Gaussian  & 0.0 ($<$0.1) & 0.0 ($<$0.1) & $-$0.0 ($<$0.01) & 0.060 ($<$0.001) & 13.1 (0.1) & 0.399 (0.001) & 0.148 ($<$0.001) \\
Laplace   & 98.4 (0.4) & 96.8 (0.6) & 3.0 (0.02) & 0.060 ($<$0.001) & 78.8 (1.0) & 0.119 (0.001) & \textbf{0.069} ($<$0.001) \\
GARCH(1,1) & 5.5 (0.7) & 1.9 (0.4) & 8.2 (0.37) & \textbf{0.031} (0.001) & 29.3 (1.4) & 0.304 (0.006) & 0.100 (0.001) \\
GRU       & 0.6 (0.2) & 0.2 (0.1) & 5.7 (0.16) & 0.036 ($<$0.001) & 17.2 (0.5) & 0.421 (0.003) & 0.134 ($<$0.001) \\
HSMM      & 82.0 (1.2) & 42.5 (1.6) & 4.8 (0.08) & 0.059 ($<$0.001) & 68.7 (1.1) & 0.176 (0.001) & 0.113 ($<$0.001) \\
HMM-NJ    & \textbf{99.3} (0.3) & \textbf{98.2} (0.4) & 8.1 (0.14) & 0.059 ($<$0.001) & \textbf{100.0} ($<$0.1) & \textbf{0.081} (0.001) & 0.072 ($<$0.001) \\
HMM-WJ    & 98.3 (0.4) & 94.8 (0.7) & \textbf{7.5} (0.09) & 0.053 ($<$0.001) & \textbf{100.0} ($<$0.1) & 0.097 (0.001) & 0.074 ($<$0.001) \\[3pt]
\midrule
\multicolumn{8}{@{}l}{\textit{Out of sample: $2025$ ($249$ days; observed $\kappa=6.867$)}} \\
Bootstrap & 97.4 (0.5) & 98.5 (0.4) & 6.0 (0.18) & 0.043 ($<$0.001) & 100.0 ($<$0.1) & 0.232 (0.002) & 0.205 (0.001) \\
Gaussian  & 62.2 (1.5) & 46.3 (1.6) & $-$0.0 ($<$0.01) & 0.043 ($<$0.001) & 36.4 (1.9) & 0.452 (0.002) & 0.235 (0.001) \\
Laplace   & \textbf{96.7} (0.6) & \textbf{98.1} (0.4) & 2.8 (0.05) & 0.043 ($<$0.001) & 99.0 (0.4) & \textbf{0.242} (0.002) & \textbf{0.207} (0.001) \\
GARCH(1,1) & 80.3 (1.3) & 72.0 (1.4) & 1.6 (0.07) & \textbf{0.026} ($<$0.001) & 96.0 (1.3) & 0.507 (0.018) & 0.232 (0.001) \\
GRU       & 71.8 (1.4) & 44.8 (1.6) & 2.9 (0.10) & 0.033 ($<$0.001) & 77.8 (2.9) & 0.488 (0.005) & 0.240 (0.001) \\
HSMM      & 96.2 (0.6) & 96.7 (0.6) & 4.1 (0.13) & 0.042 ($<$0.001) & \textbf{100.0} (0.2) & 0.287 (0.002) & 0.239 (0.001) \\
HMM-NJ    & 96.6 (0.6) & 96.1 (0.6) & \textbf{6.2} (0.15) & 0.041 ($<$0.001) & \textbf{100.0} ($<$0.1) & 0.259 (0.003) & \textbf{0.207} (0.001) \\
HMM-WJ    & 95.4 (0.7) & 95.3 (0.7) & \textbf{6.2} (0.15) & 0.040 ($<$0.001) & \textbf{100.0} ($<$0.1) & 0.275 (0.005) & 0.210 (0.001) \\
\bottomrule
\end{tabular}}
\vspace{3pt}

\parbox{0.98\textwidth}{\scriptsize
Model abbreviations are generalized autoregressive conditional heteroskedasticity
(GARCH), gated recurrent unit (GRU), hidden semi-Markov model (HSMM), hidden
Markov model with no jumps (HMM-NJ), and hidden Markov model with jumps
(HMM-WJ). Kolmogorov--Smirnov (KS) and Anderson--Darling (AD) report pass
rates at $\alpha=0.05$; coverage is the fraction
of the $99$ observed quantiles at levels $0.01,\ldots,0.99$
inside their pointwise synthetic $90\%$ envelopes.
$\kappa$ is simulated excess kurtosis. Mean absolute error in the autocorrelation
function (ACF-MAE) compares the observed ACF of $|G_t|$ with the
mean ACF over paths, using lags $1$--$252$ in training and
$1$--$248$ in the holdout. Larger values are preferred for KS, AD,
and coverage; smaller values are preferred for ACF-MAE, $W_1$, and
Hellinger distance; $\kappa$ is compared with the observed value.}
\end{table}

\begin{table}[H]
  \centering
  \caption{\textbf{Aggregate training scorecard with
    $100$ replications per ticker.} Errors, distances, and tail metrics
    are medians per method. Kolmogorov--Smirnov (KS) and
    Anderson--Darling (AD) pass fractions are pooled across replications.
    $|\hat\alpha - \alpha|$, $|\hat\beta - \beta|$, and
    $|\hat R^2 - R^2_{\real}|$: absolute deviation between ordinary
    least-squares (OLS) recovery
    and calibration (intercept error has units of $\mathrm{yr}^{-1}$).
    KS pass and AD pass: fraction of
    replications whose two-sample test against the real marginal exceeds
    $p = 0.05$. $W_1$: Wasserstein-1 distance to the real marginal.
    $\kappa$: sample excess kurtosis. Hill: tail-index estimate on the
    upper $5\%$ of $|g|$. SIM denotes the single-index model, and GARCH
    denotes generalized autoregressive conditional heteroskedasticity.
    The GARCH method covered $393$ non-market tickers; the other five
    methods covered $423$.}
  \label{tab:aggregate}
  \small
  \resizebox{\textwidth}{!}{\begin{tabular}{lrrrrrrrr}
\toprule
Method & $|\hat\alpha-\alpha|$ & $|\hat\beta-\beta|$ & $|\hat R^2 - R^2_{\real}|$ & KS pass (\%) & AD pass (\%) & $W_1$ & $\kappa$ & Hill \\
\midrule
Naive & 0.002 & 0.022 & 0.087 & 7.1 & 3.5 & 0.700 & 7.767 & 0.369 \\
Gaussian SIM & 0.048 & 0.017 & 0.008 & 0.6 & 0.5 & 0.682 & 1.427 & 0.217 \\
Hybrid & 0.002 & 0.018 & 0.021 & 73.6 & 64.6 & 0.249 & 7.165 & 0.365 \\
JumpHMM-on-residuals & 0.049 & 0.018 & 0.015 & 75.8 & 70.7 & 0.232 & 7.262 & 0.365 \\
Block bootstrap & 0.042 & 0.017 & 0.020 & 73.3 & 66.8 & 0.232 & 6.675 & 0.330 \\
GARCH(1,1)-$t$ & 0.047 & 0.017 & 0.036 & 67.4 & 61.4 & 0.267 & 6.069 & 0.317 \\
\bottomrule
\end{tabular}
}
\end{table}

\begin{table}[H]
  \centering
  \caption{\textbf{Frozen-parameter multi-asset evaluation on the
    $2025$ holdout.} Results cover $416$ non-market tickers and $100$
    replications per ticker; generalized autoregressive conditional
    heteroskedasticity (GARCH) covers $386$ tickers. Pass rates
    are first averaged within ticker and then across tickers.
    The \textit{matched in-sample (IS)} columns compare each synthetic path with a randomly
    selected contiguous $249$-day block from the $2014$--$2024$
    training history, matching the holdout sample length. $W_1$ and
    the variance ratio are medians across ticker-level medians; the
    variance ratio is synthetic variance divided by observed $2025$
    variance. The hybrid correction used the training-period market
    variance and each simulated asset draw's variance.}
  \label{tab:oos_composition}
  \small
  \begin{tabular}{lrrrrrr}
\toprule
 & \multicolumn{2}{c}{KS pass (\%)} & \multicolumn{2}{c}{AD pass (\%)} & & \\
\cmidrule(lr){2-3} \cmidrule(lr){4-5}
Method & matched IS & 2025 & matched IS & 2025 & $W_1$ (2025) & variance ratio \\
\midrule
Naive & 69.4 & 83.5 & 54.8 & 71.6 & 0.877 & 1.29 \\
Gaussian SIM & 57.4 & 73.9 & 44.9 & 64.6 & 0.866 & 0.97 \\
Hybrid & 78.8 & 87.0 & 66.2 & 79.7 & 0.716 & 0.97 \\
JumpHMM-on-residuals & 77.9 & 86.2 & 65.6 & 80.1 & 0.708 & 0.98 \\
Block bootstrap & 76.3 & 84.0 & 64.0 & 76.5 & 0.715 & 0.90 \\
GARCH(1,1)-$t$ & 74.5 & 82.6 & 62.6 & 75.0 & 0.733 & 0.91 \\
\bottomrule
\end{tabular}

\end{table}

\begin{table}[H]
  \centering
  \caption{\textbf{Out-of-sample per-ticker one-day left-tail Value-at-Risk
    coverage across the six composition methods.} For each asset and
    coverage level $\alpha$, one VaR threshold was estimated from
    $1{,}245{,}000$ daily log returns pooled across $5{,}000$ separate
    $249$-day paths with frozen parameters. Exceedances were counted
    on the observed $2025$ history. The hybrid correction retained
    the training-period market variance convention of
    Table~\ref{tab:oos_composition}. \textit{rate}: mean per-asset
    exceedance rate (nominal: $1-\alpha$). Standard deviation (SD)
    describes cross-asset dispersion in percentage points; it is not
    a standard error. Coverage summaries are descriptive.}
  \label{tab:var}
  \small
  \setlength{\tabcolsep}{5pt}
  \begin{tabular}{lrrrrr}
\toprule
 & & \multicolumn{2}{c}{$\alpha = 0.95$} & \multicolumn{2}{c}{$\alpha = 0.99$} \\
\cmidrule(lr){3-4} \cmidrule(lr){5-6}
Method & Assets & rate (\%) & SD (pp) & rate (\%) & SD (pp) \\
\midrule
Naive & 416 & 4.04 & 2.08 & 0.88 & 0.70 \\
Gaussian SIM & 416 & 4.65 & 2.22 & 1.72 & 1.08 \\
Hybrid & 416 & 5.48 & 2.45 & 1.12 & 0.81 \\
JumpHMM-on-residuals & 416 & 5.62 & 2.55 & 1.16 & 0.81 \\
Block bootstrap & 416 & 5.50 & 2.52 & 1.28 & 0.87 \\
GARCH(1,1)-$t$ & 386 & 5.07 & 2.45 & 1.20 & 0.89 \\
\bottomrule
\end{tabular}

\end{table}

\begin{table}[H]
  \centering
  \caption{\textbf{Jump settings under naive and corrected multi-asset
    composition.} Each setting used $1{,}000$ replications per asset
    across four seeds, with identical asset and market draws for the
    two composition methods. SPY was simulated in both windows,
    and the correction used each simulated market path's variance
    and each asset draw's variance. Entries are equal-weight
    cross-ticker means of per-path metrics.
    Kolmogorov--Smirnov (KS) and Anderson--Darling (AD) report
    non-rejection rates at $\alpha=0.05$. ACF$_{25}$ and ACF$_{60}$
    are absolute-return autocorrelation errors through lags $25$
    and $60$, calculated within each path before averaging over
    replications and assets; lower values indicate better
    temporal fit. $V_{\gen}=\Var(g_i)/\sigma_{\gen,i}^2$ compares
    composed variance with the active generator's variance, rather
    than observed variance.}
  \label{tab:jump_ablation}
  \footnotesize
  \setlength{\tabcolsep}{4pt}
  % Generated by 11c-Jump-Ablation-Tables.py; do not edit numbers manually.
\begin{tabular}{@{}llrrrrr@{}}
\toprule
Jump setting & Method & KS (\%) & AD (\%) & ACF$_{25}$ & ACF$_{60}$ & $V_{\gen}$ \\
\midrule
\multicolumn{7}{@{}l}{\textit{Training: $2014$--$2024$, $423$ assets, $2{,}766$ days}} \\[2pt]
Off & Naive & 6.93 & 3.38 & 0.13585 & 0.09492 & 1.33273 \\
Off & Corrected & 73.94 & 68.54 & 0.13569 & 0.09485 & 1.00011 \\
Market only & Naive & 6.28 & 3.18 & 0.13197 & 0.09208 & 1.34938 \\
Market only & Corrected & 73.64 & 67.81 & 0.12844 & 0.08969 & 1.00032 \\
Market + assets & Naive & 5.25 & 2.56 & 0.11727 & 0.08264 & 1.33445 \\
Market + assets & Corrected & 61.68 & 53.33 & 0.11581 & 0.08154 & 1.00018 \\
\midrule
\multicolumn{7}{@{}l}{\textit{Holdout: $2025$, $416$ assets, $249$ days}} \\[2pt]
Off & Naive & 82.38 & 70.14 & 0.07322 & 0.06683 & 1.36081 \\
Off & Corrected & 85.92 & 78.76 & 0.07330 & 0.06687 & 1.00041 \\
Market only & Naive & 81.76 & 69.60 & 0.07347 & 0.06698 & 1.36638 \\
Market only & Corrected & 85.88 & 78.72 & 0.07525 & 0.06796 & 1.00045 \\
Market + assets & Naive & 80.27 & 68.19 & 0.07672 & 0.06885 & 1.36032 \\
Market + assets & Corrected & 84.43 & 77.21 & 0.07832 & 0.06973 & 1.00039 \\
\bottomrule
\end{tabular}

\end{table}

\clearpage
\section*{Figures}
% Main-manuscript figures, in their original numerical order.
% Algorithms remain in the Methods section.

\begin{figure}[H]
    \centering
    \includegraphics[width=\textwidth]{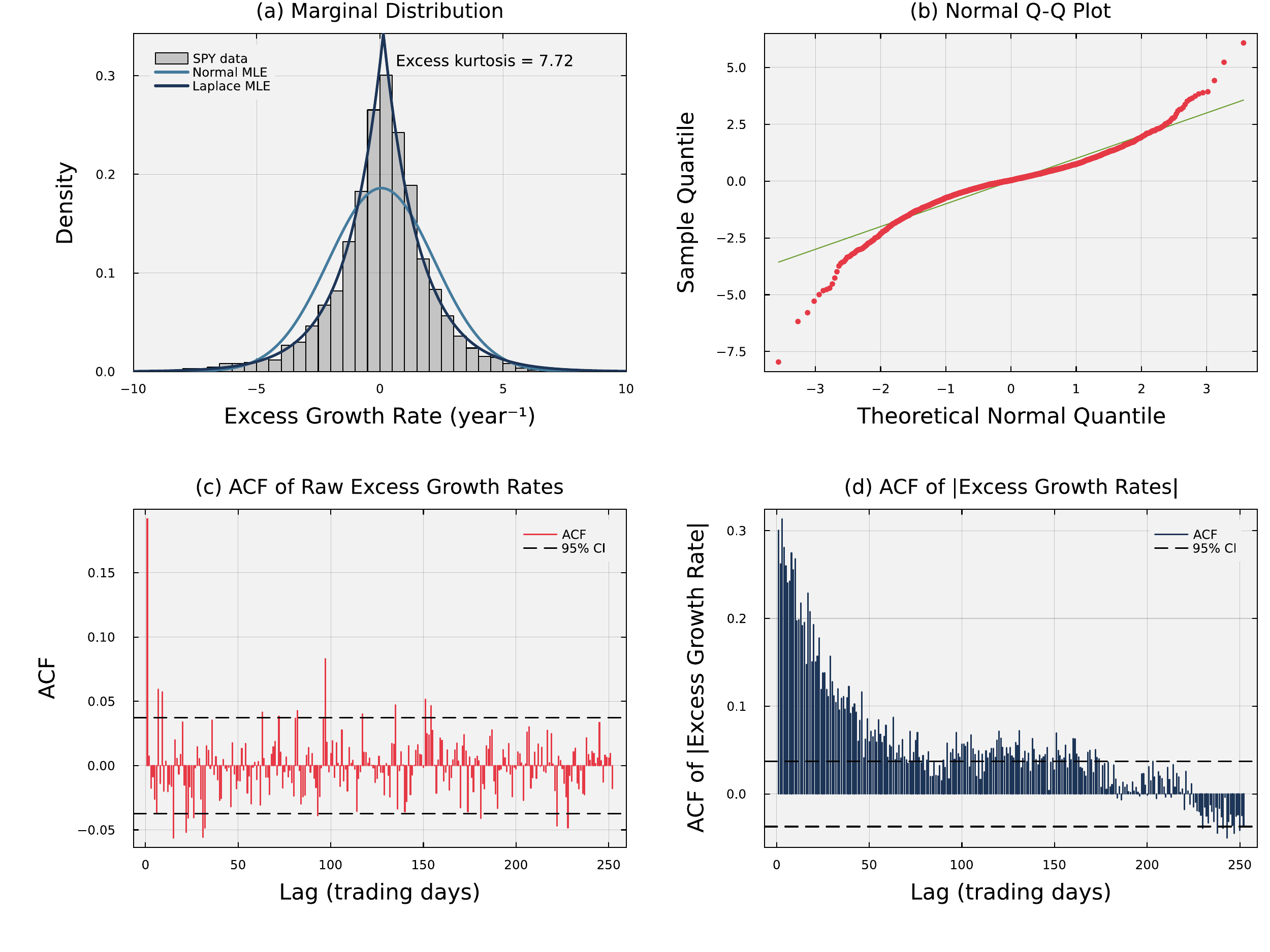}
    \caption{\textbf{Empirical stylized facts of SPY daily excess growth
    rates ($2014$--$2024$).} Panel~(a) shows the heavy-tailed return
    density; the Laplace maximum-likelihood fit tracks the empirical
    peak and tails while the Gaussian fit does not. Panel~(b) shows
    departure from the diagonal in a normal quantile-quantile plot
    at both extremes. Panel~(c) shows the
    autocorrelation function of raw excess growth rates. Individual
    correlations are small, although the Ljung--Box test detects joint
    serial dependence. Panel~(d) shows the autocorrelation function
    of absolute excess growth rates, which decays slowly across the
    reported lags.}
    \label{fig:empirical_motivation}
\end{figure}

\begin{figure}[H]
\centering
\resizebox{\textwidth}{!}{%
\begin{tikzpicture}[
  state/.style    = {circle, draw=black, very thick, minimum size=1.1cm,
                     inner sep=0pt, font=\small},
  tailbot/.style  = {state, fill=red!25,  draw=red!70!black},
  tailtop/.style  = {state, fill=blue!25, draw=blue!70!black},
  midstate/.style = {state, fill=white,   draw=black!70},
  markov/.style   = {<->, >=Stealth, thin, gray!80},
  trigger/.style  = {->, >=Stealth, thick, dashed, black!80},
  jumparc/.style  = {->, >=Stealth, very thick},
  note/.style     = {font=\footnotesize, fill=white, inner sep=2pt},
]
  \node[tailbot]              (s1)   at ( 0.0, 0) {$1$};
  \node[tailbot]              (s2)   at ( 2.0, 0) {$2$};
  \node[font=\Large]          (ld)   at ( 4.0, 0) {$\cdots$};
  \node[midstate]             (sk)   at ( 6.0, 0) {$k$};
  \node[font=\Large]          (rd)   at ( 8.0, 0) {$\cdots$};
  \node[tailtop]              (sNm1) at (10.0, 0) {$N\!-\!1$};
  \node[tailtop]              (sN)   at (12.0, 0) {$N$};
  \draw[markov] (s1)  -- (s2);
  \draw[markov] (s2)  -- (ld);
  \draw[markov] (ld)  -- (sk);
  \draw[markov] (sk)  -- (rd);
  \draw[markov] (rd)  -- (sNm1);
  \draw[markov] (sNm1)-- (sN);
  \draw[red!75!black, thick]
    plot[domain=-0.42:0.42, samples=40, smooth, variable=\t]
    ({ 0.0+\t}, {1.45 + 0.72*exp(-\t*\t/0.024)});
  \draw[red!25, thin] (-0.42,1.45) -- (0.42,1.45);
  \draw[red!75!black, thick]
    plot[domain=-0.42:0.42, samples=40, smooth, variable=\t]
    ({ 2.0+\t}, {1.45 + 0.72*exp(-\t*\t/0.024)});
  \draw[red!25, thin] (1.58,1.45) -- (2.42,1.45);
  \draw[black!55, thick]
    plot[domain=-0.42:0.42, samples=40, smooth, variable=\t]
    ({ 6.0+\t}, {1.45 + 0.72*exp(-\t*\t/0.024)});
  \draw[gray!40, thin] (5.58,1.45) -- (6.42,1.45);
  \draw[blue!75!black, thick]
    plot[domain=-0.42:0.42, samples=40, smooth, variable=\t]
    ({10.0+\t}, {1.45 + 0.72*exp(-\t*\t/0.024)});
  \draw[blue!25, thin] (9.58,1.45) -- (10.42,1.45);
  \draw[blue!75!black, thick]
    plot[domain=-0.42:0.42, samples=40, smooth, variable=\t]
    ({12.0+\t}, {1.45 + 0.72*exp(-\t*\t/0.024)});
  \draw[blue!25, thin] (11.58,1.45) -- (12.42,1.45);
  \node[draw, rounded corners=5pt, fill=yellow!20, very thick,
        minimum width=2.8cm, minimum height=1.1cm,
        align=center, font=\small] (clock) at (6.0, 5.0)
        {Poisson clock\\[1pt]prob.~$\epsilon$};
  \draw[trigger] (sk.north) -- (clock.south)
    node[note, midway, right=5pt, anchor=west] {jump triggered};
  \draw[jumparc, red!70!black]
    (clock.west)
    to[out=180, in=100]
    node[note, above, sloped, text=red!80!black,
         pos=0.45, yshift=3pt]
      {$K\!\sim\!\mathrm{Pois}(\lambda)$ forced steps}
    (s1.north);
  \draw[jumparc, blue!70!black]
    (clock.east)
    to[out=0, in=80]
    node[note, above, sloped, text=blue!80!black,
         pos=0.45, yshift=3pt]
      {$K\!\sim\!\mathrm{Pois}(\lambda)$ forced steps}
    (sN.north);
  \draw[decorate,
        decoration={brace, amplitude=6pt, mirror},
        red!70!black, thick]
    (-0.6, -0.65) -- (2.6, -0.65)
    node[midway, below=7pt, red!80!black, font=\small]
      {$\mathcal{S}_{-}$\ (bottom tail)};
  \draw[decorate,
        decoration={brace, amplitude=6pt, mirror},
        blue!70!black, thick]
    (9.4, -0.65) -- (12.6, -0.65)
    node[midway, below=7pt, blue!80!black, font=\small]
      {$\mathcal{S}_{+}$\ (top tail)};
  \draw[->, >=Stealth, gray!70, thin]
    (-1.0, -0.5) -- (13.2, -0.5)
    node[right, font=\small, text=black!70] {state index};
  \node[note, align=right, anchor=east, text=black!55]
    at (5.55, 2.72) {state emission\\[-1pt]
      $\mu_k\!+\!\sigma_k\cdot t_5$};
  \node[note, text=gray!80, anchor=west]
    at (6.55, 0.72)
    {prob.~$1\!-\!\epsilon$};
\end{tikzpicture}%
}
\caption{\textbf{Architecture of the hidden Markov model with jumps
    (HMM-WJ).} When no jump episode is active,
    the chain transitions according to the empirical transition matrix
    $\mathbf{T}$ with probability $1 - \epsilon$ (thin bidirectional
    arrows) or enters a Poisson jump episode with probability $\epsilon$
    (dashed upward arrow to the Poisson clock). During an episode, each of the
    $K \sim \mathrm{Poisson}(\lambda)$ forced steps independently draws
    the bottom tail set $\mathcal{S}_{-}$ (red, lowest-valued states)
    with probability $p_{\rm neg}$ or the top tail set $\mathcal{S}_{+}$
    (blue, highest-valued states) otherwise. Ordinary transitions then
    resume. Small bell curves above
    each state show the state-conditional location-scale Student-$t$
    ($\nu=5$) emission distribution. Tail sets are defined as the
    $N_{\rm tail}$ lowest- and highest-quantile states under the
    Laplace quantile partition.}
\label{fig:model_architecture}
\end{figure}

\begin{figure}[H]
  \centering
  \includegraphics[width=\textwidth]{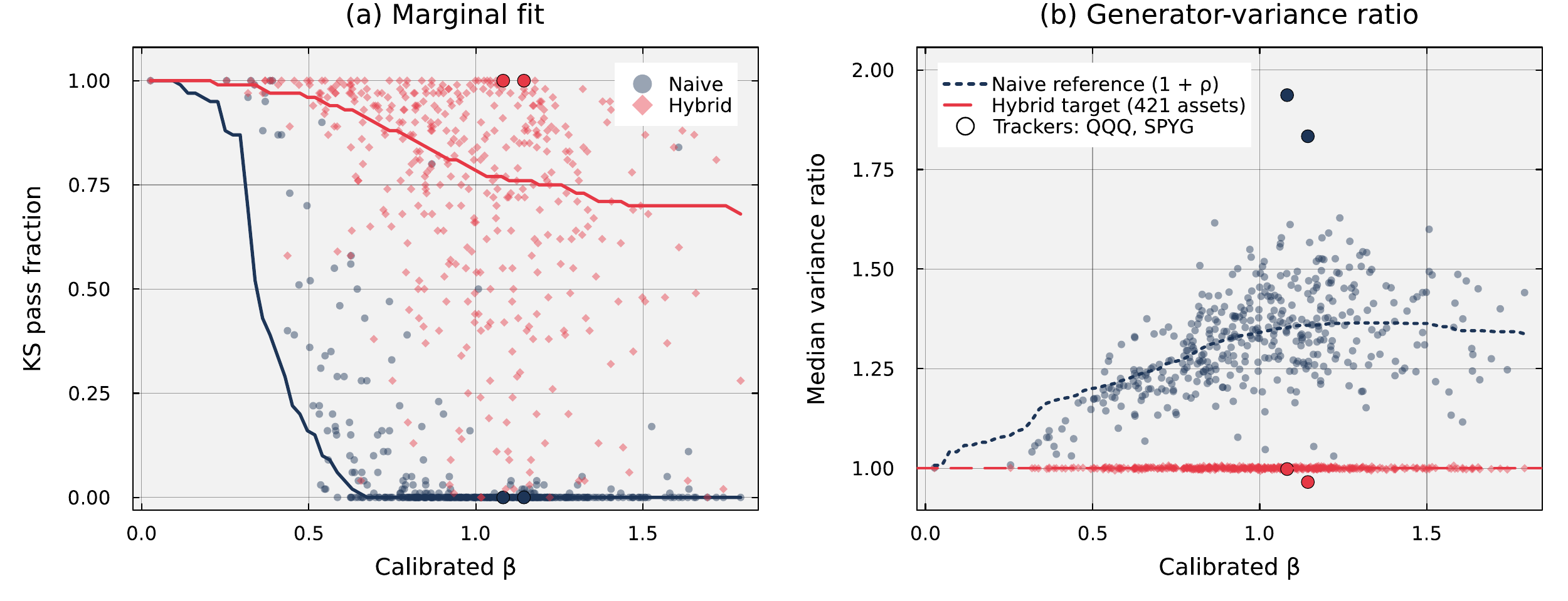}
  \caption{\textbf{Marginal fit and generator-variance preservation under paired composition.}
    We used $100$ paired paths of $2{,}766$ observations for each of
    $423$ non-market assets, with the observed $2014$--$2024$ SPY
    market path and jumps disabled. Blue circles denote naive
    composition; red diamonds denote hybrid composition. Outlined
    circles mark the tracker assets QQQ and SPYG in the corresponding
    method color. \emph{Panel (a):} each point is the fraction of paths
    passing the Kolmogorov--Smirnov (KS) test against the observed
    asset returns at $\alpha=0.05$.
    \emph{Panel (b):} each point is the median of the $100$ ratios
    $\Var(g_i)/\sigma^2_{\mathrm{gen},i}$, with the denominator
    measured from the same uncomposed generator draw. The dashed
    line marks the hybrid target of one for the $421$ non-tracker
    assets. Trackers instead target
    $\beta_i^2\sigma_m^2/R^2_{\mathrm{real},i}$ and need not lie on
    that line. The dotted curve summarizes the per-ticker medians of
    the naive reference $1+\rho_i$, where
    $\rho_i=\beta_i^2\sigma_m^2/\sigma^2_{\mathrm{gen},i}$ is computed
    per draw. This reference omits the finite-path covariance term.
    Solid curves in panel (a) and the dotted curve in panel (b) are
    Gaussian-kernel weighted medians over non-tracker assets, with
    bandwidth $0.15$ in calibrated $\beta$.
    For non-tracker hybrid paths, the median and $95$th percentile of
    $|\Var(g_i)/\sigma^2_{\mathrm{gen},i}-1|$ were $1.12\%$ and
    $3.41\%$; ticker medians therefore conceal individual-path
    variation. The annualized variance of the observed market path
    was $\Delta t\,\sigma_m^2=0.030~\mathrm{yr}^{-1}$.}
  \label{fig:preservation}
\end{figure}

\clearpage
\appendix
\section*{Online Appendix}
\renewcommand{\thesection}{S\arabic{section}}
\renewcommand{\thesubsection}{S\arabic{section}.\arabic{subsection}}
\renewcommand{\thefigure}{S\arabic{figure}}
\renewcommand{\thetable}{S\arabic{table}}
\renewcommand{\theequation}{S\arabic{equation}}
\renewcommand{\theHsection}{supp.\arabic{section}}
\renewcommand{\theHsubsection}{supp.\arabic{section}.\arabic{subsection}}
\renewcommand{\theHfigure}{supp.\arabic{figure}}
\renewcommand{\theHtable}{supp.\arabic{table}}
\renewcommand{\theHequation}{supp.\arabic{equation}}
\setcounter{section}{0}
\setcounter{figure}{0}
\setcounter{table}{0}
\setcounter{equation}{0}

\section{Empirical cross-term diagnostic}
\label{sec:supp-cov}

The variance correction of Eq.~\eqref{eq:scale} assumes
$\Cov(\tg_i^{\mathrm c},g_m)\approx0$, allowing the variance derivation to omit
the cross-term. Each generator draw is simulated independently of the
market path, so this covariance is zero in population, but on a
finite path its sample value is not. We measured the size of the
sample cross-term for three residual sources: the full-return
hidden Markov generator (JumpHMM) used by hybrid and naive composition,
the residual-fit JumpHMM generator used by the JumpHMM-on-residuals
method, and the Gaussian residual draw of the Gaussian single-index model (SIM). For each
of the $423$ non-market tickers, we sampled $R = 100$ training-length
growth-rate paths $\tg_i^{(r)}$ from each source and computed the
sample correlation of each centered draw with the observed training
growth-rate path $g_m$ of the broad-market exchange-traded fund SPY,
averaged over the $R$ draws:
\begin{equation}
  \mathrm{cov}^{\mathrm{norm}}_i
  = \frac{1}{R} \sum_{r=1}^{R}
    \frac{\Cov((\tg_i^{(r)})^{\mathrm c},\, g_m)}{\sigma_{\gen,i}^{(r)}\, \sigma_m},
\end{equation}
where $\sigma_{\gen,i}^{(r)}$ is the sample standard deviation of
draw $r$. Centering leaves covariance unchanged. The universe means were
$1.53\times10^{-4}$, $6.58\times10^{-5}$, and $2.26\times10^{-4}$ for
the full-return, residual-fit, and Gaussian sources, respectively,
and the per-ticker values of all three ranged from $-0.0061$ to
$0.0061$. For independent series with weak serial correlation and $2{,}766$
observations, the sample correlation has standard deviation near $1/\sqrt{2{,}766}
\approx 0.019$, the spread we observed across single draws, so an
average over $100$ draws has standard deviation near $0.002$. The averages were consistent with sampling noise; individual paths still had nonzero cross-terms and therefore matched their variance targets only approximately.

\section{Rebalancing under independent residuals}
\label{sec:supp-bias}

To identify the portfolio consequence of independent residuals, we derive the diversification return of a daily-rebalanced portfolio. Matching generator variance also matches empirical variance only when the generator reproduces it; neither condition restores cross-asset residual covariance.
Residual paths retain their own temporal structure but are generated independently across assets. Consider a daily-rebalanced portfolio
with constant target weights $w_i \ge 0$, $\sum_i w_i = 1$,
reset to those weights at the start of each trading day. Throughout
this derivation $g_i(t)$ is the annualized log-growth rate of
Eq.~\eqref{eq:growth_rate}, so the per-period log return on asset
$i$ is $g_i(t)\,\Delta t$ with $\Delta t = 1/252$~yr. The wealth
process is:
\begin{equation}
  W_t \;=\; W_{t-1} \sum_{i} w_i \exp\!\big(g_i(t)\,\Delta t\big),
  \label{eq:wealth-recursion}
\end{equation}
so the per-step log return on the portfolio is
$\log(W_t/W_{t-1}) = \log\!\sum_i w_i \exp(g_i(t)\,\Delta t)$. Write $g_i$ for $g_i(t)$. A
second-order Taylor expansion of the exponential about zero is
$\exp(g_i\,\Delta t) = 1 + g_i\,\Delta t + \tfrac{1}{2}(g_i\,\Delta t)^2
+ O((g\,\Delta t)^3)$. Substituting it into each summand of
$\sum_i w_i \exp(g_i\,\Delta t)$ gives:
\begin{align}
  \sum_{i} w_i \exp\!\big(g_i\,\Delta t\big)
   &= \sum_{i} w_i \!\left[\,1 + g_i\,\Delta t
        + \tfrac{1}{2}(g_i\,\Delta t)^2
        + O\!\big((g\,\Delta t)^3\big)\right]
       && \text{(Taylor)} \notag \\
   &= \sum_{i} w_i
        + \Delta t \sum_{i} w_i\, g_i
        + \tfrac{1}{2}(\Delta t)^2 \sum_{i} w_i\, g_i^2
        + O\!\big((g\,\Delta t)^3\big)
       && \text{(distribute)} \notag \\
   &= 1 + \Delta t \sum_{i} w_i\, g_i
        + \tfrac{1}{2}(\Delta t)^2 \sum_{i} w_i\, g_i^2
        + O\!\big((g\,\Delta t)^3\big).
       && \text{(}\textstyle\sum_i w_i = 1\text{)}
  \label{eq:exp-expand}
\end{align}
A second-order Taylor expansion of $\log(1+x)$ about $x=0$ gives
$\log(1+x) = x - \tfrac{1}{2}x^2 + O(x^3)$. Identifying $1 + x$ with
the right-hand side of Eq.~\eqref{eq:exp-expand}, the part beyond one
is
$x :=\Delta t \sum_i w_i\, g_i + \tfrac{1}{2}(\Delta t)^2 \sum_i w_i\, g_i^2$,
and squaring keeps only the leading $(\Delta t)^2$ term:
$x^2 = (\Delta t)^2 (\sum_i w_i\, g_i)^2 + O((\Delta t)^3)$ because
the cross term has an extra $\Delta t$ factor. Substitution gives:
\begin{align}
  \log\!\frac{W_t}{W_{t-1}}
   &= \log\!\Big(1 + x + O\!\big((g\,\Delta t)^3\big)\Big) \notag \\
   &= x - \tfrac{1}{2}\, x^2 + O\!\big(x^3\big) \notag \\
   &= \Delta t \sum_{i} w_i\, g_i
        + \tfrac{1}{2}(\Delta t)^2 \sum_{i} w_i\, g_i^2
        \notag \\
   &\quad - \tfrac{1}{2}(\Delta t)^2
        \!\Big(\sum_{i} w_i\, g_i\Big)^{\!2}
        + O\!\big((g\,\Delta t)^3\big) \notag \\
   &= \Delta t \sum_{i} w_i\, g_i
        + \tfrac{1}{2}(\Delta t)^2 \!\left[\sum_{i} w_i\, g_i^2
        \right. \notag \\
   &\hspace{8em}\left.{}- \Big(\sum_{i} w_i\, g_i\Big)^{\!2}\right]
        + O\!\big((g\,\Delta t)^3\big).
  \label{eq:rp-taylor}
\end{align}
Dividing by $\Delta t$ gives the annualized portfolio growth rate $g_p(t) := \Delta t^{-1}\log(W_t/W_{t-1})$:
\begin{equation}
  g_p \;=\; \sum_{i} w_i\, g_i
   \;+\; \tfrac{1}{2}\,\Delta t \!\left(\sum_{i} w_i\, g_i^2
        \;-\; \Big(\sum_{i} w_i\, g_i\Big)^{\!2}\right)
   \;+\; O\!\big((\Delta t)^2\big).
  \label{eq:gp-onestep}
\end{equation}

To obtain expected annualized portfolio growth, we restored the time
argument and took unconditional expectations of
Eq.~\eqref{eq:gp-onestep} under stationarity. Define the mean and
variance of each asset's annualized growth rate:
\begin{equation}
  \mu_i \;:=\; \E\!\big(g_i(t)\big), \qquad
  \sigma_i^2 \;:=\; \Var\!\big(g_i(t)\big),
\end{equation}
the covariance of the annualized rates
$\Sigma_{ij} := \Cov\!\big(g_i(t), g_j(t)\big)$, and the portfolio
variance
$\sigma_p^2 := \Var\!\big(\textstyle\sum_i w_i g_i(t)\big) =
w^\top \Sigma w$. The squared terms $\sum_i w_i\, g_i(t)^2$ and $(\sum_i w_i\, g_i(t))^2$ in the $O(\Delta t)$ correction expand using the following identity for any $X$ with finite second moment:
\begin{equation}
  \E\!\big(X^2\big) \;=\; \E\!\big(X\big)^2 \;+\; \Var(X).
  \label{eq:second-moment}
\end{equation}
Applying this identity to $X = g_i(t)$ and summing against $w_i$ gives:
\begin{align}
  \E\!\Big(\sum_{i} w_i\, g_i(t)^2\Big)
   &\;=\; \sum_{i} w_i\, \E\!\big(g_i(t)^2\big)
       && \text{(linearity)} \notag \\
   &\;=\; \sum_{i} w_i\!\left[\,\mu_i^2 + \sigma_i^2\,\right]
       && \text{(by Eq.~\eqref{eq:second-moment})} \notag \\
   &\;=\; \sum_{i} w_i\, \mu_i^2 \;+\; \sum_{i} w_i\, \sigma_i^2.
   \label{eq:exp-gi2}
\end{align}
Applying the identity to $X = \sum_i w_i g_i(t)$, whose mean is
$\sum_i w_i\mu_i$ and whose variance is $\sigma_p^2$, gives:
\begin{align}
  \E\!\Big(\Big(\textstyle\sum_{i} w_i g_i(t)\Big)^{\!2}\Big)
   &\;=\; \E\!\Big(\textstyle\sum_{i} w_i g_i(t)\Big)^{\!2}
       \;+\; \Var\!\Big(\textstyle\sum_{i} w_i g_i(t)\Big)
       && \text{(by Eq.~\eqref{eq:second-moment})} \notag \\
   &\;=\; \Big(\textstyle\sum_{i} w_i\, \mu_i\Big)^{\!2}
       \;+\; \sigma_p^2.
   \label{eq:exp-sum2}
\end{align}
We substituted Eqs.~\eqref{eq:exp-gi2}--\eqref{eq:exp-sum2} into the
expectation of Eq.~\eqref{eq:gp-onestep} and grouped the variance and
mean contributions as follows:
\begin{equation}
\begin{aligned}
  \E(g_p) \;={}& \sum_{i} w_i\, \mu_i
  + \underbrace{\tfrac{1}{2}\,\Delta t\!\left(\sum_{i} w_i\, \sigma_i^2
       - \sigma_p^2\right)}_{\displaystyle D} \\
  &+ \tfrac{1}{2}\,\Delta t\!\left[\sum_{i} w_i\, \mu_i^2
       - \Big(\sum_{i} w_i\, \mu_i\Big)^{\!2}\right]
  + O\!\big((\Delta t)^2\big),
\end{aligned}
  \label{eq:booth-fama}
\end{equation}
the Booth--Fama decomposition, to second order, of the growth of a
daily-rebalanced portfolio in annualized growth-rate units
\cite{boothFama1992, markowitz1976longRun}. The middle term, $D$, is the diversification return. The
moments $\mu_i$ and $\sigma_i^2$ of the annualized rates
have units $\mathrm{yr}^{-1}$ and $\mathrm{yr}^{-2}$ respectively;
the products $\Delta t\,\sigma_i^2$ and $\Delta t\,\sigma_p^2$ in
Eq.~\eqref{eq:booth-fama} are the annualized variances
($\mathrm{yr}^{-1}$). The third term in Eq.~\eqref{eq:booth-fama} is
the cross-sectional dispersion of the per-asset means, of order
$\Delta t\,\mu^2$, while $D$ is of order $\Delta t\,\sigma^2$. For
SPY, $\mu^2$ lies more than two orders of magnitude below $\sigma^2$
(Online Appendix~\ref{sec:supp-generator}), and we drop the third term,
treating the same ordering as typical of daily equity growth rates.
The realized annualized compound growth rate is
$\bar g_p = (T\,\Delta t)^{-1}\log(W_T/W_0) = T^{-1}\sum_{t} g_p(t)$.
If the composed growth rates form a stationary ergodic process,
$\bar g_p$ converges to $\E(g_p)$ by the ergodic theorem. This limit depends on the single-date joint law of $(g_i(t))_i$, not residual serial dependence. Finite-horizon centering and scaling are outside this stationary-process argument.

To show that $D$ is non-negative, we rewrote the bracket in $D$ as a
sum over asset pairs. Expanding the portfolio variance, using
$1 - w_i = \sum_{j \ne i} w_j$ for weights summing to one, and
collecting each pair once gives:
\begin{align}
  \sum_i w_i \sigma_i^2 - \sigma_p^2
   &\;=\; \sum_i w_i \sigma_i^2 - \sum_i w_i^2 \sigma_i^2
     - 2 \sum_{i<j} w_i w_j \Sigma_{ij}
     && \text{(expand $\sigma_p^2$)} \notag \\
   &\;=\; \sum_i w_i (1 - w_i)\, \sigma_i^2
     - 2 \sum_{i<j} w_i w_j \Sigma_{ij}
     && \text{(collect)} \notag \\
   &\;=\; \sum_i \sum_{j \ne i} w_i w_j\, \sigma_i^2
     - 2 \sum_{i<j} w_i w_j \Sigma_{ij}
     && \text{($1 - w_i = \textstyle\sum_{j \ne i} w_j$)} \notag \\
   &\;=\; \sum_{i<j} w_i w_j
     \left(\sigma_i^2 + \sigma_j^2 - 2\Sigma_{ij}\right)
     && \text{(pair terms)} \notag \\
   &\;=\; \sum_{i<j} w_i w_j \,\Var(g_i - g_j),
     && \text{(variance of a difference)}
  \label{eq:pair-identity}
\end{align}
so that:
\begin{equation}
  D \;=\; \tfrac{1}{2}\,\Delta t \sum_{i<j} w_i w_j \,\Var(g_i - g_j)
  \;\ge\; 0.
  \label{eq:dr-pairwise}
\end{equation}
For positive-weight asset pairs, $D$ vanishes only when $\Var(g_i-g_j)=0$, or when all weight is on one asset. Rebalancing converts cross-sectional dispersion into compounded growth, the variance bonus described by Markowitz~\cite{markowitz1976longRun} and the diversification return of Booth and Fama~\cite{boothFama1992}.

To separate market and residual contributions to the diversification
return, we substituted the single-index decomposition
$g_i(t) = \alpha_i + \beta_i\, g_m(t) + \eps_i(t)$ into
Eq.~\eqref{eq:dr-pairwise}, with market variance
$\sigma_m^2 := \Var(g_m(t))$, per-asset residual variance
$\sigma_{\eps,i}^2 := \Var(\eps_i(t))$, and $\Cov(g_m, \eps_i) = 0$
(the population assumption checked in Online
Appendix~\ref{sec:supp-cov}). Here the loadings are fixed and
unclipped, as in the observed-market training comparison. The
pairwise difference variance is:
\begin{equation}
  \Var(g_i - g_j) \;=\; (\beta_i - \beta_j)^2 \sigma_m^2
  \;+\; \Var(\eps_i - \eps_j).
\end{equation}
The zero market-residual covariance splits diversification return into market and idiosyncratic components:
\begin{equation}
  D \;=\;
  \underbrace{\tfrac{1}{2}\,\Delta t\,\sigma_m^2 \sum_{i<j} w_i w_j (\beta_i - \beta_j)^2}_{D_{\mathrm{mkt}}}
  \;+\;
  \underbrace{\tfrac{1}{2}\,\Delta t \sum_{i<j} w_i w_j \,\Var(\eps_i - \eps_j)}_{D_\eps},
  \label{eq:dr-sim}
\end{equation}
The market term $D_\mathrm{mkt}$ contains the weighted variance of $\beta$. Applying Eq.~\eqref{eq:pair-identity} with $\beta_i$ in place of $g_i$ gives:
\begin{align}
  \sum_{i<j} w_i w_j (\beta_i - \beta_j)^2
   &\;=\; \sum_i w_i (1 - w_i)\, \beta_i^2
     - 2 \sum_{i<j} w_i w_j \beta_i \beta_j
     && \text{(pair identity)} \notag \\
   &\;=\; \sum_i w_i \beta_i^2
     - \Big(\sum_i w_i^2 \beta_i^2
     + 2 \sum_{i<j} w_i w_j \beta_i \beta_j\Big)
     && \text{(regroup)} \notag \\
   &\;=\; \sum_i w_i \beta_i^2 - \Big(\sum_i w_i \beta_i\Big)^{\!2},
     && \text{(square of a sum)}
\end{align}
so $D_\mathrm{mkt} = \tfrac{1}{2}\,\Delta t\,\sigma_m^2 [\sum_i w_i
\beta_i^2 - (\sum_i w_i \beta_i)^2]$, half the weighted variance of $\beta$
scaled by the annualized market variance $\Delta t\,\sigma_m^2$.
Across the $423$ non-market tickers, the calibrated $\beta$
distribution spans $[0.03, 1.79]$ (Online Appendix~\ref{sec:supp-sim-derivations}). A
weighted variance is at most one quarter of the squared range, so
$D_{\mathrm{mkt}} \le \tfrac{1}{8}\,\Delta t\,\sigma_m^2\,(1.79 -
0.03)^2$. The SPY closing-price series used for the multi-asset
universe has annualized variance $\Delta t\,\sigma_m^2 =
0.030~\mathrm{yr}^{-1}$ (Fig.~\ref{fig:preservation}), at which
$D_{\mathrm{mkt}}$ is at most $0.012~\mathrm{yr}^{-1}$, or $1.2$
percentage points per year. For long-only weights on this universe, any diversification return
above this bound must include an idiosyncratic contribution $D_\eps$. In the synthetic construction, residual paths are
generated independently across assets, so
$\Var(\eps_i - \eps_j) = \sigma_{\eps,i}^2 + \sigma_{\eps,j}^2$ and,
by the same pair identity,
$D_\eps = \tfrac{1}{2}\,\Delta t \sum_i w_i (1 - w_i)\,
\sigma_{\eps,i}^2$, a weighted sum of the annualized idiosyncratic
variances $\Delta t\,\sigma_{\eps,i}^2$.

At fixed same-date moments, serial dependence does not change the second-order $D$. Under zero market-residual covariance,
Eq.~\eqref{eq:scale} chooses $s_i$ so that the composed variance $\beta_i^2\sigma_m^2+\sigma_{\eps,i}^2$ equals the
generator variance $\sigma_{\gen,i}^2$, and any residual generator
with the same contemporaneous variances and covariances gives the same $D$. As an
illustration, consider a stationary first-order autoregressive
[AR(1)] residual $\eps_i(t) = \phi\, \eps_i(t-1) + \eta_i(t)$ with
independent and identically distributed innovations $\eta_i$ whose
variance is chosen so that
$\Var(\eps_i) = \sigma_{\eta,i}^2 / (1 - \phi^2) = \sigma_{\eps,i}^2$.
The parameter $\phi$ leaves the variance at a single date unchanged,
so it leaves the second-order approximation to $\E(g_p)$ in Eqs.~\eqref{eq:booth-fama} and~\eqref{eq:dr-sim} unchanged. It does change sampling noise: for an
uncentered stationary AR(1) residual with variance $\sigma_\eps^2$,
the variance of the sample mean is approximately
$T^{-1}\sigma_\eps^2(1+\phi)/(1-\phi)$.
By contrast, omitted positive cross-asset residual covariance lowers real $\Var(\eps_i-\eps_j)$ below the independent-generator value $\sigma_{\eps,i}^2+\sigma_{\eps,j}^2$, so matching marginal variances can overstate $D_\eps$. We did not measure this excess. Isolating it requires a joint-residual comparison with fixed marginals and allocation; a gap from realized portfolio returns also includes finite-horizon sampling and second-order approximation error~\cite{boucheyNemtchinovWong2015}.

\section{Validation metric definitions}
\label{sec:supp-validation}

To distinguish marginal agreement from temporal agreement, we used
separate distributional, tail, and autocorrelation summaries
(Tables~\ref{tab:model_comparison} and~\ref{tab:jump_ablation}).
 For one observed series $x=(x_1,\ldots,x_n)$ and $R$ synthetic
paths $y^{(r)}$ of the same length, each distributional comparison
used one complete synthetic path and the fixed observed history.
The two-sample Kolmogorov--Smirnov (KS) statistic is the largest
absolute difference between their empirical cumulative distribution
functions, $D_{\mathrm{KS}}=\sup_z|F_x(z)-F_{y^{(r)}}(z)|$
\cite{kolmogorov1933,smirnov1948}. The Anderson--Darling (AD)
comparison gives greater weight to differences near the tails
\cite{andersonDarling1952}. We used the approximate two-sample KS
test and the modified, tie-aware two-sample AD calculation in
\texttt{HypothesisTests.jl}, with asymptotic $p$-values. A path
passed when $p>0.05$, and the pass percentage was $100$ times the
fraction of passing paths. Pass rates are descriptive because returns are dependent and training comparisons reuse the fitting history. Independent simulations supply independent indicators conditional on that history and fit, not new historical evidence or proof of a shared distribution.

To quantify marginal discrepancies without a pass threshold, we
computed Wasserstein-1 and histogram Hellinger scores for each path
(Table~\ref{tab:model_comparison}). Let $x_{(j)}$ and $y^{(r)}_{(j)}$
denote the $j$th order statistics in increasing order. Because each
synthetic path has the same length as the observed series, the
Wasserstein-1 distance is:
\begin{equation}
 W_1(x,y^{(r)})=\frac{1}{n}\sum_{j=1}^{n}
       |x_{(j)}-y^{(r)}_{(j)}|.
 \label{eq:wasserstein}
\end{equation}
It has the growth-rate units $\mathrm{yr}^{-1}$. For each Hellinger
comparison, we formed $50$ equally spaced bins between the minimum
and maximum of the pooled pair of samples. Let $p_b$ and $q_b$ be
the observed and simulated bin counts divided by the sample length
$n$. The histogram score is:
\begin{equation}
 H(x,y^{(r)})=\left[\frac{1}{2}\sum_{b=1}^{50}
          (\sqrt{p_b}-\sqrt{q_b})^2\right]^{1/2}.
 \label{eq:hellinger}
\end{equation}
All bins included their lower edge and excluded their upper edge, omitting the pooled maximum; the sample containing that maximum has weights summing to slightly less than one. This histogram approximation is best compared within a window because pair-specific edges and sample length affect it. Smaller scores indicate closer marginal agreement.

To describe tail shape, we computed excess kurtosis for each
synthetic path and compared its mean over paths with the observed value
(Table~\ref{tab:model_comparison}); the composition tables report
medians instead. With
$m_p(x)=n^{-1}\sum_{t=1}^{n}(x_t-\bar x)^p$, the estimator was
$\kappa(x)=m_4(x)/m_2(x)^2-3$, without a finite-sample bias
correction. The composition scorecard also used the Hill index of
absolute returns (Table~\ref{tab:aggregate}). Let
$X_{(1)}\ge\cdots\ge X_{(n)}$ be decreasingly ordered absolute
returns and $k=\max\{2,\lfloor0.05n\rfloor\}$. The implemented
upper-tail estimator is:
\begin{equation}
 \hat\xi=\frac{1}{k-1}\sum_{j=1}^{k-1}
             \log\!\left(\frac{X_{(j)}}{X_{(k)}}\right).
 \label{eq:hill}
\end{equation}
The estimate $\hat\xi$ is the reciprocal of the conventional tail
index, so larger values indicate a heavier absolute-return tail; the
threshold $X_{(k)}$ must be positive.
We also measured whether simulation envelopes bracketed $99$ observed quantiles (Table~\ref{tab:model_comparison}). For $u_j=j/100$, $j=1,\ldots,99$,
let $Q_x(u_j)$ be the observed sample quantile and
$Q_r(u_j)$ the corresponding quantile of path $r$. Let
$L_j$ and $U_j$ be the $5$th and $95$th percentiles of
$\{Q_r(u_j):r=1,\ldots,R\}$. All quantiles used linear interpolation
between order statistics, the common software default (type~7). The reported
quantile-envelope coverage is:
\begin{equation}
 C=\frac{100}{99}\sum_{j=1}^{99}
       \mathbf{1}\{L_j\le Q_x(u_j)\le U_j\}.
 \label{eq:quantile-envelope-coverage}
\end{equation}
Larger values mean that more observed quantiles lie inside the pointwise envelopes, which also depends on envelope width; this is neither Value-at-Risk exceedance coverage nor simultaneous confidence coverage.

To quantify persistence in return magnitude, we used the sample
autocorrelation function (ACF) of absolute returns. Each series was
centered by its own full-series absolute-return mean,
$\bar a_x=n^{-1}\sum_t|x_t|$. At lag $\ell$, the calculation was:
\begin{equation}
 a_\ell(x)=
 \frac{\sum_{t=1}^{n-\ell}(|x_t|-\bar a_x)(|x_{t+\ell}|-\bar a_x)}
      {\sum_{t=1}^{n}(|x_t|-\bar a_x)^2}.
 \label{eq:sample-absolute-acf}
\end{equation}
The denominator used all $n$ observations, with no lag-dependent
rescaling. Both error scores below start at lag one.
The single-asset comparison measured the error of the ensemble-average
ACF (Table~\ref{tab:model_comparison}). For a lag limit $L$, that
mean absolute error is:
\begin{equation}
 E^{\rm ensemble}_L=\frac{1}{L}\sum_{\ell=1}^{L}
 \left|a_\ell(x)-\frac{1}{R}\sum_{r=1}^{R}a_\ell(y^{(r)})\right|.
 \label{eq:acf-ensemble-error}
\end{equation}
We used $R=1{,}000$ and $L=\min(252,n-1)$: lags $1$--$252$
for the $2{,}766$-observation training series and $1$--$248$ for the
$249$-observation holdout. This evaluation lag limit is separate from
the $25$-lag window of the tuning objective in
Eq.~\eqref{eq:grid_search}. The state-resolution and emission
comparisons (Online Appendix~\ref{sec:supp-sensitivity}; Table~\ref{tab:emission_hsmm}) used the same ensemble convention.
The multi-asset jump comparison instead measured error within each
path before averaging (Table~\ref{tab:jump_ablation}). For one
asset, its mean path error is:
\begin{equation}
 E^{\rm path}_L=\frac{1}{R}\sum_{r=1}^{R}
       \left[\frac{1}{L}\sum_{\ell=1}^{L}
       |a_\ell(x)-a_\ell(y^{(r)})|\right].
 \label{eq:acf-path-error}
\end{equation}
We used $L=25$ and $60$ in both windows and averaged these scores
over assets. The fallback sensitivity check also reported this path
score with $L=25$ (Table~\ref{tab:fallback_sensitivity}). At a common
lag limit and on the same paths, the triangle inequality gives
$E^{\rm ensemble}_L\le E^{\rm path}_L$: averaging the curves can
cancel errors of opposite sign. The path score includes sampling noise that ensemble averaging can cancel. The experiments also use different lag limits, histories and assets, so their error magnitudes are not directly comparable.

To describe simulation uncertainty in the single-asset comparison,
we used a standard-error method suited to each statistic
(Table~\ref{tab:model_comparison}). For a pass fraction $\hat p$,
the reported binomial standard error was
$100\sqrt{\hat p(1-\hat p)/R}$ percentage points. This estimate is
zero when every path passes or fails; it does not establish that the
underlying pass probability is exactly zero or one. For kurtosis,
Wasserstein distance, and Hellinger score, the standard error was
the sample standard deviation of the $R$ path scores divided by
$\sqrt R$. For ensemble ACF error and quantile-envelope coverage,
we used $500$ bootstrap resamples of the $R$ path indices, each of
size $R$ with replacement. Each resample retained entire paths and
recomputed the statistic across all lags or quantile levels. The
standard error was the sample standard deviation of those $500$
recomputed values, without a further division by $\sqrt{500}$.
All these errors condition on one observed history and a fixed fitted
model; they omit uncertainty in fitting, model selection, and market
regimes. For multi-asset contrasts, we averaged within each of $1{,}000$ shared-market replications before differencing settings; the standard deviation of those differences divided by $\sqrt{1{,}000}$ preserved the market dependence within each unit (Online Appendix~\ref{sec:supp-jump-composition}).

\subsection{Six-method composition evaluation}
\label{sec:supp-composition-evaluation}

The four residual-based comparators used the observed asset's
ordinary least-squares (OLS) residuals to set their distributions
(Table~\ref{tab:aggregate}). The Gaussian single-index model (SIM)
used independent Gaussian draws with the OLS residual variance. The
residual-fit hidden Markov model, labeled JumpHMM-on-residuals after
the software package, fitted a no-jump generator to a synthetic price
path obtained by exponentiating cumulative residuals multiplied by
the daily time step. The
Politis--Romano stationary bootstrap~\cite{politisRomano1994} used a
mean block length of $50$ days in both training and holdout. The
generalized autoregressive conditional heteroskedasticity model,
GARCH(1,1)-$t$, fitted conditional variance to the residual
series~\cite{bollerslev1986}. We accepted a fit only if optimization
converged, the likelihood and coefficients were finite, the variance
intercept was positive, and the nonnegative ARCH and GARCH
coefficients summed to less than one. The Student-$t$ degrees of
freedom had to exceed two, the unconditional variance had to be finite
and positive, and a trial simulated path had to be finite and
nonconstant. All $423$ optimizations converged, but $30$ failed the
coefficient-sum condition, leaving $393$ eligible assets. All four residual
comparators entered composition without further centering or
rescaling.
The scores reflect branch-specific targets. Lower absolute OLS errors indicated closer
recovery, but on the hybrid variance-preserving branch the composed
$R^2$ approached the draw's $\rho_i$, capped at $1-f$, rather than
$R^2_{i,\real}$. Its $R^2$ error therefore mainly measured this
designed shift; loading error also included any change made by
clipping. A composed-to-generator variance ratio of one indicated
equal variances, which the $R^2$-preserving branch did not target.
 We omitted cross-sectional
rank reordering to test each ticker's marginal construction and
scored volatility clustering separately in the jump-enabled comparison.
The paired centering control isolated differences in residual-mean
constraints (Online Appendix~\ref{sec:supp-centering-control}).

\subsection{Gated recurrent unit benchmark}
\label{sec:supp-gru}

To provide an autoregressive neural benchmark, we used a gated
recurrent unit (GRU) network to predict the distribution of the next
SPY growth rate from the preceding $50$ values
(Table~\ref{tab:model_comparison}). We standardized the training
series using its mean and population standard deviation and reused
these two values unchanged to standardize inputs and to convert
generated values back to growth rates. The network had two GRU layers
with $64$ hidden units per layer, dropout probability $0.1$ between
layers during training, and separate linear heads for a conditional
mean and log variance. Both heads read the last hidden output of each
input window, and the network had $37{,}954$ trainable parameters.
For the standardized target $z_t$ that follows a window, with
predicted mean $m_t$ and predicted log variance $v_t$, the training
loss was the Gaussian negative log-likelihood with its constant term
omitted:
\begin{equation}
 \mathcal L=\frac{1}{2B}\sum_{t\in\mathcal B}
       \left[v_t+(z_t-m_t)^2\exp(-v_t)\right],
 \label{eq:gru-loss}
\end{equation}
where $\mathcal B$ is a training batch containing $B=256$ windows.
The $2{,}766$ training observations supplied $2{,}716$ overlapping
input-target pairs. Each epoch shuffled these pairs and used ten
complete batches; the final incomplete batch was dropped. The Adam
optimizer used an initial learning rate of $10^{-3}$, default moment
decay rates $(0.9,0.999)$, and no weight decay. We clipped the
gradient norm at one and halved the learning rate whenever mean
training loss had not improved for $15$ epochs, down to a minimum
learning rate of $10^{-5}$. Training ran for at most $200$ epochs. We
retained weights whenever mean training loss improved by more than
$10^{-4}$ and would have stopped after $40$ epochs without such
improvement. The fit completed $200$ epochs, selecting epoch $199$. Model selection used training loss, with no separate
validation split; the $2025$ holdout played no part in fitting or
selection.

To generate each path, we disabled dropout and drew its initial
window uniformly from the $2{,}716$ observed training input windows.
At each step, we passed the current $50$ values through the network
with a fresh zero hidden state, sampled a Gaussian value with mean
$m_t$ and standard deviation $\max\{\exp(v_t/2),10^{-6}\}$, and
rolled the window forward, dropping its oldest value and appending
the draw. A path consisted of the generated values only, with no
burn-in, and the fixed training normalization was inverted afterward.
The Python benchmark seeded NumPy and PyTorch with $1234$ before
training. Generation then used the continuing NumPy stream, without
reseeding, for $1{,}000$ training-length paths followed by $1{,}000$
holdout-length paths from the same fitted network.
Pinned dependencies, trained weights and generation records accompany the benchmark in the repository. The
benchmark is one fitted network, so its reported Monte Carlo standard
errors do not measure variation across training seeds or neural
architectures.

\section{Derivation of the single-asset generator}
\label{sec:supp-generator}

The following derivations connect the fitted generator to its marginal and temporal behavior. The hidden Markov model with jumps (HMM-WJ), its no-jump
restriction (HMM-NJ), and their symbols follow the main-text
definitions in Eq.~\eqref{eq:emission} and
Algorithm~\ref{alg:hmmwj}. Unless noted otherwise, numerical values
evaluate the closed-form expressions at the fitted SPY model of
Table~\ref{tab:hyperparams}: $N = 100$ states, Student-$t$ degrees of
freedom $\nu = 5$, jump probability $\epsilon = 10^{-4}$, mean episode
duration $\lambda = 90$ steps, tail-selection probability
$p_{\rm neg} = 0.52$, tail-set size $N_{\rm tail} = 5$, and $T = 2{,}766$ daily excess growth rates from volume-weighted average prices in $2014$--$2024$. Ljung--Box tests on raw returns at lag $20$ gave $p<0.001$ in training and $p=0.019$ in the $249$-day $2025$ holdout, despite small individual lag correlations. The training series has annualized volatility $\sigma(G_t)\sqrt{\Delta t}$, about $13.5\%$, whereas $\sigma(G_t)$ itself has units $\mathrm{yr}^{-1}$.
Table~\ref{tab:supp-generator-quantities} collects the closed-form
quantities evaluated at these values beside their sample or simulated
counterparts.

\begin{table}[H]
\centering
\caption{\textbf{Closed-form quantities of the fitted SPY generator
    against sample and simulated values.} Entries evaluate the expressions below at the fitted SPY model of Table~\ref{tab:hyperparams}; jump-path entries use the stated single-episode and finite-horizon approximations. Sample values are computed
    from the in-sample SPY series; simulated values use $400$
    in-sample-length paths for the jump-episode quantities (forced-step
    fraction, jump-path probability, and forced fraction within a jump
    path) and $1{,}000$ paths for mean sample skewness. The sup-distance entry is
    the largest gap between the no-jump model's stationary distribution
    function and the in-sample empirical distribution function; for
    comparison, the two-sample Kolmogorov--Smirnov critical value used
    by the path-level tests at $\alpha = 0.05$ for two samples of
    $T = 2{,}766$ observations is $1.358\sqrt{2/T} = 0.0365$. Observed
    autocorrelations are those of the in-sample series shown in
    Fig.~\ref{fig:empirical_motivation}d. Stationary moments describe
    the no-jump mixture unless labeled otherwise. Here $A_t=|G_t|$
    and $\theta_2$ is the transition matrix's second-largest eigenvalue
    in modulus.}
\label{tab:supp-generator-quantities}
\footnotesize
\resizebox{\textwidth}{!}{% Generated by code/spy-experiment/SI-Derivation-Checks.jl; do not edit numbers manually.
\begin{tabular}{@{}lcc@{}}
\toprule
\textbf{Quantity} & \textbf{Closed form at the fit} & \textbf{Sample or simulation} \\ \midrule
\multicolumn{3}{@{}l}{\textit{Stationary law and moments}} \\[2pt]
State occupancy count, smallest and largest (of $T = 2{,}766$) & -- & $16$, $47$ \\
Occupancy frequency $\hat\pi_k$, smallest and largest & -- & $0.0058$, $0.017$ \\
$\max_k |\bar\pi_k - \hat\pi_k|$ & $3.7 \times 10^{-4}$ & -- \\
Mean ($\mathrm{yr}^{-1}$) & $0.0636$ & $0.0631$ \\
$\sup_x |F_{\rm mix}(x) - \widehat F_T(x)|$ & $0.0045$ & -- \\
Within-state variance $V_{\rm w}$ ($\mathrm{yr}^{-2}$), share of sample variance & $0.17$, $3.6\%$ & -- \\
Variance ($\mathrm{yr}^{-2}$), Student-$t$ emissions & $4.72$ & $4.60$ \\
Variance ($\mathrm{yr}^{-2}$), Gaussian emissions & $4.61$ & $4.60$ \\
Standard deviation ($\mathrm{yr}^{-1}$), Student-$t$ emissions & $2.17$ & $2.14$ \\
Excess kurtosis, Student-$t$ emissions & $8.34$ & $7.72$ \\
Excess kurtosis, Gaussian emissions & $5.82$ & $7.72$ \\
Skewness, HMM-NJ ($1{,}000$ simulated paths) & -- & $-0.76$ \\
Skewness, HMM-WJ ($1{,}000$ simulated paths) & -- & $-0.73$ \\[4pt]
\multicolumn{3}{@{}l}{\textit{Dependence without jumps}} \\[2pt]
$|\theta_2|$, smallest lag with $|\theta_2|^\tau < 0.01$ & $0.253$, lag $4$ & -- \\
ACF of $|G_t|$ at lag $1$, no jumps & $0.26$ & $0.30$ (observed) \\
ACF of $|G_t|$ at lag $3$, no jumps & $0.021$ & $0.31$ (observed) \\
ACF of $|G_t|$ at lag $5$, no jumps & $0.001$ & $0.26$ (observed) \\
ACF of $|G_t|$ at lag $25$, no jumps & $<\!0.001$ & $0.14$ (observed) \\[4pt]
\multicolumn{3}{@{}l}{\textit{Jump-episode arithmetic}} \\[2pt]
Forced-step fraction $\pi_J$ & $0.0089$ & $0.0088$ ($400$ paths) \\
Jump-path probability & $24.2\%$ & $24.8\%$ ($99$ of $400$) \\
Forced fraction within a jump path & $3.3\%$ ($\lambda / T$) & $3.6\%$ \\
Forced-step standard deviation ($\mathrm{yr}^{-1}$) & $5.38$ & -- \\
Mean absolute return, forced $\bar a_J$ and free $\bar a$ ($\mathrm{yr}^{-1}$) & $4.96$, $1.47$ & -- \\
$\Var(A_t)$, single-episode conditioning & $3.01$ & -- \\
Predicted jump-path ACF plateau & $0.13$ & -- \\
Excess kurtosis, single-episode jump path & $7.20$ & -- \\
Variance inflation from forced steps, single-episode jump path & $17\%$ & -- \\
\bottomrule
\end{tabular}
}
\end{table}

\subsection{Generative law}
\label{sec:supp-gen-law}

The no-jump model generates a hidden state path $S_0, \dots, S_T$ on
$\{1, \dots, N\}$ and observations $G_1, \dots, G_T$ with the joint
density:
\begin{equation}
  p(s_0, \dots, s_T,\, g_1, \dots, g_T)
  \;=\; \bar\pi_{s_0} \prod_{t=1}^{T}
  \mathbf{T}_{s_{t-1},\, s_t}\; f_{s_t}(g_t),
  \label{eq:supp-joint-law}
\end{equation}
where $\bar\pi_{s_0}$ is the initial probability of state $s_0$,
$\mathbf{T}_{s_{t-1}, s_t}$ is the probability of moving from state
$s_{t-1}$ to state $s_t$, and $f_k$ is the emission density of state
$k$; bold $\mathbf{T}$ is the transition matrix and italic $T$ the
number of observations. Because $\bar\pi$ is stationary for
$\mathbf{T}$, starting from an unemitted $S_0 \sim \bar\pi$ gives
$S_1 \sim \bar\pi$, the same law as drawing $S_1$ from $\bar\pi$ and
emitting it, as Algorithm~\ref{alg:hmmwj} does. Under
Eq.~\eqref{eq:emission}, $f_k$ is the location-scale
Student-$t$ density:
\begin{equation}
  f_k(g) \;=\; \frac{1}{\sigma_k}\,
  \psi_\nu\!\left(\frac{g - \mu_k}{\sigma_k}\right),
  \qquad
  \psi_\nu(x) \;=\;
  \frac{\Gamma\!\big(\tfrac{\nu+1}{2}\big)}
       {\sqrt{\nu \pi}\; \Gamma\!\big(\tfrac{\nu}{2}\big)}
  \left(1 + \frac{x^2}{\nu}\right)^{\!-\frac{\nu+1}{2}},
  \label{eq:supp-emission-density}
\end{equation}
where $\psi_\nu$ is the density of a standard Student-$t$ random
variable with $\nu$ degrees of freedom, $\Gamma$ is the gamma
function, and $(\mu_k, \sigma_k)$ are the location and scale of state
$k$. Given the state path, emissions are independent; serial dependence therefore comes from the state path~\cite{rabinerJuang1986}.

The jump-augmented model enlarges the hidden state. Let
$J_t \in \{0, 1, 2, \dots\}$ count the forced steps remaining after
time $t$, starting from $J_0 = 0$, and let $u$ be the forced-step
state distribution, which
mixes uniform draws over the two tail sets:
\begin{equation}
  u_k \;=\;
  \begin{cases}
    p_{\rm neg}/N_{\rm tail} & k \in \mathcal{S}_{-},\\[2pt]
    (1 - p_{\rm neg})/N_{\rm tail} & k \in \mathcal{S}_{+},\\[2pt]
    0 & \text{otherwise}.
  \end{cases}
  \label{eq:supp-u}
\end{equation}
Writing $p_K(m) = e^{-\lambda} \lambda^m / m!$ for the Poisson
probability of episode length $m$ and $\mathbf{1}\{\cdot\}$ for the
indicator function, the pair $(S_t, J_t)$ evolves by the transition
kernel:
\begin{equation}
\begin{aligned}
  &\Pr\big(S_t = k,\, J_t = j \mid S_{t-1} = s,\, J_{t-1} = i\big) \\
  &\qquad=\;
  \begin{cases}
    u_k\, \mathbf{1}\{j = i - 1\} & i \ge 1,\\[6pt]
    \big(1 - \epsilon + \epsilon e^{-\lambda}\big)\,
      \mathbf{T}_{s, k}\, \mathbf{1}\{j = 0\}
      \;+\; \epsilon\, p_K(j + 1)\, u_k & i = 0,
  \end{cases}
\end{aligned}
  \label{eq:supp-kernel}
\end{equation}
and the emission draw depends only on the first coordinate,
$G_t \sim f_{S_t}$.  While an episode is
active ($i \ge 1$), the next state is an independent draw from $u$ and
the counter decrements; the tail set and the state within it are
re-drawn at every forced step. While no episode is active ($i = 0$),
the chain either takes an ordinary transition (no trigger, or a
trigger whose Poisson draw returns zero) or starts an episode of
length $j + 1 \ge 1$ whose first forced step occurs immediately. The
pair $(S_t, J_t)$ is therefore itself a Markov chain, and the HMM-WJ
generator is a hidden Markov model on the enlarged state space
$\{1, \dots, N\} \times \{0, 1, 2, \dots\}$. Algorithm~\ref{alg:hmmwj}
implements Eq.~\eqref{eq:supp-kernel} with the episode counter
truncated at the simulation horizon. It emits the initial draw from
$\bar\pi$ as the first observation and applies the kernel from the
second step onward, so the first observed day can never be a forced
step; this is the only way its path law differs from the kernel's.
When an episode ends, ordinary transitions resume from the last forced
tail state, so the chain returns to the central states through the
tail rows of $\mathbf{T}$. This return links an episode to the free steps that follow it.

\subsection{Estimators and the one-step empirical law}
\label{sec:supp-gen-estimators}

The partition, transitions, emissions and initial law are estimated from sample summaries; the fitted law need not reproduce the observed return moments. For observations $G_1, \dots, G_T$, the Laplace
log-likelihood with location $\mu_L$ and scale $b_L$ is:
\begin{equation}
  \ell(\mu_L, b_L) \;=\;
  -T \ln (2 b_L) \;-\; \frac{1}{b_L} \sum_{t=1}^{T} |G_t - \mu_L|.
  \label{eq:supp-laplace-ll}
\end{equation}
For any fixed $b_L > 0$, maximizing $\ell$ over $\mu_L$ minimizes
$\sum_t |G_t - \mu_L|$, so $\hat\mu_L$ is the sample median. Setting
the $b_L$ derivative to zero at $\hat\mu_L$ gives:
\begin{align}
  \frac{\partial \ell}{\partial b_L}
   &\;=\; -\frac{T}{b_L}
     \;+\; \frac{1}{b_L^2} \sum_{t=1}^{T} |G_t - \hat\mu_L|
     \;=\; 0
     && \text{(first-order condition)} \notag \\
  \hat b_L
   &\;=\; \frac{1}{T} \sum_{t=1}^{T} |G_t - \hat\mu_L|,
     && \text{(solve for $b_L$)}
  \label{eq:supp-laplace-mle}
\end{align}
the mean absolute deviation about the median. The state boundaries
then invert the fitted Laplace cumulative distribution function
(CDF)~\cite{kotzKozubowskiPodgorski2001}. The Laplace CDF is
piecewise exponential:
\begin{equation}
  F_L(x) \;=\;
  \begin{cases}
    \tfrac{1}{2} \exp\!\big((x - \hat\mu_L)/\hat b_L\big)
      & x \le \hat\mu_L,\\[4pt]
    1 - \tfrac{1}{2} \exp\!\big(-(x - \hat\mu_L)/\hat b_L\big)
      & x > \hat\mu_L,
  \end{cases}
  \label{eq:supp-laplace-cdf}
\end{equation}
and setting $F_L(Q_k) = k/N$ and solving the lower branch for
$k \le N/2$ and the upper branch for $k \ge N/2$ gives the closed-form
boundaries:
\begin{align}
  \frac{k}{N} &\;=\; \tfrac{1}{2}\,
    e^{(Q_k - \hat\mu_L)/\hat b_L}
  \;\Longrightarrow\;
  Q_k \;=\; \hat\mu_L + \hat b_L \ln\!\frac{2k}{N},
    && \text{(lower branch)} \notag \\
  \frac{k}{N} &\;=\; 1 - \tfrac{1}{2}\,
    e^{-(Q_k - \hat\mu_L)/\hat b_L}
  \;\Longrightarrow\;
  Q_k \;=\; \hat\mu_L - \hat b_L
    \ln\!\Big(2 - \tfrac{2k}{N}\Big).
    && \text{(upper branch)}
  \label{eq:supp-quantiles}
\end{align}
The states are equiprobable under the fitted Laplace, not under the
empirical distribution, so the in-sample state counts are not uniform:
across the $100$ SPY states, they range from $16$ to $47$ of the
$2{,}766$ observations (Table~\ref{tab:supp-generator-quantities}).
For SPY, every state therefore held at least $16$ observations with a
nonzero spread, and neither fallback rule for the emission parameters
applied.

Row counting gives the maximum likelihood estimator (MLE) for the transition likelihood conditional on the initial assigned state. For the assigned state sequence
$s_1, \dots, s_T$, let $n_{jk}$ count the transitions from state $j$
to state $k$ and let $n_{j \cdot} = \sum_k n_{jk}$ be the row total.
The likelihood of the transition parameters factorizes over rows into
multinomials, and maximizing the log-likelihood subject to each row
summing to one, with multiplier $c_j$ for row $j$, uses the
Lagrangian:
\begin{equation}
  \mathcal{L} \;=\; \sum_{j, k} n_{jk} \ln \mathbf{T}_{jk}
    \;+\; \sum_{j} c_{j} \Big(1 - \sum_{k} \mathbf{T}_{jk}\Big),
  \label{eq:supp-lagrangian}
\end{equation}
whose first-order conditions solve as follows:
\begin{align}
  \frac{\partial \mathcal{L}}{\partial \mathbf{T}_{jk}}
   &\;=\; \frac{n_{jk}}{\mathbf{T}_{jk}} - c_j \;=\; 0
     && \text{(first-order condition)} \notag \\
  \mathbf{T}_{jk}
   &\;=\; \frac{n_{jk}}{c_j}
     && \text{(solve for $\mathbf{T}_{jk}$)} \notag \\
  c_j
   &\;=\; \sum_{k} n_{jk} \;=\; n_{j \cdot}
     && \text{(row constraint $\textstyle\sum_k \mathbf{T}_{jk} = 1$)} \notag \\
  \widehat{\mathbf{T}}_{jk}
   &\;=\; \frac{n_{jk}}{n_{j \cdot}}.
     && \text{(counted estimator)}
  \label{eq:supp-markov-mle}
\end{align}
For a row with $n_{j\cdot}>0$, the first-order condition applies to
cells with positive counts. A cell with $n_{jk}=0$ adds nothing to the
log-likelihood, so any probability placed on it is taken from observed
transitions and lowers the likelihood; its maximizer is therefore
$\widehat{\mathbf{T}}_{jk}=0$, which the same formula returns.
When $n_{j\cdot}=0$, the observations do not identify that row.
The implementation sets every entry to $1/N$; this is a fallback
choice, not an estimate of an observed transition frequency.
Multiplying $\widehat{\mathbf{T}}_{jk}$ by the empirical origin
frequency $n_{j\cdot}/(T-1)$ recovers $n_{jk}/(T-1)$, the empirical
frequency of the state pair $(j, k)$. Paired with the empirical
origin frequencies, the counted estimator therefore reproduces the
in-sample joint law of $(S_{t-1}, S_t)$ exactly; everything it
asserts about longer horizons is the Chapman--Kolmogorov extension
$\mathbf{T}^\tau$ of that one-step law~\cite{bilmes1998,
grinsteadSnell1997}. The third estimator sets the emission parameters:
within each state with at least two observations and sample standard
deviation at least $10^{-12}\,\mathrm{yr}^{-1}$, $\mu_k$ and
$\sigma_k$ are the sample mean and standard deviation of the assigned
observations. Otherwise, the implementation uses the fallback rules
stated in the main text and written out in
Eq.~\eqref{eq:emission-fallback} of Online
Appendix~\ref{sec:supp-fallback}. In either case, $\sigma_k$ enters
Eq.~\eqref{eq:supp-emission-density} as the scale, so the emission
standard deviation is $\sigma_k \sqrt{\nu/(\nu - 2)}$. The law of
total variance in the next subsection quantifies the aggregate effect
of this scale convention (Eq.~\eqref{eq:supp-variance}).

The stationary initial distribution is close to empirical state occupancy for the fitted SPY chain. Let $n_k$ count the observations
assigned to state $k$ over the $T$ observations, and let
$\hat\pi_k = n_k / T$ be the occupancy frequency. The number of
transitions out of state $j$ satisfies
$n_{j\cdot} = n_j - \mathbf{1}\{s_T = j\}$, and the number of
transitions into state $k$ satisfies
$n_{\cdot k} = n_k - \mathbf{1}\{s_1 = k\}$, because only the last
observation opens no transition and only the first closes none.
Substituting both identities, with italic $T$ the number of
observations, gives:
\begin{align}
  \big(\hat\pi\, \widehat{\mathbf{T}}\big)_k
   &\;=\; \sum_j \frac{n_j}{T} \cdot \frac{n_{jk}}{n_{j \cdot}}
     && \text{(definitions)} \notag \\
   &\;=\; \sum_j \frac{n_{j \cdot}}{T} \cdot \frac{n_{jk}}{n_{j \cdot}}
     \;+\; \sum_j \frac{\mathbf{1}\{s_T = j\}}{T}
       \cdot \frac{n_{jk}}{n_{j \cdot}}
     && \text{($n_j = n_{j\cdot} + \mathbf{1}\{s_T = j\}$)} \notag \\
   &\;=\; \frac{n_{\cdot k}}{T}
     \;+\; \frac{\widehat{\mathbf{T}}_{s_T, k}}{T}
     && \text{(sum transitions into $k$)} \notag \\
   &\;=\; \hat\pi_k
     \;+\; \frac{\widehat{\mathbf{T}}_{s_T, k}
       - \mathbf{1}\{s_1 = k\}}{T}.
     && \text{($n_{\cdot k} = n_k - \mathbf{1}\{s_1 = k\}$)}
  \label{eq:supp-fixed-point}
\end{align}
The occupancy frequencies therefore satisfy the stationarity equation
$\hat\pi \widehat{\mathbf{T}} = \hat\pi$ up to a fixed-point residual
bounded by $1/T$. Because the counted chain mixes within a few steps
(Online Appendix~\ref{sec:supp-gen-nojump}), a residual of this size
moves the solution by a comparable amount. For the fitted SPY model,
the solved eigenvector $\bar\pi$ deviates from the occupancy
frequencies by at most $3.7 \times 10^{-4}$, the same order as $1/T$,
against occupancy values
between $0.0058$ and $0.017$ (Table~\ref{tab:supp-generator-quantities}). Thus $\bar\pi$ differs from empirical occupancy by order $1/T$ for this rapidly mixing fit.

\subsection{Stationary distribution and moments}
\label{sec:supp-gen-moments}

The stationary observation law of the no-jump generator follows
directly from Eq.~\eqref{eq:supp-joint-law}. Marginalizing the state
at a single time gives the mixture density and CDF:
\begin{equation}
  p(g) \;=\; \sum_{k=1}^{N} \bar\pi_k\, f_k(g),
  \qquad
  F_{\rm mix}(x) \;=\; \sum_{k=1}^{N} \bar\pi_k\,
  \Psi_\nu\!\left(\frac{x - \mu_k}{\sigma_k}\right),
  \label{eq:supp-mixture}
\end{equation}
where $\Psi_\nu$ is the CDF of the standard Student-$t$ with $\nu$
degrees of freedom. For SPY, no emission fallback was needed, and stationary weights differed from occupancies through the endpoint correction in Eq.~\eqref{eq:supp-fixed-point}. Weighting the empirical state means
by the occupancy frequencies regroups into the grand mean:
\begin{align}
  \sum_{k:n_k>0} \hat\pi_k\, \mu_k
   &\;=\; \sum_{k:n_k>0} \frac{n_k}{T} \cdot \frac{1}{n_k}
     \sum_{t \,:\, s_t = k} G_t
     && \text{(definitions of $\hat\pi_k$, $\mu_k$)} \notag \\
   &\;=\; \frac{1}{T} \sum_{t=1}^{T} G_t,
     && \text{(each $t$ counted once)}
  \label{eq:supp-mean}
\end{align}
and replacing $\hat\pi$ by $\bar\pi$ changes the value only through
the $1/T$ correction of Eq.~\eqref{eq:supp-fixed-point}: the model
mean is $0.0636~\mathrm{yr}^{-1}$ against the sample mean of
$0.0631~\mathrm{yr}^{-1}$ (Table~\ref{tab:supp-generator-quantities}).
For other assets, this identity needs a qualification. A state with a
single observation takes the full-series mean as its location
(Eq.~\eqref{eq:emission-fallback}), which need not preserve
Eq.~\eqref{eq:supp-mean}; the other fallback rule changes only the
scale (Online Appendix~\ref{sec:supp-fallback}).
Evaluated at the fitted SPY model, the largest gap between the model
CDF and the empirical CDF is
$\sup_x |F_{\rm mix}(x) - \widehat F_T(x)| = 0.0045$
(Table~\ref{tab:supp-generator-quantities}), where
$\widehat F_T$ is the in-sample empirical CDF. The path-level tests
compare a simulated path with the data, so the relevant yardstick is
the two-sample Kolmogorov--Smirnov (KS) critical value at significance
level $\alpha = 0.05$ for two samples of $2{,}766$ observations,
$1.358 \sqrt{2/T} = 0.0365$. The model distance is about one eighth of this threshold; path sampling and dependence also affect the descriptive pass rates (Table~\ref{tab:model_comparison}).

The second moment makes the role of the emission scale convention
explicit. Conditioning on the state and applying the law of total
variance gives:
\begin{align}
  \Var(G_t)
   &\;=\; \E\big[\Var(G_t \mid S_t)\big]
     \;+\; \Var\big(\E[G_t \mid S_t]\big)
     && \text{(law of total variance)} \notag \\
   &\;=\; \frac{\nu}{\nu - 2} \sum_k \bar\pi_k\, \sigma_k^2
     \;+\; \sum_k \bar\pi_k\, (\mu_k - \bar\mu)^2,
     && \text{(Student-$t$ variance)}
  \label{eq:supp-variance}
\end{align}
where $\bar\mu = \sum_k \bar\pi_k \mu_k$ is the stationary mean.
Define the within-state variance $V_{\rm w} = \sum_k \bar\pi_k
\sigma_k^2$ and the between-state variance $V_{\rm b} = \sum_k
\bar\pi_k (\mu_k - \bar\mu)^2$. The corresponding empirical
decomposition of the sample variance is $V_{\rm b} + V_{\rm w}$, up to
the $n_k - 1$ versus $n_k$ denominators of the within-state variances
and the $\bar\pi$ versus $\hat\pi$ weights, so the Student-$t$ scale convention adds
$2 V_{\rm w} / (\nu - 2)$ relative to the corresponding Gaussian mixture. At the SPY fit
(Table~\ref{tab:supp-generator-quantities}),
$V_{\rm w} = 0.17~\mathrm{yr}^{-2}$, only $3.6\%$ of the total
variance of $4.60~\mathrm{yr}^{-2}$: most dispersion lies between states. The inflation is correspondingly small, a model variance of
$4.72$ against $4.60~\mathrm{yr}^{-2}$ ($2.17$ against
$2.14~\mathrm{yr}^{-1}$ in standard deviation), and it is absent under
Gaussian emissions, whose model variance of $4.61~\mathrm{yr}^{-2}$
differs from the sample value only through those denominator and
weight corrections.

The fourth moment explains the emission comparison in
Table~\ref{tab:emission_hsmm}. Write $G_t = \mu_k +
\sigma_k X$ within state $k$, where $X$ is a standard Student-$t$
draw, and let $\delta_k = \mu_k - \bar\mu$. Expanding the fourth
central moment binomially and dropping the odd-power terms in $X$,
which vanish by the symmetry of $\psi_\nu$, gives:
\begin{align}
  \E\big[(G_t - \bar\mu)^4\big]
   &\;=\; \sum_k \bar\pi_k\,
     \E\big[(\delta_k + \sigma_k X)^4\big]
     && \text{(condition on the state)} \notag \\
   &\;=\; \sum_k \bar\pi_k \Big[
     \delta_k^4 + 4\, \delta_k^3 \sigma_k\, \E(X)
     + 6\, \delta_k^2 \sigma_k^2\, \E(X^2) \notag \\
   &\hspace{6em}
     + 4\, \delta_k \sigma_k^3\, \E(X^3) + \sigma_k^4\, \E(X^4)\Big]
     && \text{(binomial expansion)} \notag \\
   &\;=\; \sum_k \bar\pi_k \left[
     \delta_k^4
     + 6\, \delta_k^2\, \sigma_k^2\, \E(X^2)\right. \notag \\
   &\hspace{6em}\left.{}
     + \sigma_k^4\, \E(X^4)
     \right],
     && \text{($\E X = \E X^3 = 0$)}
  \label{eq:supp-fourth}
\end{align}
The standard Student-$t$ moments are finite for $\nu > 4$ and satisfy:
\begin{equation}
  \E(X^2) \;=\; \frac{\nu}{\nu - 2},
  \qquad
  \E(X^4) \;=\; \frac{3 \nu^2}{(\nu - 2)(\nu - 4)}.
  \label{eq:supp-t-moments}
\end{equation}
At $\nu = 5$ these values are $5/3$ and $25$: the within-state
kurtosis is $9$ (excess kurtosis $6$), against $3$ (excess kurtosis
$0$) for Gaussian emissions. The excess kurtosis of the stationary mixture is
$\E[(G_t - \bar\mu)^4] / \Var(G_t)^2 - 3$; evaluating it at the SPY
fit gives $8.34$ with Student-$t$ emissions and $5.82$ with Gaussian
emissions, against an observed value of $7.72$
(Table~\ref{tab:supp-generator-quantities}). The emission choice enters
Eq.~\eqref{eq:supp-fourth} through both moments in
Eq.~\eqref{eq:supp-t-moments}; the larger change by far is
$\E(X^4)$ rising from $3$ to $25$, which acts on the $\sigma_k^4$
terms that the wide tail states dominate. The simulated HMM-WJ values ($7.5$ and $5.5$) are lower than these no-jump population values because jumps lower kurtosis and finite heavy-tailed samples underestimate it (Table~\ref{tab:emission_hsmm}). Asymmetric state locations, scales and occupancies also let the mixture reproduce skewness despite symmetric individual emissions (Table~\ref{tab:supp-generator-quantities}).

\subsection{Dependence structure without jumps}
\label{sec:supp-gen-nojump}
\label{sec:supp-transitions}

The no-jump generator loses return-magnitude dependence too quickly (Table~\ref{tab:supp-generator-quantities}). Within state $k$, the sojourn time is geometric:
the chain remains for $m$ steps with probability
$\mathbf{T}_{kk}^{\,m-1} (1 - \mathbf{T}_{kk})$ and therefore stays
for $1/(1 - \mathbf{T}_{kk})$ steps on average, between one and two
days for every fitted state
under the fitted transitions. Short visits alone do not establish rapid mixing; the chain also leaves either tail set within one to two steps on average. The tail sets were $\mathcal{S}_{-}=\{1,\ldots,5\}$ and $\mathcal{S}_{+}=\{96,\ldots,100\}$. Every SPY state was populated, so empty-row fallbacks did not cause these short visits. Let $a_k =
\E\big[\,|G_t| \,\big|\, S_t = k\big]$ be the within-state mean
absolute return and $\bar a = \sum_k \bar\pi_k a_k$ its stationary
average. Because observations are conditionally independent given the
states, the lag-$\tau$ product moment reduces to a state-pair sum:
\begin{align}
  \E\big[\,|G_t|\, |G_{t+\tau}|\,\big]
   &\;=\; \sum_{j, k}
     \Pr(S_t = j,\, S_{t+\tau} = k)\; a_j\, a_k
     && \text{(conditional independence)} \notag \\
   &\;=\; \sum_{j, k} \bar\pi_j\,
     \big(\mathbf{T}^\tau\big)_{jk}\; a_j\, a_k,
     && \text{(Chapman--Kolmogorov)}
  \label{eq:supp-acf-nj}
\end{align}
so the autocorrelation function (ACF) of $|G_t|$ is:
\begin{equation}
  \mathrm{ACF}_{|G|}(\tau)
  \;=\; \frac{\sum_{j, k} \bar\pi_j\,
    \big(\mathbf{T}^\tau\big)_{jk}\, a_j\, a_k \;-\; \bar a^2}
    {\Var(|G_t|)}.
  \label{eq:supp-acf-nj-def}
\end{equation}
For an ergodic, diagonalizable transition matrix, $\mathbf{T}^\tau$ approaches the rank-one limit with rows $\bar\pi$ at geometric rate $|\theta_2|^\tau$, where $\theta_2$ is the second-largest eigenvalue in modulus; the ACF is then bounded by a constant multiple of $|\theta_2|^\tau$. Direct evaluation of Eq.~\eqref{eq:supp-acf-nj-def} gives the fitted-chain decay reported here. For the fitted SPY
matrix (Table~\ref{tab:supp-generator-quantities}), $|\theta_2| = 0.253$, and $|\theta_2|^\tau$ falls below
$0.01$ from lag $4$ onward. Evaluating Eq.~\eqref{eq:supp-acf-nj}
gives a model ACF of $0.26$ at lag $1$, close to the observed $0.30$
because the counted matrix pins the one-step state-pair law, but
$0.021$ at lag $3$ and effectively zero from lag $5$, while the
observed ACF remains near $0.26$ at lag $5$ and $0.14$ at lag $25$
(Table~\ref{tab:supp-generator-quantities};
Fig.~\ref{fig:empirical_motivation}d). Matching counted one-step transitions fixes $\mathbf{T}$ and therefore its longer-lag dependence; the small $|\theta_2| = 0.253$ motivates the added duration mechanism.

\subsection{Jump-episode arithmetic}
\label{sec:supp-gen-renewal}

Renewal arguments quantify the frequency and duration of forced-tail episodes. On each step with no active episode, an episode with at
least one forced step begins with probability
$p_e = \epsilon (1 - e^{-\lambda})$, since a trigger whose Poisson
draw returns zero is discarded; conditional on starting, the episode
length is the zero-truncated Poisson draw with mean
$\E[K \mid K \ge 1] = \lambda / (1 - e^{-\lambda})$. At $\lambda =
90$, the correction $e^{-\lambda}$ is below $10^{-39}$, so $p_e =
\epsilon$ and $\E[K \mid K \ge 1] = \lambda$ to machine precision.
The long-run fraction of forced steps, a scalar denoted $\pi_J$,
follows from the renewal-reward theorem. A renewal cycle consists of
the free steps preceding an episode and the episode itself. The number
of steps up to and including the trigger is geometric with mean
$1/p_e$, but the triggering step is itself the first forced step, so
a cycle contains on average $1/p_e - 1 = (1 - p_e)/p_e$ free steps,
and the episode contributes $\E[K \mid K \ge 1]$ forced steps in
total:
\begin{align}
  \pi_J
   &\;=\; \frac{\E[K \mid K \ge 1]}
     {\dfrac{1 - p_e}{p_e} + \E[K \mid K \ge 1]}
     && \text{(forced steps over cycle length)} \notag \\
   &\;=\; \frac{p_e\, \E[K \mid K \ge 1]}
     {1 - p_e + p_e\, \E[K \mid K \ge 1]}
     && \text{(multiply through by $p_e$)} \notag \\
   &\;=\; \frac{\epsilon \lambda}
     {1 - \epsilon + \epsilon e^{-\lambda} + \epsilon \lambda}
     && \text{($p_e\, \E[K \mid K \ge 1] = \epsilon \lambda$)} \notag \\
   &\;\approx\; \frac{\epsilon \lambda}{1 + \epsilon \lambda}
     \;=\; 0.0089.
     && \text{(drop $\epsilon e^{-\lambda}$, then $\epsilon$)}
  \label{eq:supp-rho}
\end{align}
Simulation gave $0.0088$ across $400$ paths (Table~\ref{tab:supp-generator-quantities}). The product
$\epsilon \lambda$ sets this unconditional fraction, so the fraction
alone cannot separate the two parameters; Online Appendix~\ref{sec:supp-gen-tuning} explains why conditioning
on paths with an episode emphasizes duration while leaving the
trigger probability weakly determined.

At the path level, the same arithmetic determines how episodes are
distributed across the ensemble. A simulated path contains no episode
only if every one of its $T - 1$ transition steps fails to trigger
one, so the jump-path probability at $T = 2{,}766$ is
$1 - (1 - p_e)^{T-1} = 24.2\%$; the simulated ensemble
gives $24.8\%$ ($99$ of $400$ paths;
Table~\ref{tab:supp-generator-quantities}), consistent with the $\sim\!24\%$ jump-active share in the main comparison. We call these jump paths. Using the long-run episode rate, dividing the $T - 1$ transition
steps by the mean cycle length $(1 - p_e)/p_e + \lambda$ gives about $0.274$
episodes per path in expectation. The free waiting time is approximately
$1/\epsilon = 10^4$ trading days, about $40$ trading years, far
longer than the $11$-year path, so roughly three quarters follow the HMM-NJ law, while about one jump path in eight contains multiple episodes. Ignoring horizon truncation, one episode of mean duration gives a
forced fraction near $\lambda / T = 3.3\%$; the simulated mean is
$3.6\%$ (Table~\ref{tab:supp-generator-quantities}), with multi-episode paths increasing the average.
The contrast between forced and free steps explains the added magnitude dependence. Under the forced-step distribution $u$ of
Eq.~\eqref{eq:supp-u}, the conditional standard deviation is
$5.38~\mathrm{yr}^{-1}$ against the free-step $2.17~\mathrm{yr}^{-1}$,
a variance ratio of $6.1$, and the mean absolute return is
$\bar a_J = \sum_k u_k a_k = 4.96~\mathrm{yr}^{-1}$ against
$\bar a = 1.47~\mathrm{yr}^{-1}$
(Table~\ref{tab:supp-generator-quantities}). The same split describes
the state distribution of the augmented chain. On free steps the state
is distributed as $\bar\pi$, except during the few steps after an
episode ends, when the chain relaxes from the last forced tail state
back to $\bar\pi$ at the rate $|\theta_2|$. The long-run state marginal is therefore approximately $(1 - \pi_J)\,\bar\pi + \pi_J\,u$. Its forced-state contribution has weight $O(\pi_J) \approx 1\%$; this is not an error bound for the full augmented initial law, which also sets the episode counter to zero. Thus $\bar\pi$ supplies only an approximate stationary start.

\subsection{Autocorrelation and kurtosis under jumps}
\label{sec:supp-gen-tuning}

To explain what $(\epsilon, \lambda)$ control in Eq.~\eqref{eq:grid_search}, define the episode-overlap probability: the probability $h(\tau)$ that, given a step is forced, the
step $\tau$ later belongs to the same episode. A uniformly chosen
forced step falls in an episode of length $m$ with probability
proportional to $m \Pr(K = m \mid K \ge 1)$, because an episode of
length $m$ contains $m$ forced steps, and its position within that
episode is uniform. Combining the two gives:
\begin{align}
  h(\tau)
   &\;=\; \sum_{m \ge 1}
     \frac{m\, \Pr(K = m \mid K \ge 1)}{\E[K \mid K \ge 1]}
     \cdot \frac{(m - \tau)_+}{m}
     \notag \\
   &\;=\; \frac{\E\big[(K - \tau)_+ \,\big|\, K \ge 1\big]}
     {\E[K \mid K \ge 1]}
     && \text{(sum over $m$)} \notag \\
   &\;\approx\; \Big(1 - \frac{\tau}{\lambda}\Big)_{\!+},
     && \text{($K$ concentrated near $\lambda$)}
  \label{eq:supp-h}
\end{align}
where $(x)_+ = \max(x, 0)$ and the approximation holds because the
Poisson length concentrates within a few multiples of
$\sqrt{\lambda} \approx 9.5$ around $\lambda$. The same-episode
probability therefore decays almost linearly from one to zero over
the horizon $\lambda$.

The ACF follows by conditioning on the
forced-step indicator $Z_t = \mathbf{1}\{\text{step $t$ is forced}\}$.
Forced states are independent draws from $u$, followed by independent emissions from those states. Dependence between free observations, or between an episode and subsequent free steps, decays at rate $|\theta_2|^\tau$. Neglecting this dependence and post-episode relaxation leaves covariance between the phase means. Write
$A_t = |G_t|$ for the absolute return. Under this approximation, the conditional mean is
$\E[A_t \mid Z_t] = \bar a + Z_t\, (\bar a_J - \bar a)$,
and for rare episodes the joint forced probability is
$\Pr(Z_t = 1, Z_{t+\tau} = 1) \approx \pi_J\, h(\tau)$, so:
\begin{align}
  \Cov\big(A_t, A_{t+\tau}\big)
   &\;\approx\;
     \Cov\big(\E[A_t \mid Z_t],\;
              \E[A_{t+\tau} \mid Z_{t+\tau}]\big)
     && \text{(independence given $Z$)} \notag \\
   &\;=\; (\bar a_J - \bar a)^2\,
     \Cov\big(Z_t,\, Z_{t+\tau}\big)
     && \text{(conditional mean in $Z$)} \notag \\
   &\;\approx\; (\bar a_J - \bar a)^2\, \pi_J\, h(\tau),
     && \text{(episodes rare)}
  \label{eq:supp-acf-wj}
\end{align}
so the jump mechanism adds an ACF component of height $\pi_J (\bar a_J
- \bar a)^2 / \Var(A_t)$ that decays almost linearly to zero at lag
$\lambda$, precisely the slow component the counted chain cannot
produce. For a jump path carrying a single episode, the probability
that a step is forced is $\lambda / T$ rather than $\pi_J$, because
the path contains about $\lambda$ forced steps among its $T$ steps.
Replacing $\pi_J$ by $\lambda / T$ and using the absolute-return
variance of the fitted two-phase mixture, $\Var(A_t) = 3.01$, gives a predicted height
of $0.13$ at short lags, the plateau entry of
Table~\ref{tab:supp-generator-quantities}.
Among $99$ jump-active paths out of $400$, the predicted and simulated ACF differed by $0.013$ at lag $5$ and by less than $0.01$ from lag $10$ onward. At lag $1$, the simulation gave $0.35$ against the episode contribution of $0.13$, because the short-lived free-phase contribution was omitted from the approximation.

Jump episodes lowered excess kurtosis while strengthening
absolute-return autocorrelation
(Table~\ref{tab:supp-generator-quantities}). Treating a path with one
episode as a mixture with forced-step weight $\lambda/T$, and using
$u$ in place of $\bar\pi$ for that component in
Eq.~\eqref{eq:supp-fourth}, gave population excess kurtosis of $7.20$
against $8.34$ without jumps. Forced steps increased variance by
about $17\%$ by drawing more often from the existing tail states.
The fourth central moment did not increase enough to offset the
increase in squared variance, so standardized kurtosis fell. A mixture of jump-active and no-jump groups from $4{,}000$ training-length paths, with the jump-active weight varied from zero to one, lowered ACF-MAE together with KS and AD pass rates as the weight increased. Forced steps change how often existing tail emissions are sampled, not their tail exponent, which is set by $\nu$. At the fitted jump-active share, mean path excess kurtosis was $7.5$ against the observed $7.72$ (Table~\ref{tab:model_comparison}).
Conditioning on jump paths also limits parameter identification. Within a single-episode jump path, the ACF
component of Eq.~\eqref{eq:supp-acf-wj} has height proportional to
$\lambda / T$ and decays over roughly $\lambda$ lags. Across the
$25$ lags of the objective, its amplitude and variance normalization
depend on $\lambda$; the kurtosis shift uses the same forced fraction.
 The parameter $\epsilon$ moves
the objective only through the incidence of jump paths, which the
conditioning on jump paths removes, and through the small fraction of
multi-episode paths. The objective therefore weakly identifies $\epsilon$: the tuned $\epsilon = 10^{-4}$ lies at the lower edge of the
explored range for SPY and for all three cross-asset tickers, and the
tuned optima vary with the Monte Carlo sample (Online
Appendix~\ref{sec:supp-cross-asset}).

\section{Emission distribution and semi-Markov comparison}
\label{sec:supp-emission-hsmm}

To compare the emission and duration specifications, we evaluated
three variants on SPY (Table~\ref{tab:emission_hsmm}). The first is a
hidden semi-Markov model (HSMM) baseline adapted from Bulla and Bulla
\cite{bullaBulla2006}. It partitions the series into eight
Laplace-quantile states, the count with the highest in-sample KS pass
rate among $3$, $4$, $5$, $6$, and $8$ states. For each state, we fitted a negative-binomial count distribution
by moments of the observed visit lengths when at least five visits
had mean length above $1.01$ and variance above the mean, using a
geometric fallback otherwise. The simulator added one to each
nonnegative count to obtain a positive dwell time; fitting used the unshifted visit lengths, so simulation added one step to the fitted count mean. This convention also qualifies the duration comparison. Emissions were Student-$t$ with $\nu = 5$.
The other two variants are HMM-WJ with Normal emissions and HMM-WJ
with Student-$t$ emissions, both with $N = 100$ states. The HMM-WJ variants had higher training pass rates, but their $100$ states versus the HSMM's $8$ confound a duration-only comparison; the holdout favored the HSMM on pass rates. Its training ACF-MAE equaled HMM-NJ's, whereas both HMM-WJ variants improved it (Table~\ref{tab:model_comparison}). Holding the rest of HMM-WJ fixed, changing the emission
from Normal to Student-$t$ ($\nu=5$) increased simulated excess
kurtosis from $5.5$ to $7.5$, against an observed value of $7.7$
(Online Appendix~\ref{sec:supp-generator}). The emission change improved training marginal fit but barely changed temporal error or holdout pass rates.

\begin{table}[H]
\centering
\caption{\textbf{Three-way comparison of emission distributions and
    duration specifications for SPY.} Standard
    errors in parentheses; bold marks the best value per row, which for
    excess kurtosis is the value closest to the observed one. The HSMM
    is the hidden semi-Markov baseline adapted from Bulla and Bulla
    (eight states, negative-binomial dwell times with a geometric
    fallback, and Student-$t$ emissions) \cite{bullaBulla2006}; HMM-WJ:
    hidden Markov model with jumps
    ($N = 100$). All entries are computed from $1{,}000$ simulated
    paths; pass rates use significance level $\alpha = 0.05$. Kolmogorov--Smirnov
    (KS), Anderson--Darling (AD), and mean absolute error in the
    autocorrelation function (ACF-MAE) are defined in
    Online Appendix~\ref{sec:supp-validation}. Autocorrelation error
    compares the observed curve with the ensemble-average curve,
    using lags $1$--$252$ in training and $1$--$248$ in the holdout.}
\label{tab:emission_hsmm}
\small
\begin{tabular}{@{}lccc@{}}
\toprule
\textbf{Metric} & \textbf{HSMM} & \textbf{HMM-WJ} & \textbf{HMM-WJ} \\
                & \textbf{(Student-$t$)} & \textbf{(Gaussian)} & \textbf{(Student-$t$)} \\ \midrule
States              & $8$ & $100$ & $100$ \\[4pt]
\multicolumn{4}{l}{\textit{In-sample ($2{,}766$ trading days, $2014$--$2024$)}} \\[2pt]
KS pass rate (\%)   &  $82.0$ ($1.2$)         &  $97.0$ ($0.5$)         & $\mathbf{98.3}$ ($0.4$) \\
AD pass rate (\%)   &  $42.5$ ($1.6$)         &  $90.5$ ($0.9$)         & $\mathbf{94.8}$ ($0.7$) \\
Excess kurtosis     &   $4.8$ ($0.08$)        &   $5.5$ ($0.03$)        & $\mathbf{7.5}$ ($0.09$) \\
ACF-MAE             & $0.059$ ($<\!0.001$)    & $\mathbf{0.052}$ ($<\!0.001$) & $0.053$ ($<\!0.001$) \\
Wasserstein-$1$     & $0.176$ ($0.001$)       & $0.101$ ($0.002$)       & $\mathbf{0.097}$ ($0.001$) \\
Hellinger           & $0.113$ ($<\!0.001$)    & $0.076$ ($<\!0.001$)   & $\mathbf{0.074}$ ($<\!0.001$) \\[4pt]
\multicolumn{4}{l}{\textit{Out-of-sample ($249$ trading days, $2025$)}} \\[2pt]
KS pass rate (\%)   & $\mathbf{96.2}$ ($0.6$) &  $95.0$ ($0.7$)         &  $95.4$ ($0.7$) \\
AD pass rate (\%)   & $\mathbf{96.7}$ ($0.6$) &  $95.9$ ($0.6$)         &  $95.3$ ($0.7$) \\
Excess kurtosis     &   $4.1$ ($0.13$)        &   $4.9$ ($0.08$)        & $\mathbf{6.2}$ ($0.15$) \\
ACF-MAE             & $0.042$ ($<\!0.001$)    & $\mathbf{0.040}$ ($<\!0.001$) & $\mathbf{0.040}$ ($<\!0.001$) \\
Wasserstein-$1$     & $0.287$ ($0.002$)       & $\mathbf{0.275}$ ($0.005$) & $\mathbf{0.275}$ ($0.005$) \\
Hellinger           & $0.239$ ($0.001$)       & $0.212$ ($0.001$)       & $\mathbf{0.210}$ ($0.001$) \\
\bottomrule
\end{tabular}
\end{table}

\section{State-resolution sensitivity}
\label{sec:supp-sensitivity}

To test sensitivity to state resolution, we refitted the hidden
Markov models with and without jumps (HMM-WJ and HMM-NJ) to the
$2{,}766$-day SPY training history using
$N\in\{30,60,90,100,150,200\}$ states
using $1{,}000$ paths per fit, $\alpha=0.05$ and the ensemble-ACF convention. Each refit kept five states in each
tail set and repeated the $(\epsilon,\lambda)$ search. The no-jump model's
Kolmogorov--Smirnov (KS) and Anderson--Darling (AD) pass rates changed
by less than one percentage point, and its absolute-return
autocorrelation error (ACF-MAE) was unchanged at the reported
precision. For HMM-WJ, ACF-MAE fell from $0.057$ at $N=30$ to
$0.045$ at $N=200$, while the AD pass rate fell from $97.1\%$ to
$82.8\%$; intermediate values were not monotone. Increasing $N$
narrowed the fixed-size tail sets to more extreme quantile bands and
left fewer observations for each state's emission estimates, while
retuning also changed the jump parameters. The sweep therefore
measured their combined effect, without isolating the contribution of
each change. Fallback rules kept sparsely populated states defined
(Online Appendix~\ref{sec:supp-fallback}). The primary $N=100$ was a practical compromise; this SPY-only sweep did not establish an optimum for other assets or shorter histories.

\section{Cross-asset generalization}
\label{sec:supp-cross-asset}

Separate fits to NVDA (semiconductors), JNJ (health care) and JPM (financials) tested whether the SPY findings extended to other assets. Tuning used $(n_{\mathrm{paths}} = 200, \mathrm{seed} = 1234)$ and selected $(\epsilon^{*},\lambda^{*})=(10^{-4},70)$ for NVDA and JNJ and $(10^{-4},80)$ for JPM. With $1{,}000$ evaluation paths and $\alpha=0.05$, training KS pass rates were $98.9\%$, $99.4\%$ and $99.1\%$, respectively; on the $249$-day $2025$ holdout they were $97.7\%$, $74.0\%$ and $79.2\%$. Transfer was uneven, and Monte Carlo variation in the kurtosis component of the tuning objective makes these selected grid points sample-dependent.

\section{SIM composition derivations and properties verification}
\label{sec:supp-sim-derivations}

The composition identities below use sample moments of the supplied paths. Dropping the market cross-covariance is exact only when its sample value is zero (Online Appendix~\ref{sec:supp-cov}). To
derive Eq.~\eqref{eq:scale}, first center the full-return
draw as in Eq.~\eqref{eq:center-generator}, then set
$\eps_i=s_i\tg_i^{\mathrm c}$ for $s_i\in[0,1]$ and require
that the composed series have the same variance as the generator
path:
\begin{equation}
  \Var(g_i) \;=\; \Var(\alpha_i + \beta_i\, g_m + \eps_i)
            \;=\; \Var(\beta_i\, g_m + \eps_i)
            \;=\; \sigma_{\gen,i}^2,
\end{equation}
The constant $\alpha_i$ drops out; expanding the sum gives:
\begin{equation}
  \Var(\beta_i\, g_m + \eps_i) \;=\; \beta_i^2\, \Var(g_m)
  \;+\; 2 \beta_i\, \Cov(g_m,\, \eps_i)
  \;+\; \Var(\eps_i).
\end{equation}
Substituting $\eps_i = s_i\,\tg_i^{\mathrm c}$ gives
$\Var(\eps_i) = s_i^2\,\sigma_{\gen,i}^2$ and
$\Cov(g_m,\, \eps_i) = s_i\,\Cov(g_m,\, \tg_i^{\mathrm c})$. Under the
assumption $\Cov(\tg_i^{\mathrm c},\,g_m)\approx0$, the cross term vanishes and
the variance condition becomes
$\beta_i^2\, \sigma_m^2 + s_i^2\, \sigma_{\gen,i}^2 =
\sigma_{\gen,i}^2$. Solving for $s_i^2$ gives Eq.~\eqref{eq:scale}.

To derive Eq.~\eqref{eq:r2-target}, use the same zero-cross-covariance
assumption and write the $R^2$ of the composed series of
Eq.~\eqref{eq:sim-composition} regressed on the market:
\begin{equation}
  R^2 \;=\; \frac{\beta_i^2\, \sigma_m^2}{\beta_i^2\, \sigma_m^2 + \sigma_{\eps,i}^2}.
\end{equation}
Setting $R^2$ equal to the calibrated $R^2_{i,\real}$ and solving for
the residual variance gives the target $\sigma_{\eps,\mathrm{target}}^2$
of Eq.~\eqref{eq:r2-target}. This branch applies only to assets whose
calibrated $R^2$ is at least $R^2_{\mathrm{preserve}} = 0.80$. Only QQQ and SPYG crossed this threshold; the other $421$ assets used variance preservation. Across $100$ draws per tracker, median absolute loading and $R^2$ errors were $0.005$ and $0.001$. No training path on the variance-preserving branch clipped: $\rho_i = \beta_i^2 \sigma_m^2 / \sigma_{\gen,i}^2$ stayed below $1-f=0.90$.
A Gaussian check used $\beta\in\{0.8,1.0,1.2\}$ and $R^2\in\{0.80,0.85,0.90,0.95,0.99\}$, with observed SPY and zero intercept. Reference and composed residuals were independent Gaussian draws with variance $\sigma_{\eps,\mathrm{target}}^2$ from Eq.~\eqref{eq:r2-target}. Supplying this target as $\sigma_{\gen}^2$ made the scale in Eq.~\eqref{eq:r2-rescale} one; only the composed draw was centered. All cells used the intended branch; across $100$ replications per cell, median loading and $R^2$ errors were at most $0.0011$ and $0.0017$. Independent residuals still caused finite-path differences. This checked the branch formula, not full-return reuse or temporal fit; shared SPY and unequal centering also prevent interpreting its KS pass rates as calibrated tests.

To distinguish intercept retention from mean preservation, let
$\overline{g}_m$ be the sample mean of the supplied market path.
Centering and scaling give $\overline{\eps}_i=0$ exactly, and the
implementation retains the fitted intercept $\alpha_i$ in every
branch. The resulting mean and its change from the calibrated linear
mean on the same market path are:
\begin{align}
  \overline{g}_i
    &= \alpha_i+\beta_i^{\eff}\overline{g}_m, \notag\\
  \overline{g}_i-(\alpha_i+\beta_i\overline{g}_m)
    &= (\beta_i^{\eff}-\beta_i)\overline{g}_m.
  \label{eq:supp-clipped-mean}
\end{align}
Retaining the intercept therefore preserves this linear mean when the
loading is unchanged, and a clipped path has a different mean whenever
$\overline{g}_m\ne0$. The residual floor determines the clipped loading without an offsetting intercept adjustment. In the training comparison, the OLS
calibration also implies
$\alpha_i=\overline{g}_{i,\real}-\beta_i\overline{g}_{m,\real}$,
where the bars with subscript $\real$ denote observed training means.
The difference from the observed asset mean then separates into a
change in market mean and a change in loading:
\begin{equation}
  \overline{g}_i-\overline{g}_{i,\real}
  = \beta_i(\overline{g}_m-\overline{g}_{m,\real})
    +(\beta_i^{\eff}-\beta_i)\overline{g}_m.
  \label{eq:supp-training-mean-shift}
\end{equation}
Using the observed training SPY path with $\beta_i^{\eff}=\beta_i$
therefore matches the observed asset mean exactly. The stress transformation $g_m^{(\gamma)}(t)=\overline{g}_m+\gamma[g_m(t)-\overline{g}_m]$, with $\gamma\in\{1,2,3\}$, preserved the market mean while multiplying its standard deviation by $\gamma$. Holding calibrated marginals fixed, clipping affected $0.0\%$, $76.4\%$ and $95.2\%$ of evaluated composed paths across the $423$ assets, respectively; only clipping changed means in this check. A different supplied market mean can also change asset means without clipping. Small fitted intercept errors therefore reflect the imposed mean constraint (Table~\ref{tab:centering_control}).

Retaining $\alpha_i$ also does not make the OLS intercept fitted to a
finite composed path equal $\alpha_i$ exactly. Because the composed
path is $g_i = \alpha_i + \beta_i^{\eff} g_m + \eps_i$, the OLS slope
and intercept on a finite path are:
\begin{align}
  \hat\beta_i
    &\;=\; \frac{\Cov(\alpha_i + \beta_i^{\eff} g_m + \eps_i,\, g_m)}{\Var(g_m)}
    && \text{(OLS slope)} \notag \\
    &\;=\; \beta_i^{\eff} + \frac{\Cov(\eps_i,\, g_m)}{\Var(g_m)},
    && \text{($\Cov(g_m, g_m) = \Var(g_m)$)} \notag \\
  \hat\alpha_i
    &\;=\; \overline{g}_i-\hat\beta_i\overline{g}_m
    && \text{(OLS intercept)} \notag \\
    &\;=\; \alpha_i-(\hat\beta_i-\beta_i^{\eff})\overline{g}_m.
    && \text{(Eq.~\eqref{eq:supp-clipped-mean})}
\end{align}
With independent generator and market draws, convergent sample
moments, and nonzero limiting market variance, the scaled sample
cross-covariance vanishes as $T$ grows, so $\hat\beta_i \to \beta_i^{\eff}$ and
$\hat\alpha_i \to \alpha_i$. Here $\beta_i^{\eff}=\beta_i$ unless
Eq.~\eqref{eq:clip} clips the market loading. On the
variance-preserving branch, the composed variance equals
$\sigma_{\gen,i}^2$ when the sample cross-covariance is zero: without
clipping, $\beta_i^2 \sigma_m^2 + s_i^2 \sigma_{\gen,i}^2 =
\sigma_{\gen,i}^2$ by Eq.~\eqref{eq:scale}, and under clipping,
Eq.~\eqref{eq:clip} sets $\beta_i^{\eff}$ to give the same variance.
On the $R^2$-preserving branch, the composed variance is
$\beta_i^2 \sigma_m^2 / R^2_{i,\real}$, the value implied by the
calibrated $(\beta_i, R^2_{i,\real})$ rather than by the generator
marginal. On the variance-preserving branch, clipped or not, the
residual floor guarantees only that $\Var(\eps_i)\geq
f\,\sigma_{\gen,i}^2$; it does not guarantee that adding the market
term preserves the generator's tail distribution (Online
Appendix~\ref{sec:supp-residual-choice}).

The zero-sum constraint also fixes the residual contribution to the
terminal cumulative log return. With time step $\Delta t$, and because
the centered residuals sum to zero, summing the composed growth rates
gives the exact identity:
\begin{equation}
  \Delta t\sum_{t=1}^{T} g_i(t)
  = T\Delta t\,\alpha_i
    + \beta_i^{\eff}\Delta t\sum_{t=1}^{T}g_m(t),
  \qquad \sum_{t=1}^{T}\eps_i(t)=0.
  \label{eq:supp-zero-sum-terminal}
\end{equation}
For a fixed market path and effective loading, the residual contributes
no terminal uncertainty. Partial sums before $T$ can still vary. If
clipping changes the loading between draws, terminal returns can vary
through $\beta_i^{\eff}$, but the residual sum remains zero. Centering
uses the full horizon and is not equivalent to subtracting a fitted
population mean: if the same draw is extended to a longer horizon, its
first $T$ days receive a different centering constant and scale. These
paths are therefore constrained over a specified horizon and do not
represent unconstrained cumulative residual returns.

\subsection{Role of the centered full-return draw as the residual}
\label{sec:supp-residual-choice}

The hybrid composition places $\eps_i = s_i \tg_i^{\mathrm c}$, a
centered and rescaled draw of asset $i$'s full return, in the residual
slot of Eq.~\eqref{eq:sim-composition}. The distributions differ: the residual of a single-index decomposition is the
idiosyncratic component of the return, orthogonal to the market, while
$\tg_i^{\mathrm c}$ is distributed like the centered total return. Its variance and tails include observed market co-movement, although
the draw itself is independent of the supplied market path.
The required residual variance is $\sigma_{\gen,i}^2 - \beta_i^2 \sigma_m^2$ on the unclipped variance-preserving branch. Centering
gives the residual path an exactly zero sample mean
(Eq.~\eqref{eq:center-generator}). On the unclipped variance-preserving
branch, computing $\sigma_m^2$ in Eq.~\eqref{eq:scale} from the same
market path used in the composition, as in training, gives the
residual exactly the remaining variance. The composed variance also
contains the sample cross-covariance term, which need not vanish on a
finite path (Online Appendix~\ref{sec:supp-cov}). Using the frozen
training market variance, as in the six-method holdout evaluation,
adds a further variance difference if the simulated market variance
differs from that training value. Standalone generator validation does not validate the reused regression-residual shape.

At $\rho_i = 0$, the loading is zero, Eq.~\eqref{eq:scale} gives
$s_i = 1$, and the composed path is the centered generator path
shifted to the mean $\alpha_i$. It therefore has the same centered
empirical distribution as the generator path, and it equals the
uncentered generator path only if $\alpha_i$ equals the draw's sample
mean. For $\rho_i > 0$, the independent market contribution changes
even the centered distribution. To describe that change on the
unclipped variance-preserving branch, we treat $s_i$ and $\beta_i$ as
constants rather than recomputing them from each draw, and we work
with population moments conditional on the fitted parameters, which
the sample moments of a long path approach. Write the deviation of the
composed return from its mean as $X_m + X_i$. The market contribution
is $X_m = \beta_i\, g_m^{\mathrm c}$, where $g_m^{\mathrm c}$ is the
market growth rate minus its mean, and the asset contribution is
$X_i = s_i\, \tg_i^{\mathrm c}$, the scaled asset draw. Their variances
are $v_m = \rho_i\, \sigma_{\gen,i}^2$ and
$v_i = (1 - \rho_i)\, \sigma_{\gen,i}^2$, and independence factors the fourth-moment cross terms:
\begin{align}
  \E\big[(X_m + X_i)^4\big]
   &\;=\; \E(X_m^4) + 4\,\E(X_m^3)\,\E(X_i)
     \notag \\
   &\qquad + 6\,\E(X_m^2)\,\E(X_i^2)
     + 4\,\E(X_m)\,\E(X_i^3)
     \notag \\
   &\qquad + \E(X_i^4)
     && \text{(independence)} \notag \\
   &\;=\; \E(X_m^4) + 6\, v_m\, v_i + \E(X_i^4).
     && \text{($\E X_m = \E X_i = 0$)}
  \label{eq:supp-comp-fourth}
\end{align}

\begin{figure}[H]
  \centering
  \includegraphics[width=\textwidth]{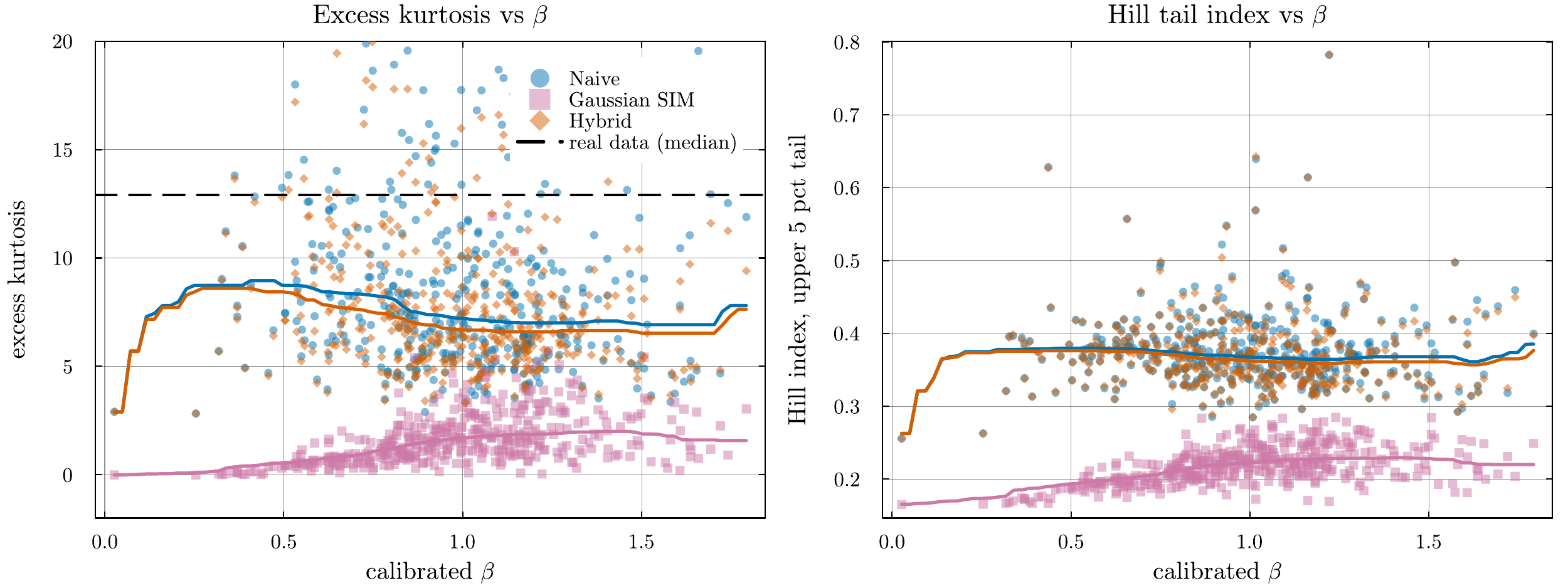}
  \caption{\textbf{Heavy-tail preservation: excess kurtosis and Hill
    tail index vs calibrated $\beta$.}
    Each point is a per-ticker median over $100$ replications.
    \emph{Left:} excess kurtosis, $y$-axis clipped to $[-2, 20]$ for
    readability. The dashed line marks the median real-data excess
    kurtosis across the universe ($\kappa_{\real} \approx 13$); the
    hybrid and naive composition methods (orange and blue) track each
    other, while the Gaussian single-index model (SIM) stays near
    $\kappa \approx 1$.
    \emph{Right:} Hill tail-index estimator on the upper $5\%$ of
    $|g|$. The hybrid and naive medians remain near $0.37$ throughout the
    $\beta$ range; Gaussian SIM sits near $0.22$. The shortfall from $\kappa_{\real}$ is present with and without
    variance correction; this similarity does not isolate the effect
    of rescaling. Both composed distributions depend on the market
    and generator variance shares (Eq.~\eqref{eq:supp-kurt-comp}).}
  \label{fig:tails}
\end{figure}

Writing $\kappa_m$ and $\kappa_{\gen}$ for the population excess kurtoses of the market and generator laws in this fixed-coefficient approximation, the component
fourth moments are $\E(X_m^4) = (\kappa_m + 3)\, v_m^2$ and
$\E(X_i^4) = (\kappa_{\gen} + 3)\, v_i^2$, where rescaling by $s_i$
leaves excess kurtosis unchanged. The excess kurtosis of the composed
return follows by dividing Eq.~\eqref{eq:supp-comp-fourth} by the
squared total variance:
\begin{align}
  \kappa(X_m + X_i)
   &\;=\; \frac{(\kappa_m + 3)\, v_m^2 + 6\, v_m\, v_i
     + (\kappa_{\gen} + 3)\, v_i^2}{(v_m + v_i)^2} \;-\; 3
     \notag \\
   &\;=\; \frac{\kappa_m\, v_m^2 + \kappa_{\gen}\, v_i^2
     + 3\, (v_m + v_i)^2}{(v_m + v_i)^2} \;-\; 3
     && \text{(collect terms)} \notag \\
   &\;=\; \kappa_m\, \rho_i^2 \;+\; \kappa_{\gen}\, (1 - \rho_i)^2.
     && \text{($v_m + v_i = \sigma_{\gen,i}^2$)}
  \label{eq:supp-kurt-comp}
\end{align}
The generator's excess kurtosis survives composition with weight
$(1 - \rho_i)^2$ and the market's enters with weight $\rho_i^2$; a check combining $200$ independent, standardized SPY-generator path pairs as $\sqrt{\rho_i}\,x+\sqrt{1-\rho_i}\,y$ matched the prediction within $0.03$ in mean excess kurtosis at $\rho_i\in\{0.1,0.3,0.5\}$. For a small market variance share, most generator excess kurtosis survives: at $\rho_i = 0.10$, it
retains weight $0.81$. As $\rho_i$ approaches one, the market contribution dominates.
The empirical quartiles group assets by loading rather than variance
share: median hybrid kurtosis falls from the lowest to the highest
$\beta$ quartile, as does naive kurtosis, and the hybrid KS pass rate
falls from $87.6\%$ to $65.9\%$ over the same range
(Table~\ref{tab:beta_bucket}). The Gaussian SIM, whose residual has no
excess kurtosis, stays far below both across the $\beta$ range
(Fig.~\ref{fig:tails}). On the clipped and $R^2$-preserving branches, the same formula uses the market's actual share of composed variance in place of $\rho_i$.

\begin{table}[H]
  \centering
  \caption{\textbf{Aggregate metrics by $\beta$ quartile, across composition methods.}
    Quartile cutoffs are computed from the calibrated $\beta$ distribution
    across the $423$-ticker universe. Pass rates pool equal numbers of paths per asset; other entries are medians across paths in the bucket. The GARCH rows use eligible fits among $393$ assets, while the other methods cover all $423$.}
  \label{tab:beta_bucket}
  \small
  \begin{tabular}{llrrrrr}
\toprule
$\beta$ bucket & Method & $|\hat\beta-\beta|$ & KS pass (\%) & $W_1$ & $\kappa$ & Var($g$) \\
\midrule
Q1 (low $\beta$) & Naive & 0.018 & 24.6 & 0.403 & 8.971 & 18.1 \\
 & Gaussian SIM & 0.015 & 0.0 & 0.620 & 0.622 & 14.0 \\
 & Hybrid & 0.016 & 87.6 & 0.177 & 8.369 & 14.6 \\
 & JumpHMM-on-residuals & 0.015 & 73.2 & 0.204 & 7.059 & 14.3 \\
 & Block bootstrap & 0.015 & 71.3 & 0.208 & 6.286 & 13.9 \\
 & GARCH(1,1)-$t$ & 0.015 & 57.8 & 0.253 & 5.255 & 13.4 \\
Q2 & Naive & 0.021 & 1.9 & 0.621 & 7.510 & 25.1 \\
 & Gaussian SIM & 0.016 & 0.2 & 0.653 & 1.509 & 18.0 \\
 & Hybrid & 0.017 & 74.9 & 0.231 & 6.889 & 19.0 \\
 & JumpHMM-on-residuals & 0.017 & 82.7 & 0.200 & 7.799 & 18.7 \\
 & Block bootstrap & 0.016 & 79.8 & 0.197 & 7.213 & 18.0 \\
 & GARCH(1,1)-$t$ & 0.015 & 76.4 & 0.227 & 6.470 & 17.8 \\
Q3 & Naive & 0.023 & 0.7 & 0.792 & 7.390 & 33.3 \\
 & Gaussian SIM & 0.017 & 1.4 & 0.688 & 1.947 & 23.4 \\
 & Hybrid & 0.018 & 66.1 & 0.279 & 6.904 & 24.4 \\
 & JumpHMM-on-residuals & 0.018 & 78.3 & 0.242 & 7.454 & 24.0 \\
 & Block bootstrap & 0.017 & 75.0 & 0.236 & 6.888 & 22.8 \\
 & GARCH(1,1)-$t$ & 0.016 & 71.0 & 0.263 & 6.265 & 21.9 \\
Q4 (high $\beta$) & Naive & 0.029 & 1.3 & 0.965 & 7.233 & 51.2 \\
 & Gaussian SIM & 0.022 & 0.6 & 0.825 & 2.054 & 36.3 \\
 & Hybrid & 0.022 & 65.9 & 0.339 & 6.739 & 37.1 \\
 & JumpHMM-on-residuals & 0.022 & 69.1 & 0.313 & 6.842 & 37.0 \\
 & Block bootstrap & 0.021 & 67.2 & 0.311 & 6.334 & 35.3 \\
 & GARCH(1,1)-$t$ & 0.021 & 64.5 & 0.361 & 6.064 & 34.7 \\
\bottomrule
\end{tabular}

\end{table}

\section{Uncertainty in the multi-asset jump comparison}
\label{sec:supp-jump-composition}

Replication-level contrasts confirmed that jump settings improved training temporal fit but worsened holdout fit (Table~\ref{tab:jump_ablation_uncertainty}). Four seeds supplied $1{,}000$ market replications per window with frozen fits. Naive and corrected composition shared draws; settings reused unchanged asset or market components, while enabled and disabled generators used separate random streams. We averaged each metric across tickers within a replication, differenced settings, and divided the standard deviation of the $1{,}000$ differences by $\sqrt{1{,}000}$; intervals used the mean plus or minus $1.96$ standard errors. This retained shared-market dependence but excluded fitting and historical-window uncertainty. All paths entered the primary scores; $247$ training-length and $22$ holdout-length market paths contained episodes, while the mean episode-containing shares among jump-enabled asset paths were $24.24\%$ and $2.46\%$. These rare holdout episodes and the transferred SPY calibration limit interpretation; KS non-rejection remains descriptive under temporal dependence.

\begin{table}[H]
  \centering
  \caption{\textbf{Monte Carlo uncertainty in the effect of jump settings
    under corrected composition.} Differences are relative to jumps
    disabled, with $1{,}000$ market replications per window.
    $\Delta$ACF$_{25}$ is the change in mean absolute error of the
    absolute-return autocorrelation function through lag $25$;
    negative values indicate improved temporal fit.
    $\Delta$KS is the change in Kolmogorov--Smirnov non-rejection
    rate in percentage points. Brackets contain approximate $95\%$
    Monte Carlo (MC) intervals conditional on the fitted models and
    observed histories.}
  \label{tab:jump_ablation_uncertainty}
  \footnotesize
  \setlength{\tabcolsep}{4pt}
  % Generated by 11c-Jump-Ablation-Tables.py; do not edit numbers manually.
\begin{tabular}{@{}lrr@{}}
\toprule
Jump setting & $\Delta$ACF$_{25}$ [95\% MC interval] & $\Delta$KS [95\% MC interval] \\
\midrule
\multicolumn{3}{@{}l}{\textit{Training: $2014$--$2024$, $423$ assets, $2{,}766$ days}} \\[2pt]
Market only & $-0.00725$ [$-0.00812, -0.00638$] & $-0.30$ [$-1.08, 0.49$] \\
Market + assets & $-0.01988$ [$-0.02063, -0.01913$] & $-12.25$ [$-13.02, -11.49$] \\
\midrule
\multicolumn{3}{@{}l}{\textit{Holdout: $2025$, $416$ assets, $249$ days}} \\[2pt]
Market only & $0.00195$ [$0.00104, 0.00287$] & $-0.04$ [$-0.41, 0.34$] \\
Market + assets & $0.00502$ [$0.00412, 0.00593$] & $-1.49$ [$-1.87, -1.12$] \\
\bottomrule
\end{tabular}

\end{table}

\clearpage
\section{Residual-centering control}
\label{sec:supp-centering-control}

To separate the variance correction from the different residual mean
constraints, we paired each native training path with an equivalently
centered version (Table~\ref{tab:centering_control}). The control used frozen models and ordinary least squares (OLS) calibrations, observed training SPY, seed $1234$, and $100$ draws of length $2{,}766$ per asset, retaining each method's native scales. With $g_i$ denoting a
native composed draw, the control removed only its residual mean:
\begin{equation}
  g_i^{\mathrm{ctrl}}(t)
  = g_i(t)-\frac{1}{T}\sum_{u=1}^{T}
    \left[g_i(u)-\alpha_i-\beta_i^{\eff}g_m(u)\right].
  \label{eq:supp-centering-control}
\end{equation}
The effective loading equaled the fitted loading except when the
hybrid rule changed it. Naive and hybrid paths already satisfied the
constraint and were reused without modification. No fits, random
draws, loadings, or scales changed between a native path and its
control. This paired translation left sample variance, recovered
loading, $R^2$, and excess kurtosis unchanged to numerical tolerance
for every evaluated path. The control assessed mean treatment; it did
not restore cumulative residual uncertainty or estimate parameter
uncertainty.
The common set contained $393$ assets with admissible GARCH fits, giving $39{,}300$ paired paths per method. The other $30$ fits had ARCH and GARCH coefficients summing to at least one. Pass rates pooled equal draw counts per asset; Wasserstein and intercept errors were medians across paths. Equivalent centering favored the residual-fit comparators while preserving the naive-to-hybrid improvement (Table~\ref{tab:centering_control}). Pass rates remain descriptive under serial dependence and the fixed residual sum.

\begin{table}[H]
  \centering
  \caption{\textbf{Paired control for residual centering on common
    training assets.} Native and centered results use the same
    $393$ assets and $100$ paths per asset. Centered paths have zero
    residual mean over $2{,}766$ observations. Kolmogorov--Smirnov (KS)
    and Anderson--Darling (AD) columns give non-rejection percentages
    at $\alpha=0.05$; higher values indicate closer empirical fit.
    $W_1$ is median Wasserstein-1 distance, and $|\hat\alpha-\alpha|$
    is median absolute ordinary least squares (OLS) intercept error, both in
    $\mathrm{yr}^{-1}$; lower values indicate closer agreement.
    Naive and hybrid already center each residual path.
    JumpHMM denotes the hidden Markov generator, with jumps disabled;
    SIM denotes the single-index model, and GARCH denotes generalized
    autoregressive conditional heteroskedasticity.}
  \label{tab:centering_control}
  \footnotesize
  \setlength{\tabcolsep}{3pt}
  % Generated by audits/2026-09-14-review/centering-correction/render_table.py
\begin{tabular}{lrrrrrrrr}
\toprule
 & \multicolumn{2}{c}{KS pass (\%)} & \multicolumn{2}{c}{AD pass (\%)} & \multicolumn{2}{c}{$W_1$} & \multicolumn{2}{c}{$|\hat\alpha-\alpha|$} \\
\cmidrule(lr){2-3}\cmidrule(lr){4-5}\cmidrule(lr){6-7}\cmidrule(lr){8-9}
Method & Native & Centered & Native & Centered & Native & Centered & Native & Centered \\
\midrule
Naive & 7.5 & 7.5 & 3.7 & 3.7 & 0.695 & 0.695 & 0.002 & 0.002 \\
Gaussian SIM & 0.6 & 0.9 & 0.5 & 0.6 & 0.675 & 0.673 & 0.047 & 0.002 \\
Hybrid & 74.4 & 74.4 & 65.1 & 65.1 & 0.246 & 0.246 & 0.002 & 0.002 \\
JumpHMM-on-residuals & 75.8 & 78.4 & 70.4 & 71.4 & 0.230 & 0.226 & 0.048 & 0.002 \\
Block bootstrap & 73.5 & 76.7 & 66.7 & 68.0 & 0.230 & 0.227 & 0.041 & 0.002 \\
GARCH(1,1)-$t$ & 67.4 & 70.0 & 61.4 & 63.0 & 0.267 & 0.262 & 0.047 & 0.002 \\
\bottomrule
\end{tabular}

\end{table}

\clearpage
\section{Pooled one-day tail thresholds}
\label{sec:supp-var-ensemble}

Pooled one-day tail thresholds used $B=5{,}000$ separately constructed $249$-day paths per asset and method (Table~\ref{tab:var}). Linear empirical quantiles pooled all $249B$ daily log returns. Each replication shared a market path across assets and methods; each asset path retained its centering and scaling rules, with hybrid scaling using training market variance. Hidden Markov models started from their fitted stationary laws with jumps disabled; GARCH retained its $100$-step warmup. Thresholds were calibrated for $423$ training assets ($393$ GARCH fits) before reading holdout data. The estimand is a uniformly selected day in this finite-horizon construction, not cumulative risk.
Pools of $100$, $500$, $1{,}000$, $2{,}500$ and $5{,}000$ paths tested numerical stability; $1{,}000$ independent paths checked simulated exceedances. Relative threshold changes divided the difference by the absolute $5{,}000$-path threshold. Doubling $2{,}500$ to $5{,}000$ changed mean observed exceedance rates by less than $0.006$ percentage points. Median absolute relative changes in $99\%$ thresholds were $0.12$--$0.30\%$ across methods, with $95$th percentiles $0.37$--$1.03\%$ and a worst case of $3.95\%$. Independent simulated exceedances were $5.03$--$5.04\%$ and $1.01$--$1.02\%$ against $5\%$ and $1\%$ targets. These checks establish simulation stability, not real-market coverage.
On the common $386$ holdout assets, corrected, residual-fit HMM, naive and Gaussian composition had mean $99\%$ exceedance rates of $1.08\%$, $1.14\%$, $0.85\%$ and $1.66\%$, respectively. Cross-asset standard deviations describe dispersion, not standard errors. Neither eligibility control nor one shared market period establishes equivalence or precise method rankings.

\section{Emission fallback inventory and sensitivity}
\label{sec:supp-fallback}

Emission fallbacks replace within-state moments when a state is sparse or nearly constant. For one asset, let $n_k$ be the
number of observations in state $k$, let $\bar G_k$ and $s_k$ be
its sample mean and standard deviation when defined, and let
$\bar G$ and $s_G$ be the corresponding full-series quantities.
The implemented emission rule was:
\begin{equation}
 (\mu_k,\sigma_k)=
 \begin{cases}
   (\bar G,s_G), & n_k<2,\\
   (\bar G_k,s_G), & n_k\ge2,\ s_k<\tau,\\
   (\bar G_k,s_k), & n_k\ge2,\ s_k\ge\tau,
 \end{cases}
 \qquad \tau=10^{-12}\,\mathrm{yr}^{-1}.
 \label{eq:emission-fallback}
\end{equation}
The scale $\sigma_k$ multiplies a standard Student-$t$ draw; it is
not the emission standard deviation. State intervals included their
lower boundary and excluded their upper boundary. A transition row
with no observed outgoing transition was set to $1/N$ in every
column. Such rows are unidentified by observed transitions. Among the $424$ full-return models, $39$ had
at least one fallback emission, affecting $50$ states. Of these,
$39$ states contained $11$--$119$ identical observations each,
seven were empty, and four contained one observation. The seven
empty states also supplied the seven uniform transition rows.
Fallback states carried $0.399$--$4.300\%$ of stationary probability
per affected asset. Sparse-state probability was at most $0.0363\%$
per asset; most fallback mass belonged to populated states with
repeated returns. SPY required no fallback, and neither did any of
the $423$ residual-fit models. Fallback mass is $100\sum_{k\in\mathcal F}\bar\pi_k$, where $\mathcal F$ is the set of fallback states. The per-asset inventory retained in the replication files uses $n_{\rm c}$ for observations in a constant state and $n_0,n_1$ for empty and singleton state counts. Its diagnostic $\Delta V$ is the percentage change in stationary mixture variance after replacing only populated constant-state scales by their empirical values; all fits used $N=100$ and $\nu=5$.

To isolate the consequence of broadening populated constant states,
we compared the fitted generators with an empirical-scale alternative
on all $39$ affected non-market assets
(Table~\ref{tab:fallback_sensitivity}). The alternative replaced
$s_G$ by $s_k$ only where $n_k\ge2$ and $s_k<\tau$; locations,
partitions, transitions, initial probabilities, and sparse-state
fallbacks were unchanged. A zero empirical scale gives a point mass
at the state mean.
We replayed the original $100$ paths of $2{,}766$ observations per
asset, paired by state sequence and random draws, with jumps disabled
and the observed training SPY path shared by both treatments.  We scored
the uncomposed generator and the naive and hybrid compositions;
each composition used its treatment's generator variance.  Pass fractions remain descriptive for these dependent series with ties.
The empirical-scale alternative improved marginal-fit scores on this
training subset (Table~\ref{tab:fallback_sensitivity}). Hybrid KS and AD pass rates rose by $10.28$ and $14.08$ percentage points, with Monte Carlo standard errors of $0.49$ and $0.60$. For the uncomposed
generators, the change in stationary mixture variance ranged from
$-0.651\%$ to $-6.520\%$ (Online Appendix~\ref{sec:supp-fallback}). The
fallback scale therefore affected both the generator's variance and
the composed marginal fit. All main comparisons retain the original
fallback rule. This paired training check did not test changes to
sparse-state estimates or validate the alternative on held-out
histories, so the alternative remains a training diagnostic.

\begin{table}[H]
 \centering
 \caption{\textbf{Paired sensitivity to the scale of populated constant states.}
 Each treatment used $100$ paired training-length paths for each of
 the same $39$ affected assets. Entries are means over these fixed
 assets and replications. The empirical-scale treatment changes only
 populated near-constant states, retaining the sparse-state fallback.
 Kolmogorov--Smirnov (KS) and Anderson--Darling (AD) pass fractions
 use $p>0.05$; their changes are percentage points. $W_1$ is the
 Wasserstein-1 distance to observed growth rates. Variance / observed
 divides each path's variance by its asset's observed training
 variance. ACF-MAE$_{25}$ is the mean absolute difference between
 each path's absolute-return autocorrelation function and the observed
 function over lags $1$--$25$, calculated before averaging paths.
 Paired change subtracts the fitted value from the empirical-scale
 value. Standard errors (SE) use the $100$ replication-level mean
 paired changes over the fixed asset set and describe simulation
 uncertainty, not uncertainty across historical markets.}
 \label{tab:fallback_sensitivity}
 \footnotesize
 \setlength{\tabcolsep}{4pt}
 \begin{tabular}{@{}llrrr@{}}
\toprule
Output & Metric & As fitted & Empirical scale & Paired change (SE) \\
\midrule
Uncomposed & KS pass (\%) & 98.64 & 99.82 & +1.18 (0.16) \\
 & AD pass (\%) & 98.90 & 99.72 & +0.82 (0.14) \\
 & $W_1$ ($\mathrm{yr}^{-1}$) & 0.1800 & 0.1696 & -0.0104 (0.0004) \\
 & Variance / observed & 1.0772 & 1.0553 & -0.0220 (0.0001) \\
 & ACF-MAE$_{25}$ & 0.14207 & 0.14183 & -0.00024 (0.00002) \\
\midrule
Naive & KS pass (\%) & 12.13 & 14.64 & +2.51 (0.22) \\
 & AD pass (\%) & 6.92 & 9.44 & +2.51 (0.23) \\
 & $W_1$ ($\mathrm{yr}^{-1}$) & 0.6234 & 0.5840 & -0.0394 (0.0002) \\
 & Variance / observed & 1.3650 & 1.3431 & -0.0219 (0.0001) \\
 & ACF-MAE$_{25}$ & 0.11663 & 0.11576 & -0.00086 (0.00002) \\
\midrule
Hybrid & KS pass (\%) & 68.64 & 78.92 & +10.28 (0.49) \\
 & AD pass (\%) & 56.15 & 70.23 & +14.08 (0.60) \\
 & $W_1$ ($\mathrm{yr}^{-1}$) & 0.2782 & 0.2474 & -0.0308 (0.0002) \\
 & Variance / observed & 1.0775 & 1.0556 & -0.0219 (0.0001) \\
 & ACF-MAE$_{25}$ & 0.10289 & 0.10145 & -0.00144 (0.00002) \\
\bottomrule
\end{tabular}

\end{table}

\end{document}